\documentclass[14pt,notitle]{article}
\begin{document}
\def\Bx{{\bf x}}
\def\BX{{\bf x}}
\def\Hx{\hat{{\bf x}}}
\def\hx{\hat{{\bf x}}}
\def\BrX{\breve{{\bf X}}}
\def\Ref#1{(\ref{#1})}
\def\emptyc{$\left.\right.$}
\newcommand{\Av}[1]{\langle #1 \rangle}
\newcommand{\Tr}{\mbox{Tr}}
\newcommand{\tr}{\mbox{tr}}
\def\DD{{\cal D}}
\def\Ap{{\bf A}_{\perp}}
\def\Ep{{\bf E}_{\perp}}
\def\Ev{E_{\Vert}}
\def\bfM{\bf}
\def\backspace{\!\!\!\!\!\!\!\!}
\thispagestyle{empty}
\
\vspace{-1cm}
\begin{center}{\Large
Moscow M. V. Lomonosov State University \\
Nuclear Physics D. V. Skolbeltsyn Institute 
}

\end{center}

\vspace{13mm}

\begin{flushright}
Manuscript Copyright 

Universal Decimal Classificator 530.145
\end{flushright}
\vspace{13mm}

\begin{center}
{\Large
TIMOSHENKO Edward Georgievich 
}
\vspace{13mm}

{\Large
Boundary Effects and Confinement\\ in the Theory of Nonabelian Gauge Fields
}

\vspace{13mm}

Speciality 01.04.02 -- theoretical physics 

\vspace{3mm}

Dissertation

for the academic degree of\\
candidate of physical--mathematical sciences\\
(equivalent to Ph. D.)

\end{center}

\vspace{3mm}

{\small
\begin{flushright}
\begin{tabular}{l}
Scientific Supervisor:\\
Associate Professor,\\
Candidate of physical-- \\mathematical sciences\\
{\normalsize Dr Sveshnikov~N.A.}\\
\end{tabular}
\end{flushright}
}

\vfill
\centerline{Moscow, 1995}
\newpage

\title{Boundary Effects and Confinement in the Theory of Nonabelian Gauge Fields}
\author{E. G. Timoshenko}
\begin{abstract}
The thesis is devoted to the problem of colour confinement in the non--Abelian Yang--Mills theory (gluon part of Quantum Chromodynamics). 
A generalisation of the 3-dimensional Fock--Schwinger gauge is proposed where the Gauss law constraint is exactly solvable. This simplifies the theory in a finite domain and incorporates the variables at the boundary into the Hamiltonian formalism.
The dependence of the partition function on the boundary value of the longitudinal component of the electric field is studied and related to the mechanism of the confinement--deconfinement transition.
The free energy density is calculated for $SU(2)$ and $SU(3)$ gluodynamics in the mean--field approximation for the collective variables. Analysis of its minima reveals a phase transition at a certain temperature, 
below which the mean collective variables have nonzero values.
This can be interpreted as a confinement--deconfinement phase transition. 
In the confinement phase the chromo--electric flux
through any element of the boundary is strictly zero. This means the singletness with respect to the group of the residual gauge transformations and hence impossibility of observing coloured objects at spatial infinity (in asymptotic states).
It is demonstrated that our confinement condition satisfies the traditional
confinement criteria. The
Wilson loop for $SU(N)$ theory is shown to satisfy the area law. The ratio of the transition temperature to the square root of the string tension coefficient is in a qualitative agreement with the result from lattice Monte Carlo simulations.
In the deconfinement phase the global symmetry $Z_{N}$ (centre of $SU(N)$) is spontaneously broken by the surface terms. The confinement phase is characterised by unbroken symmetry with all nontrivial minima having the same depth and transformable by $Z_{N}$ actions. (Translated from Russian manuscript, Moscow, 1995). 
\end{abstract}
\newpage

%
\contentsline {section}{\numberline {1}Introduction}{2}{}%
\contentsline {section}{\numberline {2}Some properties of the generalised Fock--Schwinger gauge}{12}{}%
\contentsline {section}{\numberline {3}Hamiltonian formalism of the Yang--Mills theory in a finite domain
}{18}{}%
\contentsline {section}{\numberline {4}Yang--Mills theory in the variables at infinity formalism
}{25}{}%
\contentsline {subsection}{\numberline {4.1}Some background from the formalism of the variables at infinity
}{26}{}%
\contentsline {subsection}{\numberline {4.2}Dynamics at infinity in the Yang--Mills theory
}{32}{}%
\contentsline {subsection}{\numberline {4.3}Poincar\'e algebra in the Fock--Schwinger gauge}{36}{}%
\contentsline {section}{\numberline {5}Dependence of the partition function on the variables at the boundary
}{37}{}%
\contentsline {subsection}{\numberline {5.1}Electrodynamics with an external charge}{38}{}%
\contentsline {subsection}{\numberline {5.2}Formulation in terms of the collective variables
}{41}{}%
\contentsline {subsection}{\numberline {5.3}Effective action of gluodynamics
}{43}{}%
\contentsline {section}{\numberline {6}Confinement criteria. The Wilson loop
}{53}{}%
\contentsline {section}{\numberline {7}Conclusion}{61}{}%
\contentsline {section}{\numberline {8}Appendices}{64}{}%
\newpage
\section{Introduction}

The problem of confinement in the non--Abelian gauge theory has attracted the keenest interest of researchers for several decades and it is one of the most urgent unresolved issues for the justification of the Standard Model of strong and electroweak interactions.
So far, all attempts to search for free quarks lead to the unambiguous conclusion that they are not observed in the scattering experiments of the elementary particles at energies up to the order of dozens of GeV \cite{Cutts78,Lyons80}.
Back in the octet model of quarks, it was assumed that objects with non-zero 
colour charge are bound by powerful attractive forces that do not allow them to exist in free asymptotic states, and therefore be observed experimentally.
In order to truly hold particles in a small region of of space, the interaction potential should be linearly increasing with the separation \cite{Grib92}.

Quantum chromodynamics (QCD), based on the theory of the Yang--Mills \cite{Yang54}  
non--Abelian gauge fields \cite{Frit73,Wein73} is considered a well--established model 
for the description of strong interactions for over twenty years. The discovery of the phenomenon of the asymptotic freedom 
\cite{Gros73,Polit73,Colem73} has led to notable advances in the application of the perturbation theory to describe processes at high energies \cite{Marc78,Bura80}.
However, the essential non--perturbative nature of the phenomena
at low energies has not yet allowed us to reach a clear understanding 
and the explanation of a number of problems, among which we should mention 
the confinement, the spontaneous chiral invariance breaking and the hadron mass spectrum.
Despite significant progress in the study of many important aspects 
of the confinement problem \cite{Mand80,Band81},
in the scientific community there is still no firm agreement between different approaches that attempt to derive confinement from the fundamental laws of QCD.

Allow me to present a brief overview of these approaches, without attempting to follow a chronological sequence, and only focusing on those that I consider to be of greatest theoretical interest.
Of course, the literature on the present issue is extremely extensive,
therefore my presentation does not pretend to be exhaustive.

Although the concept of confinement was originally introduced to explain
why quarks can only be observed as constituents of hadrons, 
confinement must explain, first of all, the unobservability of coloured states of quanta of the gauge field -- gluons.
The most common confinement criterion is based on the behaviour of the 
the Wilson loop, which characterises the potential energy of the interactions between two trial static sources of colour charge, though confinement of gluons does not follow from the latter.
It is reasonable to expect that the essence of the confinement mechanism
owes its origin precisely to the dynamics of gauge fields.
Therefore, we restrict ourselves to the consideration of quantum gluodynamics
and we shall not touch the question of the inclusion of dynamical quarks at present.

Numerous attempts have been made to study the confinement problem
in mathematically simpler theories, which are prototypes of 
QCD. First of all, confinement exists in gluodynamics in two
spatial dimensions where it is, however, fulfilled automatically,
since the Coulomb potential of a point charge in two dimensions is 
linearly growing. Nevertheless, a lot of 
interesting physical information on the spectrum of states of the theory has been obtained from these studies \cite{tHoo74}.

The study of compact electrodynamics in three dimensions has been intensively carried out both analytically [15-18]
and numerically \cite{Amar90}. This theory has a number of interesting properties,
such as confinement, dynamical mass generation, and a non--zero coefficient
of string tension. It is remarkable that in this theory the monopoles are instanton solutions 
and taking into account their contribution already gives the correct value of the Wilson loop. 
In other words, the condensation of instantons explains the confinement phenomenon in the compact QED$_{3}$. The latter can be interpreted as a dual Meissner effect in the theory of superconductivity and is accompanied by the appearance of Abrikosov vortices. In this case, the magnetic flux is trapped in narrow tubes connecting the monopole--antimonopole pairs. Another interesting analogy is with the theory of superfluidity. The confinement phase corresponds to the normal phase of  $^{4}He$  and has zero dielectric permittivity \cite{Poly87}. The ideas born in the framework of compact QED$_{3}$,
such as the consideration of monopoles and instantons, have subsequently played a significant role in the search for a confinement mechanism in 4-dimensional chromodynamics.

The first evidence in favour of confinement in QCD was the growth of the
the running coupling constant at large distances, 
found in starting orders of the perturbation theory, and the presence
in the theory of severe infrared divergences \cite{Itzy80,Chen84}. A physical picture that
can be associated with a linearly growing potential was proposed in Ref.  \cite{Namb74}. It was supposed that the electric flux between particles in a
quark-antiquark pair is squeezed into a narrow tube (string), the thickness of which
in the transverse direction does not depend on the distance between quarks.
The idea of the string model in QCD looked attractive also because of the agreement of the
of the spectrum of some baryons with the Regge trajectories picture \cite{Jacob74}.

S.~Mandelstam \cite{Mande76,Mande79} proposed a model
of the vacuum as a condensate of gluons and light quark--antiquark pairs
by analogy with the theory of superconductivity, where the condensation of electron pairs
leads to the Meissner effect -- the exclusion of the magnetic flux from the domain. 
In non--Abelian theory, monopoles exist as natural classical solutions and do not require their explicit introduction at the level of elementary particles. As an example of a theory in which monopole solutions are known one can cite the Georgi--Glashow model,
where the gauge symmetry is broken to an Abelian subgroup by the Higgs mechanism \cite{THoo76}. Based on the ideas of the electric--magnetic duality \cite{THoo79,Mand79}
the classification of phases in the $SU(2)$ gauge theory was given: perturbative; Georgi--Glashow one; phase with complete spontaneous symmetry breaking; confinement phase.
In the above mentioned works the evidence was obtained that the 
variational estimate for the energy of the vacuum containing a gas of monopoles is 
less than that for the perturbative vacuum. 
Nevertheless, the monopole approach has encountered a number of serious 
technical problems and the work in this direction continues
both analytically and with the help of numerical methods on the lattice
(see for example Refs.  \cite{Maeda90,Monde92,Suzu93}).

A closely related approach is based on the circumstance that the QCD vacuum
is unstable with respect to the introduction of a nonzero mean value 
of the magnetic field. G.~Savvidi \cite{Savv77,Matin78} calculated the energy correction 
in the one--loop approximation and found that it has the form $\Delta E \propto -
{\cal H}^{2} \log {\cal H}$. However, immediately there arose a question about the Lorentz
invariance of such a vacuum. 
There were considerable attempts to complicate the structure of the vacuum (periodic structure, ``spaghetti'', etc.), later called the ``Copenhagen'' vacuum,  and thus guarantee its Lorentz invariance [34-37]. 

I would like to note the work of Yu.A.~Simonov \cite{Simon88}, in which it is supposed that 
the QCD vacuum contains quasiclassical background fields, but instead of explicitly specifying the field configurations some assumptions are made about the bilocal correlators of the background fields at large distances.
Then the Fock--Schwinger covariant gauge is used to relate the field strengths to the gauge potential fields.  Although this approach is somewhat phenomenological, it reproduces reasonable values of the string tension and the magnitude of the condensates.
The approach proposed in this thesis, while different in most technical details,
shares one common idea with the former  work -- in order to understand confinement it is  more convenient to work with the effective action of the strength fields, which is particularly straightforward in the Fock--Schwinger gauge (the 3-dimensional version in our case).

It seemed very tempting to study the effect of instantons on the confinement 
 --- those are classical solutions of the Yang--Mills theory with a finite Euclidean action. One can consider as the most advanced results in this direction, apparently,
those achieved in Refs.  \cite{Calla78,Calla791,Calla792}. In four dimensions
instantons depend on the dimensional parameter $\lambda$. Assuming that it
may be arbitrary, the theory suffers from serious infrared  divergences. 
On the other hand, it is shown that the explicit introduction of a
cut--off leads to the short--range character of the instantons and their inability 
to produce the confinement phenomenon. It was proposed to consider the plasma of 
``merons'' --- long-range objects, 3-dimensional  cross sections of which are monopoles, whereas a pair of merons is topologically equivalent to an instanton. The possibility of explaining the confinement by means of instantons in QCD was questioned in Ref. 
\cite{Witt79} and apparently, although they do improve the
perturbative vacuum, the instantons alone are insufficient for the confinement.

An interesting approach is based on an attempt to apply a 
non--perturbative Ansatz in the Schwinger--Dyson (Bethe--Salpeter) equations.
The fact of the existence of confinement certainly has profound consequences on the asymptotics of the Green's functions and their analytic structure (see e.g. \ 
\cite{Oehme89,Oehme90,Oehme93} and references therein).
The interaction energy of static quarks can be represented
as the Fourier image of the temporal component of the gluon propagator
$$E(R) \sim \int d{\bf k}\, \exp (-i{\bf k}{\bf R})\, 
\Delta_{00}({\bf k}^{2}) \propto R^{\alpha-3}, 
\quad \Delta_{00}({\bf k}^{2}) \propto  k^{-\alpha},\ k\rightarrow 0.$$
The free propagator gives $\alpha =2$ and hence the Coulomb's law.
Confinement requires an infrared singularity of the propagator with
$\alpha = 4$. In finite order perturbation theory, the maximum
the singularity that can be achieved
is $k^{2}\log k^{2}$. The calculations of Ref.  \cite{Mand79ds} were performed
in the covariant gauge, and in Refs.  \cite{Anish79,Ball80} in the axial gauge.
Despite some differences in the details, they were based on a truncation of the
Dyson--Schwinger equation and neglecting some terms.
Numerical analyses of the resulting equations gave encouraging results,
but there was some disagreement between the closure schemes in the 
chain of equations in different gauges, so the method can not be considered
as systematic.

The ability to work directly with the Wilson loop $W[C]$
is an advantage of the effective loop theories derived from QCD
\cite{Migd83}. Their idea is based on obtaining integro--differential equations for $W[C]$ upon changing the shape of the loop parametrized by the
coordinates $x_{\mu}(s)$ as in Refs.  \cite{Make79,Gerv79}.
The interpretation in terms of a string is given by an approximate equation of the
type 
$$\frac{\delta^{2}W[C[x(s)]]}{\delta x_{\mu}(s)\delta x^{\mu}(s)}
=\chi^{2}\left( \frac{dx}{ds}\right)^{2}\,W[C]+ \ldots$$
Note that the loop approach deals with mathematically ill--defined products of singular
operators and is therefore ill--suited for the development of further systematic computational schemes. We note also the alternative loop approach of Refs.  \cite{rusIros90,rusIros91},
which can be very promising in combination with the 
$1/N$ decomposition ($N\rightarrow \infty,\  g^{2}N=const$),
introduced in Refs.  \cite{tHoo74N,Witt79N}.

As we have already mentioned, the Wilson criterion does not follow from the confinement of gluons in general. A more systematic approach to the phenomenon in question should explain the structure of the state space of the system, which in the confinement phase is a subspace, 
distinguished by some additional singletness condition. 
Such a condition cannot simply be the condition of global singletness. The latter does not forbid, for example, a state in which the quark
and the antiquark are separated by macroscopically large distances and can therefore be
separately observable. An attempt to construct such an algebraic theory of 
confinement was undertaken by T.~Kugo and I.~Ojima\cite{Kugo793}
on the basis of their proposed explicitly covariant operator formulation of the Yang--Mills theory \cite{Kugo78,Kugo792} in the framework of the BRST formalism,
which is a natural generalisation of the Gupta--Bleuler formalism in QED.

The states of the physical space in their formalism nullify the generators of BRST
transformations and scale transformations of the ghosts, which thereby ensures the
unitarity of the $S$-matrix. The confinement mechanism proposed by Kugo and Ojima 
is called quartet (due to the transformation properties of the asymptotic 
fields) and their confinement criterion consists of two conditions ensuring the cancellation of the ghost degrees of freedom.
The physical meaning of these conditions consists in the requirement of the absence of massless single--particle Nambu--Goldstone mode interacting with the BRST current.
Given that this criterion is fulfilled, the coloured asymptotic states are not observable and
can exist only in the unphysical sectors of the Hilbert space.
A similar formalism can be extended to the theory at non--zero temperature 
\cite{Hata93}, but in some papers there are claims that its ``kinematic'' character of confinement is trivial and the formalism itself is internally contradictory. For example, in Ref. \cite{Shinta84} it is argued that there are no representations of the extended BRST algebra
satisfying the Kugo--Ojima conditions. It is known that the quartet mechanism works well in the perturbative Higgs phase \cite{Kugo791}, but it apparently cannot be considered as a general confinement criterion.

The theory of gauge fields on a lattice \cite{Wils74,Kogu75} has proved to be one of the most powerful tools for studying the confinement problem. First of all, it was found that 
lattice theory in the strong coupling approximation gives the area law
in the Wilson loop, i.e. possesses confinement.
The investigation of this approximation at non--zero temperatures showed that the colour is not confined at high temperatures, and that both regimes are separated by a phase transition \cite{Poly78,Sussk79}.    
The real interest is indeed in the continuum limit of lattice theory,
which must be carried out in the neighbourhood of the zero coupling constant.
For this reason, the strong coupling approximation does not allow us to draw conclusions 
about the continuous theory. Significant progress in the study of lattice theory
at weak coupling was obtained based on the Monte Carlo method 
\cite{Creu802}. It has allowed the authors to estimate the values of the phase transition temperature and the string tension in the continuous limit through the application of the
renormalisation group techniques [66-69]. 
A natural order parameter in the lattice theory is  
the mean Polyakov line, 
$${\cal P}({\bf x}) = \mbox{tr\ T-}\exp\left( \int_{0}^{\beta}dt\,
A_{0}(t,{\bf x})\right),$$
and the correlator of two Polyakov lines
at different points allows us to determine the correlation length and hence 
deduce the type of the phase transition \cite{Mari89pr,Svet86}.
Work in this direction has revealed the important role of the global group $Z_{N}$ of 
the centre of $SU(N)$, which was first realised in the  approach based on the ideas of duality.
There is a simple connection of the confinement--deconfinement phase transition
with the general theory of critical phenomena \cite{Svet82}, which is accompanied by a
spontaneous breaking of $Z_{N}$ symmetry in the deconfinement phase.
It is found that the phase transition is a transition
of the second order in $SU(2)$ and of the first order in $SU(3)$
gauge theories \cite{Svet822,Enge82,Jaco86,Mari89}. At present moment these results are considered to be firmly established in the lattice theory
(see e.g. \ \cite{Kars92} for an overview of the current situation),
but the order of the phase of the transition can change in the continuum limit.
The main efforts of the groups presently performing calculations on increasingly
powerful supercomputers, are focused on problems of the next order of complexity --- inclusion of quarks and the computation of the spectrum of bound states in the theory.

Despite astonishing advances in computational methods, there is still considerable dissatisfaction among theorists.
First of all, it is not certain whether the qualitative properties of the phase transition persist 
in the continuous limit  (not to mention quantitative ones, although by applying reasonable renormalisation methods some characteristic quantities have been found to agree with known experimental values). The situation improves markedly with increasing accuracy, nevertheless, 
the main problem is that numerical values are obtained without understanding the underlying physics of the phenomena, and so far none of the of the proposed theoretical schemes have been able to demonstrate their full agreement with the lattice data, or predict a result that would subsequently find a good confirmation numerically.

For these reasons, novel approaches to the explanation of confinement are being offered yet again. In our opinion, a generic variational approach may turn out to be promising as such. The main difficulty there consists in the proper choice of the trial functional and the 
the possibility of exact calculation of the mean values with its weight. In Ref.  \cite{Koga941} the trial states are explicitly gauge invariant and reduce to Gaussian ones at zero coupling constant. It is found that the energy minimum is reached at a value of the variational
parameter away from the perturbative one. Another possibility 
\cite{rusSvesVar} is to work in terms of only the physical degrees of freedom and a
positively defined Hilbert space, thereby  reducing the arbitrariness in the choice of trial functions and giving legitimacy to the to the standard variational inequalities (Bogoliubov--Gibbs, Feynman, etc.).

The work of Ref. \cite{Khve94} emphasises the importance of explicitly resolving the
Gaussian constraint in order to describe the confinement. Note that the way of resolving the 
constraint (in the formalism without fixing a gauge) and the choice of variables 
in that paper differs from the one used by us here.

As a criticism of some claims about the triviality of confinement
(for example, in view of the possibility to work completely in terms of 
of gauge--invariant objects) \cite{Lunev92,Lavel93}, and thereby
the very absence of the problem, we note that they do not explain at least two facts firmly established on the lattice: the linear interaction potential of a 
quark--antiquark and the absence of confinement at high temperatures.

To conclude the review, allow me to focus on the mechanism of the so--called 
$A_{0}$-condensation. Although, in our opinion, it has not led to major advances
and its very methodology is not quite justified, nevertheless, the main object of its study coincides with the that of our present approach, namely the effective 
action expressed in terms of the temporal component of the gauge field $A_{0}$ 
is the main descriptor of the confinement--deconfinement phase transition. 
The main idea of this approach is to explain the non--zero mean value of the Polyakov line in the deconfinement phase by a non--zero mean value of $A_{0}$,
to which we attribute the role of the Higgs field spontaneously breaking 
$Z_{N}$ \cite{Dahl85,Mandul88} symmetry. As analytical methods
were plagued by severe infrared divergences there was 
a hope that a non--zero mean of this field could serve as a natural infrared regulariser [84-87]
and might help remedy the problem.
The results of calculations of the effective potential in the first loops
carried out in these studies found a non--zero mean of the field in the deconfinement phase.
In Ref.  \cite{Belya91} it was stated that this result is not gauge invariant and that for the
the effective action, expressed in terms of the Polyakov loop, the condensation
is absent, which in turn was criticised in Ref.  \cite{rusSkal92}. 
For our presentation the ultimate resolution of this controversial issue is of no direct interest. 
In any case, the perturbative nature of those calculations,
the lack of a consistent treatment of the surface terms and dealing with the infrared divergence in a rather careless manner --- these are the defficiences which allow us to question the validity of those results. 
As we shall see below, the very fact that there is a non--zero 
mean $A_{0}$ for a theory that correctly accounts for the surface terms, does not imply the occurrence of a spontaneous symmetry breaking, but instead corresponds to the
confinement phase. We emphasise that the formal neglect of the surface 
terms leads to physically wrong results and mathematically 
incorrect computations, which is probably signalled by the difficulties encountered by the authors.
The works on $A_{0}$ condensation are therefore of educational rather than
practical value to us. In Ref.  \cite{Anish84} and its follow--up papers the 
interpretation of $A_{0}$ condensate as the imaginary part of the chemical
potential was given and it was also found that inserting the global singlet projector
inside the trace does not change the value of the partition function.
The mechanism of occurrence of the non--zero mean Polyakov line in the deconfinement phase in our approach is explained by the spontaneous breaking of 
$Z_{N}$ symmetry by the surface term (which is zero in the
confinement phase).

Thus, the development of the approach explaining the 
confinement--deconfinement phase transition in QCD from the first principles 
is the most important task of the theory of non--Abelian gauge fields.
To justify the advantages of the present approach I may note the following circumstances.
The existence of confinement means the presence of long--range interactions in the system,
and this is the origin of the linearly growing potential.
The reason for this phenomenon in a local theory is well understood.
Indeed, due to the local gauge invariance, the system possesses the 
Gaussian constraint. By choosing a physical gauge, this can be resolved,
expressing the auxiliary component of the electric field through the 
physical components of the fields in the form of an integral operator. Substituting the latter expression into the Hamiltonian of the system leads to some effectively long--ranged and 
nonlocal equations of motion for the physical fields.

On the other hand, the nonlocality is the source of the of the boundary nontriviality of the theory. The surface terms, containing the auxiliary component, cease to be surface terms
after resolving the constraint.
The boundary condition for this component acquires the status of a constraint,
since it requires the nullification of a certain integral construction
of the physical variables. Therefore, there is a need to develop a consistent scheme for accounting of any surface terms.
The presence of non--physical degrees of freedom in a covariant formulation of the 
Yang--Mills theory makes the analysis of the boundary conditions problem noticeably more complicated due to the additional arbitrariness introduced by them.
Thus it seems reasonable to work in some physical gauge,
thereby ensuring the positive definiteness of the Hilbert space of states.

It is well known that due to the masslessness of gluons the Yang--Mills theory 
suffers from serious infrared divergences. 
In order to handle these carefully it is necessary to introduce an infrared regularisation at the first stage, which is then removed in the final result. The simplest and most physically straightforward way to accomplish this is to regularise the system by confining it inside a finite region. Note that such a procedure is always necessary for construction of the theory
at finite temperature, since, by the translational invariance of the system
the logarithm of the partition function is proportional to the volume of space.
Thus, we naturally arrive at the formulation of the Yang--Mills theory 
in a finite region with taking into account the non--trivial values of those variables at the boundary, which in the infinite limit must give rise to non--zero surface terms.

The generalised Fock--Schwinger gauge is singled out for the analysis of our problem
due to its following remarkable properties:\\
1) The Gaussian constraint can be solved in an explicit form here;\\
2) By matching the gauge choice with the shape of the boundary of the domain
the structure of the surface terms is significantly simplified;\\
3) In this gauge it is possible to relate the gauge potential  fields to the strength fields 
by means of a linear differential relation; \\
4) In view of the above points, the Hamiltonian of the system turns out to be
a fourth--order polynomial with respect to the canonical variables;\\
5) Only the longitudinal components of the chromo--electric and --magnetic
fields contain non--Abelian and non--local structures. 
Due to this property, the functional integral can be easily rewritten in
terms of only two collective variables functionally conjugated to
these components;\\
6) The mean--field approximation is nontrivial for the collective variables due to the 
quasi--classical character of the background strength fields.

Despite the presence in our confinement mechanism of a background constant magnetic field at temperatures below the critical one, this circumstance does not create problems with the theory invariances, since in the zero temperature limit the mean 
field value vanishes and the theory becomes fully Poincar\'e and gauge invariant as it should.
Indeed, the contribution of the surface terms leads in the
confinement phase to the insertion into the mean of an invariant object --- the singlet projector of the group of the large gauge transformations at infinity.

As an obvious disadvantage of the Fock--Schwinger gauge we should remark
that the non--covariant character of the gauge condition and the presence of non--locality
lead to a substantial complication of the renormalisation procedure.
This issue will be the subject of later research and is not discussed in the thesis. We avoid the use of a formal systematic renormalisation procedure by working in terms of 
renorm--invariant variables and a fixed ultraviolet cut--off. The result for the observed quantity 
$\xi=T_{c}/\sqrt{\chi}$
turns out to be finite due to the exact cancellation of divergences in it.

The outline of the thesis is as follows.

Chapter 2 introduces notations, describes the general
properties and  basic formulae of the generalised Fock--Schwinger gauge such as:
the explicit solution of the Gaussian constraint, the linear differential relations between different
variables (e.g.\ gauge fields and strengths), the 
explicit formulae for transformations between this and an arbitrary gauge, and others.

Chapter 3 is devoted to the development of the Hamiltonian formalism for the  
Yang--Mills theory in a finite domain, taking into account the contribution of the surface terms, detailed analysis of the equations of motion, symmetry transformations
and the problem of the boundary conditions choice.

Chapter 4 presents an alternative approach to the dynamics of the Yang--Mills theory
in the formalism of variables at infinity, the presentation of which is preceded by 
by some general information from the algebraic Quantum Field Theory (QFT).
The equivalence of this formalism to the one outlined in Chapter 3 is shown, the 
dynamics at infinity is also derived.

Chapter 5 studies the dependence of the partition function of the Abelian and non--Abelian
gauge theories on the variable at the boundary $|\chi|$. 
A formulation of the theory in terms of the collective variables
conjugate to the longitudinal components of the strength fields is derived.
The effective action of the theory in terms of the collective variables is calculated in the mean field approximation and based on its analysis a novel interpretation of the confinement--deconfinement phase transition mechanism is proposed.

In Chapter 6 it is shown that our confinement criterion satisfies the traditional criteria: the area law in the confinement phase and the spontaneous breaking of $Z_{N}$ symmetry in the deconfinement phase. 
We calculate the ratio of the phase transition temperature to the square root of the 
of the string tension coefficient.

The main results of the thesis are formulated in the Conclusion.
\newpage

%
\section{Some properties of the generalised Fock--Schwinger gauge}
\label{G}

In this Chapter we consider the procedure for fixing the generalised Fock--Schwinger (FS) gauge in the Yang--Mills theory,
determine the boundary conditions on the gauge fields, which prohibit the residual transformations, obtain certain identities experessing some physical poperties in this gauge through the others, and also find the relations between the FS and Coulomb gauges.

We proceed from the necessary notations. Let
$\Bx$ denote the 3-dimensional radius vector, $ x =|\Bx|$ its length and $\Hx = \Bx/x$ the unit vector in its direction. We introduce the orthogonal projector, as well as the longitudinal and perpendicular components of vectors in
${\bf R}^{3}$:
\begin{eqnarray}
&& P = {\bf 1} - \Hx \otimes \Hx,
\qquad {\bf a}_{\perp} = P{\bf a}, \quad {\bf a}_{\Vert} = (1-P){\bf a}.
\label{G-1}
\end{eqnarray}
The Fock--Schwinger gauge is defined by the condition 
\begin{equation}
\Hx\,{\bf A}(t,\Bx)=0,\qquad \mbox{i.e.}\ {\bf A}={\bf A}_{\perp}.
\label{G-2}
\end{equation}

The covariant 4-dimensional version of this gauge $x_{\mu}A^{\mu}=0$
was introduced in the classical works \cite{Fock37,Schw51}, and this has found
numerous applications in the non--Abelian gauge theory \cite{Dura82,Shif80}.
In Ref.  \cite{Shut89} it was noted that this gauge, along with the rather popular axial gauge,
allows one to resolve the Gaussian constraint explicitly, does not contain the ghost fields, and moreover is invariant with respect to spatial rotations, which is essential for the analysis of the system eigenstates. However, the literature for some time lacked any substantial study of its main properties. Such analysis was curried out by us in papers 
\cite{SveTim92PL,SveTim91Pr}, 
and then extended to the generalised FS gauge \cite{Tim94Pro}, which we shall introduce now.

Let $V_{R}$ be a regular domain in ${\bf R}^{3}$, topologically equivalent to a ball, with a
smooth boundary $\partial V_{R}$. It is convenient to choose such a curvilinear coordinates system $\BX$ in domain $V_{R}$ that on the surface  $\partial V_{R}$ the first component is constant and equal to the parameter 
 $R$, which will also serve as an infrared cut--off in our treatment, i.e.\ 
$\partial V_{R} =\{\BX: X_{1}(\Bx) = R =const\}$.  Recall that we denote the Cartesian coordinates by lowercase letters  $\Bx$. The field of vectors normal to the boundary, taken for all values of $R$, forms a smooth (differentiable) vector filed in the whole domain.

The local orthonormal curvilinear basis can be presented in the following way
\begin{equation} \label{G-3}
e^{(k)}_{i} = \frac{1}{h_{k}}\,\frac{\partial x_{i}}{\partial X_{k}}\,,
 \quad h_{k} = \left( \sum_{i = 1}^{3} \left( \frac{\partial x_{i}}
{\partial X_{k}} \right)^{2} \right)^{1/2}, 
\quad h \equiv \prod_{i=1}^{3}h_{i}.
\end{equation}
In our notations we do not distinguish the upper and lower indices and summation over the repeated indices is implicitly assumed, unless otherwise stated.  The components of vectors in the curvilinear basis, in order to distinguish them from those in the Cartesian basis, are denoted by letters enclosed in parentheses,
\begin{equation} \label{G-4}
A_{(k)} = e^{i}_{(k)}\,A_{i}, \quad \partial_{(k)} \equiv \frac{1}{h_{k}}\,
\frac{\partial}{\partial X_{k}}.
\end{equation}
The vector ${\bf e}_{(1)}$, obviously, defines the field of normals described above. 
It is natural to introduce the (2+1) decomposition of vector components onto the longitudinal and transverse components (denoted by the Greek letters): 
$i \rightarrow (1, \alpha ),\ \alpha = 2,3$.
For the vector notation for the ``Greek'' curvilinear components
we use the following symbols 
 $\BrX = (X_{2},X_{3})$.

The gauge field theory in a finite domain
$V_{R}$ acquires the simplest form in a special gauge, which is consistent with the shape of the enclosing boundary. Namely, we shall require that the gauge field component along the normals field is equal to zero in every point:
\begin{eqnarray} \label{G-5}
{\bf e}_{(1)}(\Bx)\,{\bf A}(t,\Bx) = 0,\ \, {\bf A}={\bf A}_{\perp},
\quad {\bf A}_{\perp}= P {\bf A},\ \, P ={\bf 1} -{\bf e}_{(1)}
\otimes {\bf e}_{(1)}.
\end{eqnarray}
Such a gauge naturally generalises the originally proposed gauge of Ref. 
\Ref{G-2}, and thus is called the generalised Fock--Schwinger gauge. 

A remarkable property of this gauge is that the Gauss constraint 
\begin{equation} \label{G-5p}
\nabla_{i} E_{i}=0
\end{equation}
can be resolved in it explicitly as follows 
\begin{equation}
\label{G-6}
E_{(1)} = - \frac{h_{1}}{h} \int_{X^{(0)}_{1}}^{X_{1}}
dX_{1}'\, (h\Phi_{\perp})(X_{1}',\BrX),
\end{equation}
expressing the longitudinal composition of the electric field through the transverse part of the constraint 
\begin{equation} \label{G-6p}
\Phi_{\perp} \equiv \nabla_{i} E_{\perp\,i} =
\frac{h_{\alpha}}{h}\nabla_{(\alpha)}\left(
\frac{h}{h_{\alpha}}E_{(\alpha)}\right).
\end{equation}
The lower integration limit 
 $X^{(0)}_{1}$ is a constant, which, as we shall see below, is reasonable to choose at the origin of the coordinate system.
In deriving  \Ref{G-6} we have used the explicit expression for the differentiation operator
 $\partial {\bf e}_{(1)} 
= h^{-1}\partial_{(1)}h$ and the fact that the non--Abelian part of the constraint does not contain any longitudinal vector components due to the gauge condition.

In complete analogy, we can resolve the identity 
\begin{equation}\label{G-7}
\nabla_{i}G_{i} = 0, \quad G_{i} \equiv \nabla_{j} F_{ij},
\end{equation}
expressing $G_{(1)}$ via the transverse components of this vector,
\begin{equation} \label{G-8}
G_{(1)} = - \frac{h_{1}}{h} \int_{\bar{X}^{(0)}_{1}}^{X_{1}}
dX_{1}'\, (h\ \nabla_{i} G_{\perp\,i})(X_{1}',\BrX).
\end{equation}
Next, one of the components of the Bianchi identities can be written in the form
\begin{equation} \label{G-9}
\nabla_{i} B_{i} =0, \quad B_{k} = \frac{1}{2}\epsilon_{ijk}F_{ij},
\end{equation}
which analogously allows us to express the longitudinal components of the magnetic field 
\begin{equation} \label{G-10}
B_{(1)} = - \frac{h_{1}}{h} \int_{X^{(0)}_{1}}^{X_{1}}
dX_{1}'\, (h\ \nabla_{i} B_{\perp\,i})(X_{1}',\BrX).
\end{equation}
It can be shown that the remaining two components of the Bianchi identities 
\begin{equation} \label{G-11}
\epsilon_{ijk}\,e^{i}_{(\alpha)} \nabla_{j} E_{k} = 0
\end{equation}
are equivalent to the following identities
\begin{equation} \label{G-12}
\frac{1}{h_{\alpha}}\frac{\partial}{\partial X_{1}} \left(
h_{\alpha} E_{(\alpha)} \right) = \nabla_{(\alpha)} \left(
h_{1} E_{(1)}\right),
\end{equation}
linking the transverse electric field components with the covariant derivative of the longitudinal one. 

A very attractive property of the Fock--Schwinger gauge, 
due to which its covariant version has often been used,
is the identity relating the gauge potential field to the strength field.
Such a relation also holds in the generalised FS gauge.
It is a direct consequence of the gauge condition 
and the definition of the strength tensor \cite{Tim94Pro} 
\begin{equation} \label{G-13}
\frac{1}{h_{\alpha}}\frac{\partial}{\partial X_{1}} \left(
h_{\alpha} A_{(\alpha)} \right) = h_{1} F_{(\alpha)(1)}.
\end{equation}
The integral form of the identity 
\begin{equation} \label{G-14}
A_{(\alpha)}=\frac{1}{h_{\alpha}}\int_{\tilde{X}_{1}^{(0)}}^{X_{1}}
dX_{1}'\,(h_{1}h_{\alpha}F_{(\alpha)(1)})(X_{1}',\BrX),
\end{equation}
as it is easy to understand, breaks the symmetry with respect to the
residual gauge transformations allowed by the gauge condition 
\Ref{G-5}.
The choice of boundary conditions in the previous integral relations 
does not affect the residual symmetry, since they relate the
quantities that are transformed by it uniformly.

The fixation of some boundary condition in \Ref{G-14} turns out to be very restrictive.
Let us choose $\tilde{X}_{1}^{(0)}=x^{0}_{1}$ at the origin point  ${\bf x}^{0}$,
i.e.\ we impose a boundary condition of the form
\begin{equation} \label{G-15}
\lim_{X_{1}\rightarrow x_{1}^{0}}(h_{\alpha}A_{(\alpha)})({\bf X}) = 0.
\end{equation}
The following  is a valid
\begin{quotation}\noindent
Statement. {\it
For any gauge field there exists a unique element of the group of gauge transformations, which transforms it into a field satisfying the
conditions (\ref{G-5},\ref{G-15}), all solutions of which are connected only by homogeneous
transformations, i.e. these conditions ensure the unique choice of a Lie algebra element for each gauge orbit.
}
\end{quotation}

Let $\tilde{{\bf A}}$ be an arbitrary gauged field.
The transition to the generalised FS gauge is performed by means of the gauge
transformation
 $U({\bf X})$:
\begin{equation}\label{G-16}
\backspace
{\bf A}({\bf X}) = U^{-1}\,(\,\tilde{{\bf A}}({\bf X})-g^{-1}\,\partial \,)
\,U({\bf X})\,, 
\quad {\bf E}({\bf X}) = U^{-1}\tilde{{\bf E}}({\bf X})\,U({\bf X}),
\end{equation}
which can be found as the solution of the equation 
\begin{equation}\label{G-17}
\frac{1}{h_{1}}\frac{\partial }{\partial \,X_{1}}U({\bf X}) =
 g ({\bf e}_{(1)}\,\tilde{{\bf A}})({\bf X})\,U({\bf X})\,.
\end{equation}
The initial condition  for now is chosen in the simplest way
$U(X_{1}=0,\BrX)=1$.
The solution of this can be written via the Dyson P-exponent 
\begin{equation}\label{G-18}
\backspace
U({\bf X}) = P\exp\int_{0}^{1}d\alpha\,R(\alpha, {\bf X}),
\ \, R(\alpha,{\bf X}) = g X_{1}(h_{1}{\bf e}_{(1)}\tilde{{\bf A}})
(\alpha X_{1},\BrX).
\end{equation}
It can be shown \cite{SveTim91Pr} that for the components of the gauge field 
the transformation formulae can be explicitly expressed as
\begin{eqnarray}
&&{\bf A}^{b}({\bf X}) =\tilde{{\bf A}}^{a}({\bf
X})\,P\exp\int _{0}^{1}d\alpha \,(-gt^{abc}\,R^{c}(\alpha,{\bf X})\,)- \nonumber\\
 &&- g^{-1}\int _{0}^{1}d\beta \,\partial R^{a}(\beta ,{\bf X})\,P\exp\int _{0}^{\beta }d\gamma \,
(-gt^{abc}\,R^{c}(\gamma ,{\bf X})\,).\,
\label{G-19}
\end{eqnarray}

By choosing the residual transformation, it is possible to satisfy 
the necessary boundary condition \Ref{G-15}, which plays a key role
in proving the uniqueness of the gauge field.
Indeed, suppose the existence of two distinct fields ${\bf A}'$ and 
${\bf A}''$ satisfying (\ref{G-5},\ref{G-15}) and such that
${\bf A}'' \neq U^{-1} {\bf A}' U$, $ \forall \ U = \mbox{const}$.
Then there must exist a gauge transformation linking them
\begin{equation} \label{G-20}
{\bf A}''({\bf X})  =  U^{-1}({\bf X})\  ({\bf  A}'({\bf
X})-g^{-1}\,\partial \,)\ U({\bf X})\,.
\end{equation}
Multiplication by ${\bf e}_{(1)}$ gives
$\frac{1}{h_{1}}\frac{\partial \,U({\bf X})}{\partial \,X_{1}} =  0\,$,\
i.\ e.\ $U = U(\BrX)$.

The set of transformations of this type forms the group of residual 
of $G_{res}$ gauge transformations of the FS gauge.
The boundary condition \Ref{G-15} forbids such transformations. Indeed,
in the limit ${\bf X} \rightarrow {\bf x}^{0}\,$ 
in \Ref{G-20} the gauge field will have a singularity and we come to a 
contradiction with the boundary condition, which thus proves the  uniqueness.

As a simple example, we consider the ellipsoidal coordinates
$1 \le X_{1} <\infty$, $-1 \le X_{2} \le 1$, $0 \le X_{3} \le 2\pi$,
where $X_{3}=\phi$ - polar angle, $X_{1} =(r_{1}+r_{2})/2a$,
$X_{2}=(r_{1}-r_{2})/2a$. 
This coordinate system is defined by two points,
located at a distance $\pm a$ from the centre along the axis $z$, and 
${\bf r}_{1}$, ${\bf r}_{2}$-- radius--vectors from these points 
to the point of observation. The origin of the coordinates is at the point $x^{0}_{1} = 1$,
$x^{0}_{2} = 0$. The Lam\'e parameters in the ellipsoidal coordinates are equal to
\begin{equation}\label{G-21}
h_{1}^{2} =a^{2}\frac{X_{1}^{2}-X_{2}^{2}}{X_{1}^{2}-1},\ 
h_{2}^{2} =a^{2}\frac{X_{1}^{2}-X_{2}^{2}}{1-X_{2}^{2}},\ 
h_{3}^{2} =a^{2}(X_{1}^{2}-1)(1-X_{2}^{2}).
\end{equation}
In the limit $a=0$, the ellipsoidal region becomes a ball,
the coordinate system becomes spherical, and the
the gauge becomes the traditional Fock--Schwinger gauge.
In spherical coordinates $X_{1}=r$, $X_{2}=\phi$, $X_{3}=\theta$
the Lam\'e parameters have a particularly simple form
\begin{equation}\label{G-21p}
h_{1}=1,\quad h_{2}=X_{1}\sin X_{3},\quad h_{3} = X_{1}. 
\end{equation}
An important property of the spherical FS gauge, which simplifies the formulae considerably
is that ${\bf e}_{(i)}$ is independent of $X_{1}$.
The normal vector is then equal to the unit radius--vector ${\bf e}_{(1)}=\Hx$.

Finally, I shall consider the question of equivalence of the 
Fock--Schwinger and the Coulomb gauges in the Abelian theory in infinite volume, and also
establish explicit transformation formulae between them.
It is useful to introduce the longitudinal and transverse components of vectors in the momentum space,
denoted by the symbols $\perp,\ \Vert$ as upper indices of the vectors
\begin{equation}\label{G-22}
Q = {\bf 1} -\partial \otimes \Delta ^{-1} \partial \,, 
\qquad  {\bf  a^{\perp }}  =  Q{\bf  a}
\,,\quad {\bf a^{\Vert }} = (1-Q){\bf a}\,.
\end{equation}
By the inverse Laplace operator here we mean the integral operator
\begin{equation}\label{G-23}
[\,\Delta ^{-1}\,{\bf f}\,]\ (x) =  -\int d{\bf  x'}\,
(\,4\pi \mid {\bf x-x'}\mid\,)^{-1}
{\bf  f}({\bf x'})\,.
\end{equation}
The Coulomb gauge
\begin{equation}\label{G-23p}
\partial {\bf A} = 0,\qquad \mbox{i.e.}\ {\bf A}={\bf A}^{\perp}
\end{equation}
is naturally singled out in the Abelian theory in the infinite volume,
since its variables diagonalise the free Hamiltonian.
We need to introduce the following integral operators in ${\bf R}^{3}$
\begin{eqnarray}
&&[\,\mbox{K}\,{\bf f}\,]\ ({\bf x}) = {\bf f(x)}-\,{\bf \partial }\int
^{x}_{0}dy\,{\bf \hat{x}f}(y\hat{{\bf x}})\,,
\label{G-24}\\
&& [\,\mbox{G}\,{\bf f}\,]\ ({\bf x}) =  {\bf f(x)}-\hat{{\bf x}}\,x^{-2}\int
^{x}_{0}y^{2}dy\,[\,{\bf \partial \,f}\,]\,(y\hat{{\bf x}}), \label{G-25}
\end{eqnarray}
satisfying the simple algebra \cite{SveTim91Pr}
\begin{eqnarray}
&&\mbox{KP} = \mbox{P} \,,\quad \mbox{PK} = \mbox{K} \,,\quad \mbox{QK} = \mbox{Q} \,,\quad \mbox{KQ} = \mbox{K}\,,
\label{G-26}\\
&&\mbox{GP} = \mbox{G}\,,\quad \mbox{PG} = \mbox{P} \,,\quad \mbox{QG} = \mbox{G} \,,\quad \mbox{GQ} = \mbox{Q}\,.
\label{G-27}
\end{eqnarray}

Let ${\cal X}_{\Vert}$, ${\cal X}_{\perp}$ and ${\cal X}^{\Vert}$, 
${\cal X}^{\perp}$ denote the longitudinal and transverse subspaces of 
${\bf R}^{3}$ in the sense of transversality in the coordinate and the momentum spaces.
Consider the Gauss law contraint in the presence of an external charge density
\begin{equation}\label{G_27p}
\partial {\bf E} = \rho.
\end{equation}
The total strength is gauge--invariant in the Abelian case 
and is expressed through the corresponding transverse components in both gauges
\begin{equation}\label{G-28}
{\bf E}= {\bf E}_{\perp} - \frac{\Hx}{x^{2}}\int_{0}^{x}y^{2}dy\,
(\partial {\bf E}_{\perp}-\rho)(y\Hx) = {\bf E}^{\perp}+ 
\partial\Delta^{-1}\rho.
\end{equation}
Under the assumption of boundary conditions of the form:
\begin{equation}\label{G-29}
\lim_{x\rightarrow 0}x\,{\bf A}_{\perp}({\bf x}) = 0, \quad 
\lim_{x\rightarrow 0}x^{2}E^{\perp}({\bf x}) =0
\end{equation}
we can prove that the mappings defined by the operators
$$
\begin{array}{lllll}
K: & {\cal X}^{\perp}\rightarrow{\cal X}_{\perp}, &\qquad& Q:  &  
{\cal X}_{\perp}\rightarrow{\cal X}^{\perp} \\
P: & {\cal X}^{\perp}\rightarrow{\cal X}_{\perp}, &\qquad& G:  &  
{\cal X}_{\perp}\rightarrow{\cal X}^{\perp} \\
\end{array}
$$
are one-to-one transformations between the variables of the two gauges 
\begin{equation}\label{G-29p}
\begin{array}{lllllll}
{\bf A}_{\perp} &=& K{\bf A}^{\perp}, &\qquad & 
{\bf A}^{\perp} &=& Q {\bf A}_{\perp}, \\
{\bf E}_{\perp} &=& P({\bf E}^{\perp}+\partial\Delta^{-1}\rho), &\qquad & 
{\bf E}^{\perp} &=& G( {\bf E}_{\perp}-\partial\Delta^{-1}\rho).\\
\end{array}
\end{equation}
 These linking formulae for the gauge field follow directly from the 
relations (\ref{G-16},\ref{G-17}) (or \Ref{G-28}) and similar relations for the transition to the Coulomb gauge, and for the electric field from \Ref{G-28} formula.
The functional Jacobian of the transition from one set of variables to another is 
independent of the variables themselves and equal to some normalisation constant 
${\cal D}A_{\perp}\,{\cal D}E_{\perp} = N\,
{\cal D}A^{\perp}\,{\cal D}E^{\perp}$ \cite{SveTim91Pr}.
Using the algebra of operators (\ref{G-26},\ref{G-27}) we can see that
this transformation is canonical, i.e. \ it preserves the Poisson brackets
\begin{equation}
\{\,A_{\perp }^{i}({\bf x}),E_{\perp }^{j}({\bf y})\,\} = \,(\,\mbox{KQP}\,)_{ij}({\bf x},{\bf
y}) = \,\mbox{P}_{ij}\,\delta ({\bf x}-{\bf y}).
\end{equation}

In Appendix A we present the formula for the propagator in the FS gauge, which is derived using the connection formulae we have used for the Coulomb gauge variables.
We emphasise that the latter ones are in agreement with the FS gauge variables in the sector restricted by a fixed boundary condition of the form \Ref{G-29}.
By choosing a different lower limit in the formula \Ref{G-24}
one can establish the kind of transformation to an arbitrary sector of the FS gauge.
In other words, the formulation of the theory in the FS gauge has additional
degrees of freedom numbering different topological sectors,
which are transformed into each other by the action of the elements of the residual symmetry group.  This fact does not lead to any physical 
consequences in the Abelian theory, where the integral over the orbits of the group 
gives only a normalisation factor in the functional integral,
but may turn out to be essential in the non--Abelian theory.


\section{Hamiltonian formalism of the Yang--Mills theory in a finite domain
}
\label{H}

The standard assumption of the Hamiltonian formalism in the field theory
is the absence of surface terms arising from integration 
by parts. For problems in a finite domain this is ensured by a proper choice 
of the boundary conditions satisfied by the fields at the surface. In the limit of infinite volume
it is sufficient to assume that the fields decrease sufficiently fast at the 
spatial infinity. Such requirements are certainly
well justified for many problems in the field theory.
Theory with a gauge symmetry, however, has a number of specific
properties. First of all, in its covariant formulation it contains 
superfluous (unphysical) degrees of freedom. The standard assumptions about
the behaviour of these components at infinity is, strictly speaking, not obvious.
Indeed, if one chooses a physical gauge, the extra component of the
electric field after the resolution of the Gaussian constraint is expressed in the form of a 
volume integral of the physical variables. 
In other words, this {\it turns the surface terms into volume expressions},
and therefore there is no reason to neglect them.
The boundary condition on such a variable, therefore, rather
has the meaning of an additional constraint (the constraint at the surface in the original
formulation). But as we know, constraints do require a more 
careful analysis. Since, as a rule, we are not able to explicitly resolve
such a relation, the equations of motion for the physical variables must be derived without
its use, and only then we should impose it on the equations of motion. 
A simple way to account for a surface constraint is to 
add it to the Hamiltonian with a new Lagrange multiplier. 

Thus, it is necessary to derive the equations 
of motion while taking into account the surface terms, but in addition, the values of the 
fields at the boundary should appear as Hamiltonian variables on equal footing.
Moreover, the addition of the surface term to the Hamiltonian, necessary for 
taking into account the surface constraint, acquires a significant physical meaning.
These features of the theory appear rather unusual. Nevertheless, 
such a formalism can be quite easily constructed for the 
Yang--Mills theory, which is the main purpose of this Chapter.

On the other hand, we can avoid these problems completely if we work with the Hamiltonian after excluding the non--physical variables. But in this case, the functional becomes
effectively a nonlocal expression and the theory contains nonlocal 
long--range interactions. The Hamiltonian formalism can be made 
well--defined in the traditional sense by introducing 
into the Hamiltonian a smooth cut--off function with a bounded support 
\cite{Morc85,Morc87,Morc87jmp}. Due to this function all surface terms 
turn to zero, but at a price --- the equations of motion contain a 
dependence on the cut--off function. In the limit, when the support of the function 
becomes infinite, the equations of motion of the local variables 
still contain ill--defined delocalised variables.
Thus, the problem of accounting for infinity persists,
although it takes a slightly different form. 

The second approach described above is quite feasible for the Yang--Mills theory
\cite{SveTim92PL,SveTim91Pr}
and is discussed in the next Chapter, but the first one has clear advantages in its simplicity and elegance.
Moreover, the system regularised by a smooth cut--off does not have a direct physical meaning, while the problem about a system inside a  finite domain is quite natural. 
In the infinite limit both approaches, of course, become equivalent, but the procedure of taking the limit presents certain difficulties by itself.

The problem of accounting for the boundary in the Hamiltonian formalism 
naturally arises, for example, when considering surface waves
in hydrodynamics \cite{rusZakh68,Busl86,Lewi86}, where the surface terms have a
direct physical meaning. In the work of Regge and Teitelboim \cite{Regg74} it was shown that the surface terms can play an essential role in the theory of gravitation, where this
question is still a subject of active study \cite{Solov92,Bala92}.

For a certain class of field models, in Ref. \cite{Solov93} a generalisation of the Hamiltonian formalism was proposed, the main idea of which
is to modify the Poisson brackets by the addition of some surface terms so that they are strictly  (and not with accuracy up to the surface terms, as in the traditional formalism), satisfy the Jacobi identities.

We restrict ourselves to the formulation for the case when the Hamiltonian of the system 
contains derivatives of canonical variables of at most first order.
Consider the class of local functionals of the canonical variables
\begin{equation} \label{H-1}
F = \int_{V} d{\bf x}\ f[\varphi^{A};\, \varphi^{A}_{,i}] .
\end{equation}
Let the canonical variables possess the canonical Poisson bracket 
\begin{equation} \label{H-2}
\{ \varphi^{A}({\bf x}), \varphi^{B}({\bf x}') \} = I^{AB} \,
\delta({\bf x},{\bf x}'),
\end{equation}
where $I^{AB}$ is the standard samplectic matrix with the property: 
$I^{2} = - 1,\ I = const$.
Variation of this functional with respect to a change of the canonical variables can be written as
\begin{equation} \label{H-3}
\delta F = \int_{V} d{\bf x}\ \biggl(\,{\cal E}_{A}(F)\,\delta\varphi^{A}
+ \partial_{i}\bigl({\cal E}^{i}_{A}(F)\,\delta\varphi^{A}\bigr) \,\biggr),
\end{equation}
where the Euler derivatives of the zeroth and first orders are equal to respectively 
\begin{equation} \label{H-4}
{\cal E}_{A}(F) = \frac{\partial f}{\partial \varphi^{A}} -
\partial_{i} \frac{\partial f}{\partial \varphi^{A}_{,i}} \,,
\quad {\cal E}_{A}^{i}(F) = \frac{\partial f}{\partial \varphi^{A}_{,i}}.
\end{equation}
We define the Poisson bracket of two arbitrary functionals $F$ and $H$ as:
\begin{eqnarray}
\{ F, H \} &=& \int_{V} d{\bf x} \biggl(\biggr.
{\cal E}_{A}(F)\,I^{AB}\,{\cal E}_{B}(H)  \nonumber \\
&+& \partial_{i}\bigl(  {\cal E}_{A}^{i}(F)\,I^{AB}\,{\cal E}_{B}(H)
+ {\cal E}_{A}(F)\,I^{AB}\,{\cal E}^{i}_{B}(H)  \bigr)  \nonumber \\
&+& \partial_{i} \partial_{j} \bigl( {\cal E}_{A}^{i}(F)\,I^{AB}\,
{\cal E}_{B}^{j}(H) \bigr)  \biggl.\biggr).   \label{H-5}
\end{eqnarray}
If $H$ is the Hamiltonian of the system, then the time evolution of functional  $F$
is governed by the differential equation 
\begin{equation}\label{H-5p}
\dot{F}= \{ F, H \}.
\end{equation}
This Poisson bracket has the required properties
of antisymmetry, completeness and satisfies the Jacobi identities of \cite{Solov93}.
We can see that the equations of motion of the form (\ref{H-5},\ref{H-5p})
clearly follow from the variational principle of action with full accounting for the surface terms
\begin{eqnarray}
S &=& \int_{V} d{\bf x} \left( \frac{1}{2} \dot{\varphi}^{A}\,I^{-1}_{AB}\,
\varphi^{B} - h(\varphi^{A}; \varphi^{A}_{,i} ) \right), \label{H-6} \\
\delta S &=& \int_{V} d{\bf x}\ \biggl(\biggr. \dot{\varphi}^{A}\,I^{-1}_{AB}\,
\delta \varphi^{B} - {\cal E}_{B}(H) \delta\varphi^{B} \nonumber \\
&-& \partial_{i} \bigl({\cal E}^{i}_{B}(H) \delta\varphi^{B}\bigr) \biggl.\biggr).
\label{H-6p}
\end{eqnarray}

Let us apply the described formalism to the Yang--Mills theory.
We consider the Hamiltonian of the system in a finite region 
 $V$
\begin{eqnarray} 
{\cal H}_{V} &=& \int_{V} d{\bf x}\ \biggl(\biggr. \frac{1}{2}
(E^{a}_{i})^{2} + \frac{1}{4} (F^{a}_{ij})^{2} \nonumber \\
&-& A_{0}^{a} \,\bigl( \partial_{i}E^{a}_{i} - g t^{abc}A^{b}_{i}
E^{c}_{i} \bigr) \biggl.\biggr). \label{H-7}
\end{eqnarray}
When carrying out the reduction of the system to physical variables 
in the generalised Fock--Schwinger gauge we have seen
that the variable $E_{(1)}$ is not independent. Therefore 
any boundary condition on it is an additional constraint.
We modify the Hamiltonian $H_{V} = {\cal H}_{V} + \Delta H_{V}$
by the addition of a surface term
containing this constraint with the surface Lagrange multiplier $A_{0}(R\Hx)$,
\begin{equation}  \label{H-8}
\Delta H_{V} = \int_{\partial V} d\BrX\ A^{a}_{0}
\left( \frac{h}{h_{1}} E^{a}_{(1)} + \chi^{a} \right),
\end{equation} 
where $\chi = \chi(\BrX)$ is an arbitrary function specifying the
boundary condition, which we do not specify yet.
Note that without the addition of this surface term, the equations of motion in a
finite domain are not localisable.

Assuming the usual canonical Poisson bracket
$I_{A^{a}_{i} E^{b}_{j}} = \delta^{ab} \delta_{ij}$,
it is easy to obtain the following equations of motion 
\begin{eqnarray}
&&\int_{V} d{\bf x}\,g^{a}_{i} \dot{A}^{a}_{i} = \int_{V} d{\bf x}\,g^{a}_{i} 
( E^{a}_{i} + \nabla_{i} A^{a}_{0} ), \label{H-9} \\
&&\int_{V} d{\bf x}\,g^{a}_{i} \dot{E}^{a}_{i} = \int_{V} d{\bf x}\,g^{a}_{i} 
( - g\,t^{abc} E^{b}_{i} A^{c}_{0} + \nabla_{j} F^{a}_{ij} )  \nonumber\\
&& -\int_{\partial V} d\BrX\,g^{a}_{(\alpha)}\ 
\frac{h}{h_{1}} \biggl(\biggr. \frac{1}{h_{\alpha}} \partial_{(1)}
(h_{\alpha}A^{a}_{(\alpha)})
- \frac{1}{h_{1}} \partial_{(\alpha)} (h_{1}A^{a}_{(1)}) \biggl.\biggr).
\label{H-10}
\end{eqnarray}
In deriving the surface term we have used the following identities for the derivatives of the basis vectors 
\begin{eqnarray}
&& e^{(k)}_{j}\partial_{(k)} e^{(\alpha)}_{j}=(1-\delta^{k\alpha})
\frac{1}{h_{k}h_{\alpha}}
\frac{\partial h_{k}}{\partial X_{\alpha}}, \label{H-11} \\
&& e^{(\beta)}_{j} (\partial_{(\alpha)}e^{(1)}_{j}-\partial_{(1)}e^{(\alpha)}_{j})
=\delta^{\alpha\beta}\frac{1}{h_{1}h_{\alpha}}\frac{\partial h_{\alpha}}
{\partial X_{1}}. \label{H-12}
\end{eqnarray}
Integrals of the test functions cannot be removed here, i.e.\ the time evolution is defined only for distributions. The equation of motion for the variable canonically conjugate to
 $A_{0}$ gives the Gaussian constraint 
\begin{eqnarray}
0 &=& \int_{V} d{\bf x}\,g^{a}\, \nabla_{i}E^{a}_{i} \nonumber \\
&-& \int_{\partial V} d\BrX\,g^{a} \left( 
\frac{h}{h_{1}} E^{a}_{(1)} + \chi^{a}\right). \label{H-13}
\end{eqnarray}
Due to the independence of the values of test functions inside and on the boundary this splits into the two independent conditions 
\begin{eqnarray}
\nabla_{i} E^{a}_{i} &=& 0,\label{H-14} \\
\frac{h}{h_{1}} E^{a}_{(1)}(R,\BrX) &=& -\chi^{a}(\BrX).
 \label{H-15}
\end{eqnarray}
Now we are going to fix the generalised Fock--Schwinger gauge \Ref{G-5}.

Note that Eq.  \Ref{H-9} is essentially local, so we can treat it as an equation for a function rather than that for a distribution.
We can apply to the both sides of the transverse part the operation 
$\frac{1}{h_{1}h_{\alpha}}
\frac{\partial}{\partial X_{1}}h_{\alpha}$. By means of the identities 
\Ref{H-12} and a relationship, which is the longitudinal part of 
\Ref{H-9},
\begin{equation}\label{H-16}
\frac{1}{h_{1}}\frac{\partial A_{0}^{a}}{\partial X_{1}} = - E_{(1)}^{a}
\end{equation}
we obtain the following equation of motion 
\begin{equation}\label{H-17}
\dot{F}^{a}_{(\alpha)(1)} = -gt^{abc}F^{b}_{(\alpha)(1)}A^{c}_{0},
\end{equation}
where for brevity we have utilised identity 
 \Ref{G-13}. Similarly, the equation of motion is local for the longitudinal component in
\Ref{H-10} and we get
\begin{equation}\label{H-18}
\dot{E}_{(1)}^{a} = - gt^{abc}E^{b}_{(1)}A^{c}_{0} + G^{a}_{(1)}.
\end{equation}
By means of direct, though somewhat cumbersome calculations, by using Eqs.
(\ref{H-11},\ref{H-12}), 
one can prove the following identity for the longitudinal component of vector
 $G_{i}=\nabla_{j}F_{ij}$,
\begin{equation}\label{H-19}
G_{(1)} = -\frac{h_{1}h_{\alpha}}{h}\nabla_{(\alpha)}\left(
\frac{h}{h_{1}h_{\alpha}}F_{(\alpha)(1)}\right).
\end{equation}

We are now ready to consider the problem of choosing the boundary conditions.
In Eq. \Ref{G-10}, following from Bianchi's identities, as well as in 
the relation expressing the longitudinal component of the electric field via the
local Gaussian part \Ref{G-6}, we require the usual conditions of regularity of the fields at the origin ${\bf x}^{0}$ 
(i.e.\ we choose the constant  $X_{1}^{(0)}=x_{1}^{0}$):
\begin{eqnarray}
&&\frac{h}{h_{1}} B_{(1)}({\bf X}) \biggl|_{X_{1}=x_{1}^{0}} \biggr. = 0,
\label{H-20} \\
&&\frac{h}{h_{1}} E_{(1)}({\bf X}) \biggl|_{X_{1}=x_{1}^{0}} \biggr. = 0,
\label{H-22}
\end{eqnarray}
In relations \Ref{G-8} for $G_{(1)}$, and in the non--local part of the constraint \Ref{H-15} we choose the constant $\tilde{X}_{1}^{(0)}=R$, which gives the boundary conditions at the surface: 
\begin{eqnarray}
&&\frac{h}{h_{1}} G_{(1)}({\bf X}) \biggl|_{X_{1}=R}\biggr. = 0, \label{H-23}\\
&&\frac{h}{h_{1}} E_{(1)}({\bf X}) \biggl|_{X_{1}=R}\biggr. = 
-\chi(\BrX), \label{H-24}
\end{eqnarray}
i.e.\ thanks to Eq.  \Ref{H-19}, 
\begin{equation}\label{H-25}
F_{(\alpha)(1)}(R,\BrX)=0.
\end{equation}
With this boundary condition on $F_{(\alpha)(1)}$, the equations of motion for the
physical variables (\ref{H-9},\ref{H-10}) become local.
Lagrange multiplier $A_{0}$ is determined from Eq.  \Ref{H-16}
\begin{equation}\label{H-26}
A_{0}({\bf X}) = A_{0}(R,\BrX) -
\int_{R}^{X_{1}}dX_{1}'\,(h_{1}\,E_{(1)})(X_{1}',\BrX)
\end{equation}
only up to the surface Lagrange multiplier,
which, in general, remains arbitrary.

This arbitrariness reflects the symmetry of the original action with respect to the
time--dependent gauge transformations and is analogous to the arbitrariness
in the choice of $A_{0}$ in the Coulomb gauge. It is fixed by setting the boundary
conditions with respect to time, such as the periodic ones in the partition function at finite
temperature. Therefore, in the temperature theory, in general, one cannot
turn the surface Lagrange multiplier to zero by the choice of the gauge 
transformations, i. e .\ such a theory contains a natural order parameter $A_{0}(\BrX)$. In our consideration, it will be assumed only that
this quantity is a constant in time.

From equations (\ref{H-17},\ref{H-18}) we see that all boundary conditions,
other than surface constraint, are time--preserving and, thus,
are admissible. However, from the formula \Ref{H-18} we obtain the 
law of variation of $\chi(\BrX)$ in time
\begin{equation}\label{H-27}
\dot{\chi}^{a}(\BrX)=gt^{abc}\,A^{b}_{0}(R,\BrX)\,\chi^{c}(\BrX).
\end{equation}
In the particular choice of the remaining arbitrariness, when both the colour vectors
$\chi(\BrX)$ and $A_{0}(R,\BrX)$ have the same (opposite) direction
at the initial time instant, $\chi$ is constant.
In the general case, the boundary condition cannot be fixed
and varies according to the formula \Ref{H-27}.

The contradiction is easily eliminated by the expansion of the phase space of the system.
Suppose that $\chi$ are additional  Hamiltonian variables. Their Poisson brackets with ordinary variables are equal to zero, and among themselves are defined by the canonical Poisson bracket
 \cite{Kiri76}
\begin{equation}\label{H-28}
\{\chi^{a}(\BrX),\chi^{b}(\BrX')\}=
\omega^{ab}[\chi]\,\delta(\BrX,\BrX') \equiv 
gt^{abc}\,\chi^{c}(\BrX)\,\delta(\BrX,\BrX').
\end{equation}
Then for arbitrary functionals $F$, $G$ the Poisson bracket is
\begin{equation}\label{H-29}
\{F,G\} = \int_{\partial V}d\BrX\,\omega^{ab}[\chi]
\frac{\delta F}{\delta \chi(\BrX)}\,
\frac{\delta G}{\delta \chi(\BrX)}
\end{equation}

The Jacobi identity for such brackets follows from the Jacobi identity in Lie algebra.
The phase space $\Gamma_{\chi}$ of these variables, 
according to the symplectic construction of Berezin--Kirillov \cite{Kiri76}, 
are the orbits of the co--adjoint
representations of the gauge group $SU(N)$.
On the example of the group $SU(2)$ we note that the symplectic spaces
$\Gamma_{\chi}$ are spheres of radius $|\chi|$ in the co--adjoint
the representation of the group. The radius of the sphere is the integral of the motion and
represents an external parameter of the theory, as well as the frequency of the
of rotations of the dynamics on the boundary $|A_{0}(R,\BrX)|$. 

The symplectic matrix $\omega^{ab}$ is non--degenerate on the orbits, and the differentiable
form $\prod_{\BrX}(\omega^{-1})_{ab}d\chi^{a}(\BrX)
\bigwedge d\chi^{b}(\BrX)$
 is a symplectic matrix invariant with respect to the Hamiltonian phase flow
(Liouville's theorem).
Then the equation \Ref{H-27} naturally coincides with the Hamilton equation
for the given variables, and it is expressed through the action of the group in the co--adjoint
representation.
The total phase space of the system is a direct product 
$\tilde{\Gamma} = \Gamma\times\Gamma_{\chi}$,
where $\Gamma$ is the phase the space of localised variables.

The system has residual invariance
with respect to the group $G_{res}$ of residual gauge transformations.
The generators of these transformations are the surface integrals
\begin{equation}\label{H-30}
Q_{V}(g) = \int_{\partial V} d\BrX\,g^{a}(\BrX)\,  
\left((\frac{h}{h_{1}} E^{a}_{(1)})(R,\BrX)+\chi(\BrX)\right).
\end{equation}
Indeed, they possess the algebra
\begin{equation}\label{H-31}
\{Q_{V}(g),Q_{V}(g')\} = Q_{V}(-[g,g']),
\end{equation}
where $[g,g']^{a}=gt^{abc}g^{b}g^{c}$ denotes the colour commutator. The 
Poisson bracket of the generators with the
Hamiltonian, as we have just proved, is zero
\begin{equation}\label{H-32}
\{Q_{V}(g),H_{V}\}=0.
\end{equation}
The action of the generators on the gauge field has the form
\begin{equation}\label{H-33}
\{Q_{V}(g),A_{(\alpha)}({\bf X})\} = \nabla_{(\alpha)} g(\BrX).
\end{equation}

For every fixed $A_{0}(R,\BrX)$, we have a certain
representation of the dynamics. 
This representation, however, does not have a fixed value of
$\chi(\BrX)$, which evolves in time.
The action of the symmetry group $G_{res}$ transforms
one dynamics representation into another by changing the direction of the colour vector
$A_{0}(R,\BrX)$. Thus, a minimal representation of the Hamiltonian, which is invariant with respect to symmetry transformations and time
evolution, turns out to be a direct integral of all possible representations with 
different directions of $A_{0}$. 
The values of the vector lengths $\chi$ and $A_{0}$ are thus
are fixed in each such representation and are natural
order parameters numbering the irreducible representations of the dynamics.

To conclude this Chapter, I would like to consider the expression of the Hamiltonian
systems in terms of the physical variables
\begin{eqnarray}
H_{R} &=& \frac{1}{2}\int_{V}d{\bf X}\,h\,(E_{(\alpha)}^{2}+B_{(k)}^{2}) \nonumber\\
&+& \int_{V}d{\bf X}\,h\,\Phi_{\perp}\sigma
+\int_{\partial V}d\BrX\,A_{0}(R,\BrX)\chi(\BrX), \label{H-35}\\ 
\sigma({\bf X}) &=& -A_{0}(R,\BrX)+
\int_{R}^{X_{1}}dX_{1}'\,(h_{1}E_{(1)})(X_{1}',\BrX), 
\label{H-36}
\end{eqnarray}
where $E_{(1)}$ is given by formula in Eq. \Ref{G-6}.

The variable at the boundary $\sigma_{R}(\BrX) = -A_{0}(R,\BrX)$
plays an important role in the following considerations.
Introducing this notation is useful because it allows us to rewrite the bulk part of the 
the Hamiltonian as a local expression by introducing the collective variable  $\sigma$.


{
%

\renewcommand{\a}{\alpha}\renewcommand{\b}{\beta}\newcommand{\g}{\gamma}
\renewcommand{\d}{\delta}\newcommand{\e}{\epsilon}\newcommand{\ve}{\varepsilon}
\newcommand{\z}{\zeta}\newcommand{\h}{\eta}\newcommand{\q}{\theta}
\newcommand{\vq}{\vartheta}\renewcommand{\i}{\iota}\renewcommand{\k}{\kappa}
\renewcommand{\l}{\lambda}\newcommand{\m}{\mu}\newcommand{\n}{\nu}
\newcommand{\x}{\xi}\renewcommand{\o}{o}\newcommand{\p}{\pi}
\renewcommand{\r}{\rho}\newcommand{\vr}{\varrho}\newcommand{\s}{\sigma}
\newcommand{\vs}{\varsigma}\renewcommand{\t}{\tau}\renewcommand{\u}{\upsilon}
\newcommand{\f}{\phi}\newcommand{\vf}{\varphi}\renewcommand{\c}{\chi}
\newcommand{\y}{\psi}\newcommand{\w}{\omega}\newcommand{\G}{\Gamma}
\newcommand{\D}{\Delta}\newcommand{\Q}{\Theta}\renewcommand{\L}{\Lambda}
\newcommand{\X}{\Xi}\renewcommand{\P}{\Pi}\renewcommand{\S}{\Sigma}
\newcommand{\U}{\Upsilon}\newcommand{\F}{\Phi}\newcommand{\Y}{\Psi}
\newcommand{\W}{\Omega}                                    
\newcommand{\cedilla}[1]{\c{#1}}
\newcommand{\be}[1]{\begin{#1}}
\newcommand{\ee}[1]{\end{#1}}
\newcommand{\eq}{equation}
\newcommand{\en}{eqnarray}
\newcommand{\si}{\vec{\sigma}_{\infty}(\hat{{\bf x}})}
\renewcommand{\Tr}[1]{\mbox{Tr}_{#1}}
\renewcommand{\tr}[1]{\mbox{tr}_{#1}}
\newcommand{\iy}{\infty}
\newcommand{\dd}{{\cal D}}
\newcommand{\la}[1]{\label{#1}}
\newcommand{\nn}{\nonumber}
\renewcommand{\v}[1]{\hat{#1}}
\newcommand{\rar}{\rightarrow}
\newcommand{\Rar}{\Rightarrow}
\newcommand{\lar}{\leftarrow}
\newcommand{\Lar}{\Leftarrow}
\newcommand{\lra}{\leftrightarrow}
\newcommand{\av}[1]{\langle #1 \rangle}
\newcommand{\ex}[1]{\mbox{#1-exp}}
\renewcommand{\j}[1]{{\cal #1}}
\newcommand{\pp}{_{\perp}}
\newcommand{\fr}[2]{\frac{#1}{#2}}
\newcommand{\ti}[1]{\tilde{#1}}
\newcommand{\mb}[1]{\mbox{#1}}
\newcommand{\rf}[1]{(\ref{#1})}
\newcommand{\pa}[1]{\partial_{#1}}
\newcommand{\na}[1]{\nabla_{#1}}
\newcommand{\mo}[1]{\mid\,#1\,\mid}
\newcommand{\bu}{\bullet}
\section{Yang--Mills theory in the variables at infinity formalism
} \label{V}

In the previous Chapters we have considered the construction of a
a closed Hamiltonian formalism for the 
Yang--Mills theory in a finite domain, allowing for
non--trivial values of the variables at the boundary.
The phase space of the system has been extended
by introducing additional degrees of freedom at the boundary.
In doing so, the Poisson bracket had to be modified 
by adding some surface terms so that it still satisfied the Jacobi identities.
As has been shown, the Poisson bracket thus introduced follows from the
variational principle of action in a finite region, and is therefore uniquely defined. 

The choice of the surface part 
of the Hamiltonian was dictated by the requirement of localisation of the dynamics, and this
requirement was satisfied by the presence of additional
surface constraints. It is natural to expect that in the infinite
limit the dynamics in the finite region tended to be the same dynamics
which arises in the limit of the smooth cut--off formulation of the system.
The latter regularisation is mathematically well--defined,
is unique and contains no additional surface degrees of freedom.
The procedure for taking its limit turns out to be essentially nontrivial for
systems with long--range interactions and must be realised in a weak topology,
i.e., \ with reference to some narrow class of admissible physical states.

The mathematically rigorous basis for the present consideration 
is given by the formalism of variables at
infinity, proposed by Morcchio and Strocchi in the framework of the algebraic
QFT \cite{Morc85,Morc87,Morc87jmp}, the formulation of which we shall preface with some definitions and the results of a mathematical nature \cite{Emch72,Brat79}
necessary for understanding the substance of the subject.

\subsection{Some background from the formalism of the variables at infinity}

In the context of algebraic QFT (Quantum Field Theory), observables correspond 
to self--adjoint elements of the $C^{*}$ algebra ${\cal R}$ (a complex Banach algebra with
involution and the property
$\mid\mid R\mid\mid^{2} = \mid\mid R^{*}R\mid\mid
\quad \forall \ R\in{\cal R}$),
and the states --- positive ($\f(A^{*}
A) \geq 0$) linear functionals $\f$ over it, i.e., the elements of the dual
of the space ${\cal R}^{*}$. 
In QFT, a quasi--local algebra is usually used
${\cal A} = \overline{\bigcup_{V \in \Im}{\cal A}^{V}}$
defined as
$C^{*}$-inductive limit of algebras ${\cal A}^{V}$ associated to finite
regions from the set $\Im = \{V: V \in {\bf R}^{3}\}$. The dynamics is
the one-parameter group $\a^{t}_{V}$ of automorphisms of ${\cal A}$,
generated by the truncated Hamiltonian $H_{V} \in {\cal A}$:
\be{\eq} \la{HaM}
\a^{t}_{V}[A] \equiv e^{i\,tH_{V}}\,A\,e^{-i\,tH_{V}}\,, \quad \forall\
A \in {\cal A}\,.
\ee{\eq}
Any $C^{*}$ algebra can be considered as a $W^{*}$ algebra (von Neumann algebra ${\cal N}$, which is a subalgebra of the algebra ${\cal B}({\cal H})$
of all bounded operators in some
Hilbert space ${\cal H}$ and such 
that ${\cal N}'' = {\cal N}$. Here the dash corresponds to taking the commutant
${\cal N}$, i.e., the set of all operators in ${\cal B}({\cal H})$,
commuting with all elements of ${\cal N}$.

      There is a canonical GNS (Gelfand--Naimark--Segal) construction
to produce a vector representation of $\p_{\vf}$ of the algebra ${\cal R}$,
corresponding to any state $\vf \in {\cal R}^{*}$ in a certain
Hilbert space ${\cal H}_{\vf}$. Thus the vector $\F \in
{\cal H}_{\vf}$ associated with $\vf$ and defined as 
$\av{\vf,R} =
(\F,\p_{\vf}(R)\F) \quad \forall \,R \in {\cal A}$
turns out to be cyclic
in this representation. 
Universal representation
$\p_{u} = \bigoplus _{\vf
\in {\cal R}^{*}}\p_{\vf}$ 
is the direct sum of GNS representations in the Hilbert space
${\cal H}_{u} = \bigoplus _{\vf
\in {\cal R}^{*}}{\cal H}_{\vf}$
over all states over the algebra.

The universal von Neumann algebra ${\cal R}'' \equiv \p_{u}({\cal R})''$,
endowed with the strong topology, is a Banach dual space 
to the set ${\cal F}$, called the pre--dual, and forming a closed
subspace of $({\cal R}'')^{*}$. The elements of ${\cal F}$, which are called
the normal states over ${\cal R}''$, can otherwise be further
characterised as density matrices, i.e.
$$\exists \,\r \in \j{B}(\j{H})\,,\quad \Tr{\j{H}}(\r) < \iy\,,
\quad \av{\vf,A} = \Tr{\j{H}}(\r\,A)\ \ \forall \,A \in \j{R}''\,.$$
Two representations $\p_{1}$ and $\p_{2}$ of $C^{*}$ of the algebra $\j{R}$ are
quasi--equivalent (we write this as $\p_{1} \approx \p_{2}$),
if any state $\w \in \j{R}^{*}$ representable as 
normal on $\p_{1}(\j{R})'$ :
$$\exists\, \r \in \p_{1}(\j{R})''\ \  \mbox{so that}
\quad \av{\w,A} = \Tr{\j{H}_{1}}(\r\,A)\ \ \forall \,A \in \j{R}\,,$$
is just as normally representable on $\p_{2}(\j{R})'$ and vice versa. Simply put,
representations of $\p_{1} \approx \p_{2}$, if they have the same set of
of algebra states admitting a density matrix on their bicommutants.

A primary (factor) representation is a representation,
quasi--equivalent to
irreducible. Primary representations will play an important role
due to the following property. Centre
 $\j{Z}_{\p} = \p(\j{R})' \bigcap
\p(\j{R})$ of any primary representation $\p$ is trivial and is of the form
$\j{Z}_{\p} = \{\l\,{\bf 1},\ \ \l \in {\bf C}\}$.
The same property holds for irreducible representations also, but they, unlike the primary ones, are obliged to be associated with pure states.

We shall consider systems with long--range interactions for which 
the limit of the dynamics (\ref{HaM}) at $V \rar \iy$ does not exist in the topology of the
norms. In this case the dynamics in the infinite volume can be formulated using the construction of Morchio and Strocchi [28-31].

   Let $\j{F}$ be some class of states from $\j{A}^{*}$ satisfying the
the properties:
\begin{description}
\item \ \ i) $\j{F}$ is closed with respect to linear operations;
\item \ \ ii) $\j{F}$ is norm--closed and separable: from $\f(A) = 0 \ \ \forall\,
\f\in\j{F}\ \ \Rightarrow A = 0$;
\item \ \ iii) From $\f\in\j{F}\ \Rightarrow \ \f _{AB}(\,.\,) \equiv
\f(A\,.\,B)\in\j{F} \quad \forall A, B\in\j{A}$.
\end{description}
We denote by $w_{\j{F}}$ the weak topology defined by $\j{F}$ on $\j{A}''$,
and let $\j{M}$ be an extension of $\j{A}$ in $w_{\j{F}}$-topology\footnote{$\j{A}''$
as a Banach space is isomorphic to the doubly dual 
 $\j{A}^{**}$.}
$$\j{A}\,\subset\,\j{M}\,\subset\,\j{A}''\,.$$
In abstract terms, the $W^{*}$-algebra $\j{M}$ is dual to the Banach space $\j{F}$.
Then there exist weak limits 
\be{\eq} \la{w_Lim}
w_{\j{F}}\mbox{ -}\lim_{V \to \iy}\a^{t}_{V}[A] \equiv \a^{t}[A] \in\j{M}
\quad \forall\ A\in\j{A}
\ee{\eq}
and $\a^{t}$ can be extended to a single parametric automorphisms group
$\j{M}$ \cite{Morc87,Morc87jmp}.

     The dynamical system is the triple $(\,\j{M},\ \j{F},\ \a^{t}\,)$.
It is  important to note that in the presence of long--range interactions 
the algebraic dynamics can only be defined on an algebra $\j{M}$
with a nontrivial centre $\j{Z} = \j{M}\,\bigcap \,(\cap_{V}\j{A}')$,
generated by the variables at infinity.

    We say that the factor representation $\p$ associated with a state
of $\j{F}$ and stable with respect to $\a^{t}$, leads to an effective localisation of the
of dynamics if there exists an effective localisation subalgebra $\j{A}_{l}\,
\subset \,\j{M}$ with the property
\begin{description}
\item \ \ i) $\j{A}_{l}$ is exactly representable in $\p$;
\item \ \ ii) $\j{A}_{l}$ is weakly dense in $\j{M}$;
\item \ \ iii) there exist automorphisms $\a^{t}_{\p} \quad \j{A}_{l}$
such that
$$\forall \f\in\p: \ \ \f(\a^{t}[A]) = \f(\a^{t}_{\p}[A])\quad
\forall\,A\in\j{A}_{l}.$$
\end{description}
Hence the centre of $\j{A}_{l}$ is trivial and  there are no delocalised variables in it.

     The symmetry of the system $(\,\j{M},\ \j{F},\ \a^{t}\,)$
is naturally defined as the group $\b^{\vs}$ of automorphisms of $\j{M}$ obtained by the
of the centre of $\j{A}_{l}$ and commuting with the dynamics\footnote{For this it is
sufficient to be commuting with the dynamics at finite cut--off.}
 $\b^{\vs}\,\a^{t} = \a^{t}\,\b^{\vs}$.
For physical reasons, it would seem sufficient to restrict ourselves to the 
representation of the effective
localisation and the dynamics of $\a^{t}_{\p}$. 
However, the presence of the $\b^{\vs}$ symmetry introduces additional complexity,
because it does not commute with $\a^{t}_{\pi}$, although it does commute with
algebraic dynamics.
In typical examples, it turns out
that $\a^{t}_{\p}$ does not leave invariant representations of the form $\p_{(\b
^{\vs})^{*}\f}$.
 Therefore, we have to consider a set of states
$\j{S} = \{\,\f^{\vs} = (\b^{\vs})^{*}\f\,\}$,
which is already stable under the
by the action $(\a^{t}_{\p})^{*}$, and to introduce a representation $\P_{\j{S}}$ which is a direct sum of all representations $\P_{\j{S}}$
by the direct sum of all representations $\j{A}_{l}$ obtained by the GNS construction
based on the states $\f_{\vs}$.

The representation $\P_{\j{S}}$ is reducible and has a nontrivial centre
$\j{Z}_{\j{S}}$ generated by variables at infinity. Dynamics
$\a^{t}_{\p}$ uniquely extends up to an action on $\j{Z}_{\j{S}}$,
whose nontrivial form leads to some interesting physical 
effects. Note
also that it is the dynamics of $\a^{t}_{\p}$ that arises in the context of considering the
of the given representation of $\P_{\j{S}}$.

      Allow me to explain the necessity of introducing such mathematical structures.
If the interaction is not decreasing  fast enough at infinity, 
there is no limit (by the norm) of the dynamics as  the finite cut is removed. 
This leads to the fact that the time derivative of the dynamics, i.e., the equations of motion of the local observables, symbolically written in the form
\be{\eq} \la{EqMoT}
\frac{d}{dt}\,\a^{t}_{R}[\vf(x)] = F[\vf(x), \vf_{R}], \quad \vf(x)\in\j{A}
\ee{\eq}
contain delocalised variables $\vf_{R}$ in the limit $R \rar \iy$.
For example:
\begin{description}
\item \ \ $1^{\circ}$ $\vf_{R} = -\int_{0}^{\iy}dx\,f_{R}'(x)\,\vf(x),$
where $f_{R}(x)$ is a smooth function such that
\be{\eq}  \la{f_r}
f_{R}(x) = \left\{
\begin{array}{lll}
1 , & x\leq R'& \qquad R' = (1-\varepsilon )R   \\
0 , & x\geq R''& \qquad R'' = (1+\varepsilon )R  \,.
\end{array}
\right.
\ee{\eq}

Such variables arise in field--theoretic models.
\item \ \ 
$2^{\circ}$ $\vf_{V} = \frac{1}{V}\int_{V}dx\,\vf(x)$
 --- ergodic averages. This example is typical of models in the mean--field approximation.
fields, where such constructions can be present directly in the
the Hamiltonian $H_{V}$.
\end{description}
The limits of these variables are obviously well defined in the weak sense. Let
$\f_{\l}$ is a state such that $\av{\f,\vf(x)} = \l$. Then
$$w\mbox{-}\lim_{R} \vf_{R} \equiv \lim_{R}\av{\f^{\l},\vf_{R}} = -\int
_{0}^{\iy}dx\,f_{R}'(x)\l = \l,$$
and analogously for $\vf_{V}$.
Therefore, the weak limit of the dynamics
$w\mbox{-}\lim_{R}
\a_{R}^{t} \equiv \a^{t}$ 
exists.

The dynamics of the $\a^{t}_{\p}$ of the representation with an effective localisation of the
variable at infinity, corresponding to some observable,
contains quite important information about the spectral properties of the theory
as this is related, by the generalised Goldstone's theorem \cite{Morc85},
to the spectrum of the commutator of the symmetry generator with the observable in question. 
To find the dynamics at infinity,
in general, it is necessary to solve the full dynamical problem.
Let  $\{\vf^{i}(x)\}$  be the set of local variables of the theory, and $\{\vf^{A}\}$
variables from this set, for which in the equations of motion
\be{\eq} \la{Qasd}
\fr{d}{dt}\a^{t}_{R}[\vf^{i}(x)] = F^{i}[\vf^{j}(x),\vf^{A}_{R}]
\ee{\eq}
contain delocalised contributions.

 It is necessary to introduce a set of variables at infinity 
$\vf^{A}_{\iy}$ and a set of primary states 
$\{\f_{\l}\}$ such that $(\a^{t}_{\p})^{*}\f_{\l} = \f_{\l(t)}$
is contained in this set, each of which possesses the property
\be{\eq} \la{Xa}
\av{\f_{\l},\,\vf^{A}(x)} = \l^{A}\,.
\ee{\eq}
        By definition, in order to obtain 
$\a^{t}_{\p}[\vf_{\iy}]$ it is necessary, by solving Eq.  \rf{Qasd} find
$\a^{t}_{R}[\vf^{A}]$, consider the vacuum state $\f_{\l_{0}}$,
cyclic for representation $\p$, and compute the dynamics 
$\a^{t}_{\p}[\vf_{\iy}]$ dependent on the initial condition $\vf_{\iy}(t=0) =
\l$
\be{\eq} \la{asm}
\a^{t}_{\p}[\vf_{\iy}]^{a}\,[\l] = \l^{a}(t)
= \lim_{R' \to \iy}\lim_{R \to \iy}
\av{\f_{\l_{0}},\,\b^{\l}_{R'}\a^{t}_{R}[\vf^{A}(x)]}\,.
\ee{\eq}
Unfortunately, this path can only be followed in its entirety in precisely solvable cases. 
Let us consider the equation of motion \rf{Qasd}, remove the
the truncation in the weak topology of the representation $\p$ associated with the vector
$\f_{\l_{0}}$ in it. In this case we have
\be{\eq} \la{Qasm}
\fr{d}{dt}\a^{t}_{\p}[\vf^{i}(x)] = F^{i}[\vf^{j}(x),\l^{A}_{0}]\,.
\ee{\eq}
The last equation has the property
\be{\eq} \la{QQasm}
\av{\f_{\l_{0}},\,\fr{d}{dt}\a^{t}_{\p}[\vf^{A}(x)]} = 0
\ee{\eq}
reflecting the stability $\p:\ \ (\a^{t}_{\p})^{*}\f_{\l_{0}} = \f_{\l_{0}}$.
For this the functional $F$ must be special, for instance,
$$F^{A}[\vf^{i}(x),\l_{0}^{B}] = F^{A}[\vf^{B}(x) - \l^{B}_{0}]\,.$$
However, for $\f_{\l}\,(\l \not= \l_{0})$ we have
\be{\eq} \la{QQQQ}
0 \not= \fr{d}{dt}\av{(\a^{t}_{\p})^{*}\f_{\l},\,\vf^{A}(x)} =
\av{\f_{\l},\,F^{A}[\vf^{i}(x),\l^{B}_{0}]}\,.
\ee{\eq}
A closed equation, obviously, arises if $F^{A}$ is linear in 
$\{\vf^{B}\}$ and does not depend on $\{\vf^{i},\,\,i \not= B\}$.

 Often one can restrict oneself to the Hartree-Fock approximation. Let the dynamics
has the form \rf{Qasm}, then in this approximation
\be{\eq} \la{HHFF}
\fr{d}{dt}\a^{t}_{\p}[\vf^{A}_{\iy}] = F^{A}[\vf^{B}_{\iy},\l^{B}_{0}]\,.
\ee{\eq}
We can generalise the whole construction by replacing the definition of the states \rf{Xa} with the condition
\be{\eq} \la{XXa}
\lim_{\mo{{\bf x}} \to \iy}\av{\f_{\l},\,\vf^{A}(x)} = \l^{A}
\ee{\eq}
for the case where the states $\l \not= \l_{0}$ are not translationally--invariant 
$(\a^{{\bf x}})^{*}\f^{\l} \not= \f^{\l}$. The right--hand side of Eq.  \rf{QQQQ} 
might simplify in the limit 
$\mo{{\bf x}} \rar \iy$ due to the properties of clustering
(weakening of correlations) with respect to translations of the form
\be{\eq} \la{Klust}
\lim_{\mo{{\bf x}} \to \iy}\f_{\l}(\a^{{\bf x}}[A]\,\a^{{\bf x}}[B]) =
\lim_{\mo{{\bf x}} \to \iy}\biggl(\f_{\l}(a^{{\bf x}}[A])
\,\f_{\l}(\a^{{\bf x}}[B])\biggr)\,.
\ee{\eq}

We now turn to the problem of the equilibrium states for systems with a
non--trivial dynamics of variables at infinity.
It is well known that the trace of the density matrix of translationally-- invariant
systems do not exist, because its logarithm is proportional to the space volume. 
Therefore, we first calculate the mean and other thermodynamic
characterisation using local Gibbs states, $\w_{\b}^{V}$,
\be{\en} \la{LocGib}
&& \w_{\b}^{V}(A) = \Tr{}(\r_{\b}^{V}\,A)\,,\qquad \r_{\b}^{V} = Z_{\b\,V}^{-1}
\,e^{-\b\,H_{V}}\,, \nn\\
&& Z_{\b\,V} = \Tr{}e^{-\b\,H_{V}}\,.
\ee{\en}
and then proceed to
the thermodynamic limit, pushing $V$ and, if necessary, other parameters
to infinity. 
When $V$ is finite
the trace can be computed over various spaces of unitary--equivalent
representations of the algebra of permutation relations, which at $V \rar
\iy$ become unitarily non--equivalent and give different answers.

            The idea of the algebraic approach to the construction of the equilibrium state of a
system consists in establishing a connection with its dynamics.
If the dynamics is defined as the 
automorphisms $\a^{t}$ group of the algebra of observables $\j{A}$, consider the
equation of state $\w_{\b}\in\j{A}^{*}$
\be{\eq} \la{KMS}
\w_{\b}(\,A\,\a^{t}[B]\,)\biggl|\biggr._{t = i\,\b} = \w_{\b}(\,B\,A\,)\,,
\qquad \forall \ A, B \in\j{A}\,,
\ee{\eq}
called the KMS (Kubo--Martin--Schwinger) condition. Note firstly that this condition is trivially
satisfied for local Gibbs states \rf{LocGib}. In known cases,
when there is  a reasonable way of calculating the thermodynamic limit and
a well--defined limit state is obtained, it satisfies the
KMS condition with respect to the limit dynamics \cite{Brat79,Bona89}.
In addition, the KMS states
under sufficiently weak additional conditions, have a number of remarkable
properties, allowing their decomposition into simpler states. The 
KMS states that do not admit further decomposition are called
extreme and are interpreted as pure thermodynamic phases.

Thus, suppose that there exists a state $\av{...}_{\b}\in
\j{M}^{*}$,
satisfying the KMS condition on the dynamics of $\a^{t}$ as the
of automorphisms of $\j{M}$. For such a state, there is a theorem about
central decomposition at infinity:
\be{\eq} \la{DecInf}
\av{w\,A}_{\b} = \int_{\aleph} \m_{\b}(d\lambda)\ (\lambda,w)_{\b}\,\av{\lambda,\,A}_{\b}\,,
\qquad \forall \,A\in\j{A},\ \ w\in\j{Z}(\j{A}'')\,.
\ee{\eq}
It asserts the existence of a central measure $\m_{\b}$ concentrated on the set $\aleph$
of all KMS primitive states, performing a decomposition
of the mean $\av{...}_{\b}$ of any element of a bicommutant algebra of the form $wA$
by the product of the average $\av{\lambda,\,A}_{\b}$ of $A\in\j{A}$ over the
near--margin KMS state with some functional on
the centre of the bicommutant $(\lambda,\,w)$.

     In any primary KMS state, the variables at infinity
are fixed on their expectations. In order to satisfy the condition
KMS, the measure $\m_{\b}$ must have certain properties of invariance.
States
$\av{\lambda,\,.}_{\b}$ as usual one can obtain via limits of the local Gibbs states:
\be{\eq} \la{LimGib}
\av{\lambda,\,A}_{\b} = \lim_{R \to \iy}\Tr{\j{H}_{\lambda}}(\r_{\b\,R}A),
\ee{\eq}
where the trace is computed over a space characterised by a fixed
value of the mean $\lambda$.
A necessary condition for a state to be KMS is that it is t-invariant
\be{\eq}  \la{t_Inv}
\w_{\b}(\a^{t}[A]) = \w_{\b}(A)\,,
\ee{\eq}
which must be satisfied. We have
\be{\eq}  \la{wwwww}
\av{\lambda,\,\a^{t}[A]}_{\b} = \av{(\a^{t}_{\p})^{*}\lambda,\,A}_{\b}\,.
\ee{\eq}
The measure $\m_{\b}$ must be arranged to compensate for the time dependence
of states \rf{wwwww}. The $(\lambda\,.\,)$ functional is normalised to unity and we are not interested in its explicit form while calculating the average of the local operators.
If the dynamics of the variables at infinity $(\a^{t}_{\p})^{*}
[\lambda_{\iy}]$
is Hamiltonian, then there exists a Liouville measure
of their phase space, $(d\lambda)$,
conserved by the Hamiltonian phase flow.
Then states of the form
\be{\eq} \la{KMSs}
\w_{\b}(A) = \int (d\lambda)\,\av{\lambda,\,A}_{\b}\,
\ee{\eq}
with $\av{\lambda,\,A}_{\b}$ defined using the thermodynamic limit
of the local Gibbs states \rf{LimGib} are t-invariant. 
The last decomposition means that the equilibrium state is actually defined
using the given representation of $\Pi_{S}$ and the integral over
$(d\lambda)$ accounts for the dynamics at infinity.
Similarly, it can be
demonstrated that the states \rf{KMSs} also satisfy directly the KMS condition.

%

\subsection{Dynamics at infinity in the Yang--Mills theory
}
The evolution $\alpha ^{t}_{R}$  for the Yang--Mills system in the
Fock--Schwinger gauge is generated by the truncated Hamiltonian of the form
\begin{equation}
\label{Hr}
H_{R} = \frac{1}{2}\int d{\bf x}\,f_{R}(x)\,[\ \Ep^{2}+
\Ev^{2}+\frac {1}{2}F_{ij}^{2}\ ]\,,
\end{equation}
where $\Ev$ is given by the formula \Ref{G-6}, and the smooth function $f_{R}(x)$
satisfies the properties \rf{f_r}. The canonical commutator 
is given by the formula
\begin{\en}
&& \backspace\backspace
[\,A_{\perp}(g_{1})\ ,\ E_{\perp}(g_{2})\,] = i\,(g_{1}\ ,\ g_{2})\,, \nn\\
&& \backspace\backspace
(g_{1}\ ,\ g_{2}) = \int d{\bf x}\,g_{(1)\,i}^{a}({\bf x})\,P_{ij}(\hat{{\bf
x}})\,g_{(2)\,j}^{a}({\bf x}), \ \  
\ g_{(A)}\in {\cal G}({\bf R}^{3})\ (A = 1,2).\,
\end{\en}
Since the surface terms are now zero due to the cut--off function,
the equations of motion follow from \rf{Hr} directly
$\Ap(t) = \alpha _{R}^{t}[\Ap],\ \Ep(t) = \alpha _{R}^{t}[\Ep]$:
\be{\en}
&& \dot{A}^{a}_{\perp\,k} = f_{R}\,E^{a}_{\perp\,k}-
P_{kl}\nabla _{l}\tilde{\sigma }^{a}_{R}\,, \la{EEqq} \\ 
&&  \dot{E}^{a}_{\perp\,k} =
gt^{abc}E^{\perp\,b}_{k}\tilde{\sigma }^{c}_{R}
-P_{kl}\nabla _{i}f_{R}\,F^{a}_{il}\,, \\
&& 
\tilde{\sigma } ^{a}_{R}({\bf x}) = \int _{x}
^{\infty }dy\,f_{R}(y)\frac{1}{y^2}\int
_{0}^{y}z^{2}\,dz\,\Phi_{\perp} ^{a}(z\hat{{\bf x}})\,.
\la{EEqqQ}
\ee{\en}
It is straightforward to verify the following commutation relation 
\begin{equation}               \label{FF}
[\,\Phi_{\perp}^{a}({\bf x})\ ,\ 
\Phi_{\perp}^{b}({\bf x}')\,] = -i\,gt^{abc}\Phi_{\perp}^{c}({\bf
x})\,\delta ({\bf x}-{\bf x}')\,,
\end{equation}
in deriving which we have used the identity 
 $\nabla _{i}\nabla _{j}F_{ij} = 0$.
Next we calculate the time derivative of $\Phi_{\perp}$
\be{\en}
\dot{\Phi}^{a}_{\perp}(\Bx) &=&\frac{g}{2}t^{abc} [\Phi_{\perp}^{b}(\Bx),
\int_{0}^{\infty}y^{2}dy\,\Phi_{\perp}^{c}(y\Hx)]_{+}\int_
{\mbox{{\small max}}(x,y)}^{\infty} \frac{dz}{z^{2}}\,f_{R}(z) \nn \\
&+& \frac{1}{x^{2}}\frac{\partial}{\partial x}\left(x^{2}f_{R}(x)\hat{x}_{j}
\nabla_{i}F_{ij}^{a}\right). \label{PhiDot}
\ee{\en}
Thus the integral with an arbitrary function of spherical angular coordinates  
$\vs$  will have zero time derivative 
\begin{equation}
\label{cH}
[\,H_{R}\ ,\ \Phi_{\perp}(\varsigma)\,] = 0 \ \ \forall \ 
\varsigma(\hat{{\bf x}})\,,
 \qquad \Phi_{\perp} (\varsigma) \equiv \int d{\bf x}\,\Phi_{\perp} ^{a}({\bf
x})\,\varsigma ^{a}({\hat {\bf x}}).
\end{equation}
Therefore, $\Phi_{\perp} (\varsigma)$  generate the group $G_{res}$ 
of residual gauge transformations 
$\beta ^{\varsigma}$,
commuting with the cut--off dynamics 
$\alpha ^{t}_{R}\beta ^{\varsigma} =
\beta ^{\varsigma}\alpha ^{t}_{R}$, 
and their action on the gauge field has the form
\begin{equation}
-i\,[\,\Phi_{\perp} (\varsigma)\ ,\ \Ap^{a}({\bf x})\,] =
\nabla \varsigma^{a}(\hat{{\bf x}})\qquad\forall \ 
\varsigma^{a}(\hat{{\bf x}})\,,
\end{equation}
From Eq. \Ref{PhiDot} after integrating by parts there follows the equation of motion for the variable $\tilde{\s}_{R}$
\be{\en}
\dot{\tilde{\s}}_{R}^{a}(\Bx) &=& - \frac{g}{2}t^{abc}\int_{x}^{\infty}dy
\,[\tilde{\s}_{R}^{b\,'}(y\Hx),\tilde{\s}_{R}^{c}(y\Hx)]_{+} \nn\\
&+&\int_{x}^{\infty}dy\,f_{R}^{2}(y)\,\hat{x}_{j}\nabla_{i}F_{ij}^{a}(y\Hx).
\la{SiDot}
\ee{\en}
The quasi--local algebra ${\cal A}_{l}$ is generated by the fields $\Ap$, $\Ep$,
but the equations of motion also contain a composite variable, 
which we rewrite as,
\begin{\en}
\backspace\backspace &&
\tilde{\sigma } ^{a}_{R}({\bf x}) = f_{R}({\bf x})\sigma ^{a}({\bf x})
-\sigma ^{a}_{R}({\bf
x})\,,\quad \sigma ^{a}_{R} = - \int _{x}^{\infty }dy\,f_{R}\,'(y)
\sigma ^{a}(y\hat{{\bf x}})\,,
\\ \backspace\backspace &&
\sigma ^{a} = -(\,\Delta _{x}^{-1}\,\Phi ^{a}\,)({\bf x}) = \int
_{x}^{\infty }\frac{dy}{y^{2}}\,\int _{0}^{y}z^{2}\,dz\,
\Phi_{\perp}^{a}(z\hat{{\bf x}})\,.
\end{\en}
In the $R\rightarrow \infty $ limit, the construction of $\sigma ^{a}_{R}$ 
will contain increasingly delocalised contributions due to the 
the carrier of the function $\mbox{supp}\,f_{R}'(y)=[R',R'']$.
The representation of the algebra $\pi _{\sigma }$ is fixed by the requirement that the field 
$\sigma ({\bf x})$ had an average value in it
\begin{displaymath}
\phi _{\sigma }(\sigma ^{a}({\bf x})) = \phi _{\sigma }
(\alpha ^{t}[\sigma ^{a}({\bf x})]) = \sigma ^{a}
\end{displaymath}
fixed in time.  For ${\bf R}^{3}$-translation invariant
states $\sigma ^{a}$ must be a constant.

Let us introduce a minimal set of ${\cal S}$-primary states over
${\cal A}_{l}$ containing $\phi _{\sigma \hat{\eta }}\ 
(\,\sigma  = \sigma \hat{\eta }\,)$ and stable under 
$(\alpha ^{t}_{\pi })^{*}\,,\ (\beta ^{\varsigma})^{*}$. We denote as $\Pi _{{\cal S}}$
the representation of  ${\cal A}_{l}$, given by the direct some of all representations from
 ${\cal S}\,$, and as 
${\cal Z}_{{\cal S}}$ the center $\Pi _{{\cal S}}({\cal A}_{l}'')$. Here
\begin{displaymath}
{\cal S} = \{\phi _{\sigma _{\infty }(\hat{{\bf x}})}\,,\  \forall \ 
\sigma _{\infty }(\hat{{\bf x}}) \}
\end{displaymath}
with an arbitrary function  $\sigma _{\infty }(\hat{{\bf x}})$ taking values in
$Lie-G_{res}$. The algebra ${\cal A}_{l}$ is presumed to be weakly asymptomatically Abelian 
with respect to translations in the representation 
 $\pi$
\be{\eq} \la{AsAb}
w\mb{-}\lim_{\mo{{\bf x}} \to \iy}\p\left(\,[\a^{{\bf x}}[A],\,B]\,\right) = 0\,,
\qquad \forall\ A, B\in\j{A}_{l}\,.
\ee{\eq}
Then the variables at infinity 
\begin{displaymath}
\sigma ^{a}_{\infty }(\hat{{\bf x}}) = - w\mbox{-}\lim _{R\rightarrow
\infty }\int _{0}^{\infty }dy\,f_{R}\,'\,\sigma ^{a}(y\hat{{\bf x}})
\end{displaymath}
exist in the weak topology 
 $\Pi _{{\cal S}}$ for translationally invariant states $\phi $:
\begin{equation}
\pi (\sigma _{\infty }) = \lim _{R\rightarrow \infty }
\phi (\alpha _{R\hat{{\bf x}}}[\sigma ]) = \phi (\sigma ) = \sigma \,,
\end{equation}
and also $\forall \ A\in {\cal A}_{l}$
\begin{equation}
[\,\sigma ^{a}_{\infty }(\hat{{\bf x}})\ ,\ A\,] = 
- w\mbox{-}\lim _{R\rightarrow \infty }\int
_{0}^{\infty }dy\,f_{R}'(y)\,\int _{0}^{\infty }
\frac{z^{2}\,dz}{\mbox{max}(y,z)}\
[\,\Phi_{\perp} ^{a}(z\hat{{\bf x}})\ ,\ A\,] = 0\,.
\end{equation}
The last relation is valid due to the finite region of integration 
by virtue of (\ref{AsAb}) and, in addition, by the property of the support 
$\mbox{supp}\ f_{R}\,'$.
Hence, the variables at infinity belong to the centre of
$\sigma _{\infty }(\hat{{\bf x}})\in {\cal Z}_{{\cal S}}$. 
The symmetry $\ \beta ^{\varsigma}$ commutes with the
Hamiltonian and acts on the variables at infinity via the
co--adjoint representation $Ad^{*}$
\begin{equation}
-i\,[\,\Phi_{\perp} (\varsigma)\ ,\ \sigma _{\infty }^{a}\,] 
= gt^{abc}\, \varsigma ^{b}(\hat{{\bf
x}})\,\sigma_{\infty}^{c}(\hat{{\bf x}})\,.
\end{equation}

The effectively localised dynamics  $\a^{t}_{\pi}$ of representation $\pi$ has the form
\begin{\en}   \la{e1e}
\dot{A_{\perp\,k}^{a}} &=& E_{\perp\,k}^{a}-P_{kl}\nabla _{l}
(\,\sigma -\pi (\sigma _{\infty })\,)^{a}\,,
\\ \la{e2e}
\dot{E}_{\perp\,k}^{a} &=& gt^{abc}\,E_{\perp\,k}^{b}\,
(\,\sigma -\pi (\sigma _{\infty })\,)^{c}
-P_{kl}\nabla _{i}F_{il}^{a}\,,
\end{\en}
and from Eq  \Ref{SiDot} it follows that 
\be{\en}
\dot{\s}^{a}(\Bx) &=& 
gt^{abc}\,\pi(\s_{\infty})^{b}\s^{c}
- \frac{g}{2}t^{abc}\int_{x}^{\infty}dy
\,[\s^{b\,'}(y\Hx),\s^{c}(y\Hx)]_{+} \nn\\
&+&\int_{x}^{\infty}dy\,\hat{x}_{j}\nabla_{i}F_{ij}^{a}(y\Hx).
\la{PiSiDot}
\ee{\en}
However, note that from Eq. \Ref{PhiDot} there follows the equation of motion for the charge 
 $Q^{a}_{\iy}(\Hx) \equiv 
\int_{0}^{\iy}y^{2}dy\,\Phi^{a}_{\perp}(y\Hx)$, which tells about the non--commutativity of the dynamics  $\a^{t}_{\pi}$ with the symmetry transformations,
\be{\eq} \la{QDot}
\dot{Q}_{\iy}^{a}(\Hx) =  gt^{abc}
Q^{b}_{\iy}(\Hx)\,(\s^{c}_{\iy} - \pi(\s_{\iy})^{c}).
\ee{\eq}
We used the assumption that the variable at infinity,
corresponding to the variable $x^{2}\hat{x}_{j}\nabla_{i}F_{ij}$,
is zero, which is similar to condition \Ref{H-23} in the formalism with the boundary 
terms.
Eq.  \Ref{QDot} is the analogue of \Ref{H-27} and it is natural to require,
the constraint, $Q_{\iy}=-\chi$,
where $\chi$ is some additional variable at infinity.
This is achieved by a similar addition of a surface term of the form
\Ref{H-8}, where $\tilde{\s}_{R}$ plays the role of $A_{0}(\Hx)$.
For any finite $R$ such a term is zero, nevertheless the
limiting expression in the weak topology is finite. These properties of the absence of
weak continuity are typical for theories with long--range interactions
and can be characterised as a {\it bifurcation} of an infinite dynamical system
\cite{Morc85}.
In the primary representation of $\pi$, the variable at infinity 
$\chi(\Hx)$ is fixed on its expectation, whose colour direction must coincide with $\pi(\s_{\iy})$.
The dynamics $\a^{t}_{\p}$ at infinity is obtained from \Ref{QDot} 
by calculating the limit
$x \rar \iy$ averages over the states $\f_{\si}(\,.\,) = \av{\si,\,.\,}$
characterised by the property
\be{\eq}   \la{DCXZ}
\lim_{\mo{{\bf x}} \to \iy}\f_{\si}(\vec{\s}({\bf x})) = \si
\ee{\eq}
In the Hartree--Fock approximation, meaning the condition of
clustering (weakening of correlations) of the form 
\Ref{Klust} and this is zero.
We finally write
\begin{equation} \la{EvV}
\dot{\sigma }^{a}_{\infty }(\hat{{\bf  x}})  =  -gt^{abc}\,
\sigma ^{b}_{\infty }(\hat{{\bf
x}})\,\pi(\sigma_{\iy}) ^{c}\,,\quad \sigma ^{a}_{\infty }
(\hat{{\bf x}})\in {\cal Z}_{{\cal S}},
\end{equation}
which is the equation of colour rotations
around the vector
 $\pi(\s_{\iy})$.

To conclude this section, we emphasise 
that the formalism of variables at infinity
leads to the same physical conclusions as the Hamiltonian formalism
with surface terms in the infinite limit.
For this reason, we shall hereafter freely use both terms ``variables
at infinity'' and ``variables at the boundary'' as equivalent, implying
that they appear in somewhat different mathematical contexts.




\subsection{Poincar\'e algebra in the Fock--Schwinger gauge}

      The obvious shortcomings of the Fock--Schwinger gauge are
the lack of explicit Poincar\'e- and even ${\bf R}^{3}$- translational covariance.
This in turn complicates the renormalisation procedure. Nevertheless,
the theory still possesses space--time symmetries, though
this fact is not as obvious as in the covariant gauges. The existence of these
symmetries of the Yang--Mills system imply the existence of operators $Q^{A}_{R}
\in \j{A}_{l}$ generating symmetry transformations
$$ \a^{q}[A] = w\mb{-}\lim_{R\to\iy}e^{i\,Q_{R}q}\,A\,e^{-i\,Q_{R}q}\,,
\qquad A\in\j{A}_{l}\,.$$
At the formal level, the existence of the Poincar\'e algebra in physical
gauge was demonstrated in Ref.  \cite{Best90}. There it was
shown that the operators
\be{\en}
P_{0} = H = \fr{1}{2}\int d{\bf x}\,({\bf E}^{2} + {\bf B}^{2})\,,
\qquad P_{i} = \int d{\bf x}\,T_{0i}\,,   \nn \\
M_{ij} = \int d{\bf x}\,(x_{i}T_{0j} - x_{j}T_{0i} + E_{i}^{a}A_{j}^{a}
- E_{j}^{a}A_{i}^{a})\,,     \nn \\
M_{0i} = t\,P_{i} - \fr{1}{2}\int d{\bf x}\ x_{i}\,({\bf E}^{2} + {\bf
B}^{2})\,, \qquad T_{0i} = - {\bf E}^{a}\pa{i}{\bf A}^{a},
\ee{\en}
restricted on the surface of the constraint and the gauge condition, satisfy
the commutation relations of the Poincar\'e group.

  The presence of variables at infinity in the FS gauge leads to
additional complexities. Therefore, it is necessary to
make sure that the translational ${\bf R}^{3}$ invariance
still exists, since its presence is necessary for a consistent formulation of the theory.
Note that in the Fock--Schwinger gauge one can write
\be{\eq} \la{P_i}
P_{i} = - \int d{\bf x}\,{\bf E}\pp^{a}\pa{i}{\bf A}\pp^{a}\,,
\ee{\eq}
and $P_{0} = H$, where $H$ is the Hamiltonian of the system. 

       In order to make sense of the $P_{\m}$ operators we need to introduce infrared
regularisation by means of the function $f_{R}(x)$. For functions $f_{R}(x)$
depending only on the radius it turns out that the commutation relations will be
fulfilled  in the weak sense
$$ \lim_{R,\,R'\to\iy} \f(\,[P_{\m}^{R}, P_{\n}^{R'}]\,) = 0\,,
\qquad \forall\, \f\in\p\,.$$
For the commutator of the spatial components of the momentum with each other this
formula is trivially verified. It is much more interesting to consider the commutator
$[H,P_{i}]$. We proceed with the QED case, where the variables at infinity
are absent in the equations of motion. As pointed out in Chapter \ref{G},
physical quantities in the FS gauge can be rewritten in terms of the
diagonalising Coulomb variables. In this case, the 
algebraic properties (\ref{G-26},\ref{G-27}) of the 
transition operators are used and the integration by parts is performed, which is admissible due to the presence of a spherically symmetric cut--off function. It turns out that in the limit of the removed cut--off
$$P_{i}^{FS} = P_{i}^{Cuol}\,,$$
as this happens for the Hamiltonian \Ref{H}.

   We now turn to the case of Yang--Mills fields. We write the expectation 
of the commutator as
$$\lim_{R\,R'}i\,\f([H_{R'},P_{i}^{R}]) =
\lim_{R\,R'}\fr{d}{dt}\biggl|\biggr._{{}_{t=0}}
\f(\a^{t}_{R'}[P_{i}^{R}]) = \lim_{R}\fr{d}{dt}
\biggl|\biggr._{{}_{t=0}}\f(\a^{t}_
{\p}[P_{i}^{R}])\,,$$
$P_{i}^{R}\in\j{A}_{l}$.
All we need is the derivative
$\fr{d}{dt}\a^{t}_{\p}$, which we calculate using (\ref{e1e},\ref{e2e}).
It consists of
of the local part and the contribution containing $\p(\s_{\iy})$. After a simple but
somewhat long transformations, the first summand can be written as the
integral of the gradient of a scalar even function, which is therefore
nullified. The remaining construction
\be{\eq}  \la{DDd}
\d\biggl(\fr{d}{dt}\a^{t}_{\p}[{\bf E}^{a}\pa{i}{\bf A}^{a}]\biggr) =
E^{a}_{k}\na{k}\pa{i}\,\p(\s_{\iy})\,,
\ee{\eq}
is obviously zero if and only if $\p(\s_{\iy})$ is a constant.
This is the necessary and sufficient condition for the dynamics,
which is representation--dependent, to preserve the translational invariance
$\a^{t}_{\p}\a^{{\bf x}} = \a^{{\bf x}}\a^{t}_{\p}$
in the considered representation $\pi$.

}

\section{Dependence of the partition function  on the variables at the boundary
}\label{P}

In this Chapter, we study the dependence of the partition function of the 
Abelian and non--Abelian gauge theories on the value of the variable
at the boundary $|\chi(\Hx)|$ following our work \cite{SveTim95IJMP}. 
In the framework of the Hamiltonian formalism of Chapter 
\ref{H} we have established that the surface Lagrange multiplier
$A_{0}(R\Hx)$ is not determined from the equations of motion.
The functional integral includes all its possible values.
The action of the system without the surface term is invariant 
with respect to time--dependent gauge transformations.
When taking into account the non--zero surface term with $|\chi|$,
this invariance is absent. 
The direction of the vector $\chi$, as we have shown, must be treated
as a Hamiltonian variable and has nontrivial 
\Ref{H-27} dynamics, while its length is fixed and plays the role of an external 
parameter characterising different phases of the system.   

In the theory at finite temperature, the possibility of time--dependent gauge transformations is furthermore limited by the the requirement of preservation of the periodic boundary conditions in time for the gauge fields, which will be discussed in Chapter 
\ref{W}. The question about physically feasible values of $|\chi|$
at a given finite temperature is reduced to finding the most
statistically probable values at which the partition function 
is maximal, and to the vacuum values at which the energy is minimal for zero temperature.

\subsection{Electrodynamics with an external charge}\label{P-E}

We shall begin our study with a simple case that admits an explicit solution, namely the 
electrodynamics with an external charge density.
We can work in the Coulomb gauge here.
Note that from the relations \Ref{G-28} and \Ref{G-29p} there follows the expression
of $\hx E^{\Vert}(R\hx)$ in the Gauss law
through the physical transverse variables of the Coulomb gauge.
\begin{equation}
R^{2}\hx E^{\Vert}(R\hx)- \int_{0}^{R}y^{2}dy\,\rho(y\hx) =
\hat{\Delta} \int_{0}^{R} (R-y)dy\,\hx{\bfM E}^{\perp}(y\hx)\,,
\label{BouRel}
\end{equation}
where we used the notation for the spherical part of the Laplacian 
 $\Delta = x^{-1}\partial^{2}_{x}\,x + x^{-2}\hat{\Delta}$.

The partition function of the theory in the spherical domain
can be represented as the following functional integral
\begin{eqnarray}
&& Z = \int {\cal D}{\bfM A} {\cal D}{\bfM E}\,\delta(\partial {\bfM A})
\delta(R^{2} E_{\Vert}(R\hat{{\bfM x}}) + \chi(\hat{{\bfM x}})) 
\nonumber\\ &&
\exp \int_{\Lambda}d^{4}x\left(i{\bfM E}\dot{{\bfM A}} - \frac{1}{2}{\bfM E}
^{2} + \frac{1}{2}{\bfM A}\Delta{\bfM A} - \frac{1}{2\epsilon^{2}}(\partial
{\bfM E} - \rho)^{2}\right)\,, \label{Z}
\end{eqnarray}
where we introduced the notation $\Lambda=[0,\beta] \times V$
(we also use the notation $\partial\Lambda=[0,\beta]\times\partial V$.)
The Gaussian constraint is regularised by using a small parameter $\epsilon$,
which must be set to zero at the end of calculations. This Gaussian integral
is computed in the standard way by shifting the integration variables.
To calculate the integral over ${\bf E}$, we introduce a new integration variable,
${\bf E}_{1}$,	
\begin{equation}
{\bfM E} = {\bfM E}_{1} + {\cal E}\,, \qquad {\bfM E}_{1}(R\hat{{\bfM x}}) 
= 0\,. \label{EEp}
\end{equation}
Here, the new variable ${\bfM E}_{1}$ satisfies the zero boundary condition 
and ${\cal E}$ is chosen so that there is no linear term on ${\bfM E}_{1}$.
This gives the following equation for determining ${\cal E}$
\begin{eqnarray}
&&i\dot{{\bfM A}} - {\cal E} + \frac{1}{\epsilon^{2}}\partial(\partial{\cal E})
- \frac{1}{\epsilon^{2}}\partial \rho = 0\,, 
\label{eqE} \\
&& R^{2}\hat{{\bfM x}}{\cal E}(R\hat{{\bfM x}}) + \chi(\hat{{\bfM x}}) = 0\,.
\label{boE}
\end{eqnarray}
The last boundary condition follows from the second delta function in \Ref{Z}.
It is possible to decompose this vector into longitudinal and transverse components in the
momentum space ${\cal E} = {\cal E}^{\perp} - \partial\varphi$.
Then the transverse component is ${\cal E}^{\perp} = i\dot{{\bfM A}}$,
and the equation for $\varphi$ takes the form
\begin{eqnarray}
&& (\Delta - \epsilon^{2}) \varphi = -\rho\,, \label{eqFi} \\
&& R^{2}\frac{\partial\varphi}{\partial R} = \chi(\hat{{\bfM x}})\,.
\label{boFi}
\end{eqnarray}
The partition function \Ref{Z} can be represented as a product
\begin{eqnarray}
&&\backspace \backspace Z = Z_{1} \tilde{Z}\,, \nonumber\\
&&\backspace \backspace 
Z_{1}= \int {\cal D}{\bfM A}^{\perp}{\cal D}{\bfM E}
\,\exp\int_{\Lambda} d^{4}x \,
\left(-\frac{1}{2}\dot{{\bfM A}}^{2} +
\frac{1}{2}{\bfM A}\Delta{\bfM A} - \frac{1}{2}{\bfM E}_{1}^{2}
- \frac{1}{2\epsilon^{2}}(\partial {\bfM E}_{1})^{2} \right)\,, \nonumber \\
&& \backspace\backspace 
\tilde{Z} = \exp \beta \left( -\frac{1}{2}\int
_{\partial V} d\hat{{\bfM x}}\,
\chi(\hat{{\bfM x}}) \varphi(R\hat{{\bfM x}}) - \frac{1}{2}\int_{\partial V}
d{\bfM x}\,\rho({\bfM x})\varphi({\bfM x}) \right)\,, \label{Zt}
\end{eqnarray}
where $\varphi$  is the solution of (\ref{eqFi},\ref{boFi}).

The solution (\ref{eqFi},\ref{boFi}) is obviously the sum of the homogeneous part,
$\phi$ satisfying the nontrivial boundary condition, and the inhomogeneous part satisfying the zero boundary condition 
\begin{equation}
\varphi = \phi - G \bullet \rho\, \qquad G = (\Delta - \epsilon^{2})^{-1}\,,
\label{FiFi}
\end{equation}
where $G$ is the Green's function with the Neumann zero boundary condition.
We introduce the following notations
\begin{eqnarray}
\backspace\tilde{Z} &=& \tilde{Z}_{\chi}\tilde{Z}_{\rho\rho}\tilde{Z}_{\rho\chi}\,, 
\nonumber \\
\backspace\tilde{Z}_{\chi} &=& \exp \left(- \frac{\beta}{2}\int_{\partial V} d\hat{{\bfM x}}\,
\chi(\hat{{\bfM x}}) \phi(\hat{{\bfM x}}) \right)\,, \nonumber \\
\backspace
\tilde{Z}_{\rho\rho} &=& \exp \left(\frac{\beta}{2}\int_{\partial V} d{\bfM x}d{\bfM y}\,
\rho({\bfM x}) G({\bfM x},{\bfM y})\rho({\bfM y})\right)\,, \label{Zrho} \\
\backspace
\tilde{Z}_{\rho\chi} &=& \exp\left(- \frac{\beta}{2}\int_{\partial V} d{\bfM x}\, 
\phi({\bfM x}) \rho({\bfM x}) \right) 
\exp \left(\frac{\beta}{2} \int_{\partial V} d\hat{{\bfM x}}\,
\chi(\hat{{\bfM x}})\,(G\bullet\rho)(\hat{{\bfM x}})\right). \nonumber
\end{eqnarray}
It is natural to consider this problem in terms of spherical coordinates.
The solution of \Ref{eqFi} regular inside the sphere is given by the formula
\begin{eqnarray}
&& \phi_{lm} = C_{lm}\,\sqrt{\frac{\pi}{2\epsilon r}}I_{l+1/2}(\epsilon r)\,
\label{Film} \\
&& I_{n-1/2}(z) = \sqrt{\frac{2}{\pi z}} z^{n} \left(\frac{1}{z}\frac{d}{dz}
\right)^{n}\,\cosh z\,.
\end{eqnarray}
The constant $C_{lm}$ is thus defined by equation (\ref{boFi}).
The role of the regulariser now $\epsilon$ becomes clear. The solution for
the zero mode will be
\begin{equation}
\phi_{00} = C_{00} \frac{\sinh \epsilon r}{\epsilon r}\,,
\qquad C_{00} = \frac{\chi_{00}}{R^{2}\epsilon (\cosh \epsilon R/
\epsilon R - \sinh \epsilon R/(\epsilon R)^{2})}\,,
\label{Fioo}
\end{equation}
and it has a $1/\epsilon^{2}$ singularity when $\epsilon$ tends 
to zero. At the same time, the solutions for the other modes are regular in this limit
and tend to 
\begin{equation}
\phi_{lm} = C_{lm}\,r^{l}\,,\qquad C_{lm} = \frac{\chi_{lm}}{lR^{l+1}}\,.
\end{equation}
In the consideration of the zero mode one should be very careful
and hold a finite $\epsilon$. The Green's function for the zero mode is defined
as a solution to the equations
\begin{eqnarray}
&&\left(\frac{1}{r}\frac{\partial^{2}}{\partial r^{2}}r - \epsilon^{2}
\right) G_{00}(r,r') = \frac{\delta(r-r')}{rr'}\,, \\
&&\frac{\partial G_{00}(r,r')}{\partial r}\biggl.\biggr|_{r = R} = 0\,
\end{eqnarray}
and can be easily found in the explicit form
\begin{eqnarray}
 G_{00}(r,r') &=& \frac{1}{\epsilon r r'}\biggl(\biggr. \frac{1}{2} \sinh \epsilon
|r-r'| - \frac{1}{2}\sinh \epsilon(r+r') \nonumber\\
&+& \frac{\sinh \epsilon R - \cosh \epsilon R/\epsilon R}
{\cosh\epsilon R - \sinh \epsilon R/\epsilon R}
\sinh \epsilon r \sinh \epsilon r' \biggl.\biggr)\,.
\end{eqnarray}
The leading terms in the Laurent decomposition at small $\epsilon$ have the form
\begin{eqnarray}
 G_{00}(r,r') &\simeq& -\frac{3}{\epsilon^{2}R^{3}} + \frac{9}{5 R}
- \frac{1}{2} \frac{r^{2} + r'^{2}}{R^{3}} -\frac{1}{\max (r,r')}\,, \\
 \phi_{00} &\simeq& \left(\frac{3}{\epsilon^{2}R^{3}} - \frac{3}{10 R}
+ \frac{r^{2}}{2 R^{3}}\right) \chi_{00}\,.
\end{eqnarray}
Substituting these expressions into the formulae (\ref{Zrho})
gives for the non--zero modes in the limit $\epsilon = 0$
\begin{equation}
\tilde{Z}_{\chi\,l>0} = 
\exp\left(-\frac{\beta}{2R}\sum_{lm > 0} \frac{|\chi_{lm}|^{2}}{l}\right)\,.
\end{equation}
For simplicity, we assume that the charge distribution $\rho$ is
spherically symmetric. In this case it gives a contribution only to the 
zero mode. We also introduce the notations for the integrals
\begin{eqnarray}
{\cal Q}_{00} &=& \int_{0}^{R} r^{2}\,dr\,\rho_{00}(r)\,,\\
{\cal G}_{00} &=& \int_{0}^{R} r^{4}\,dr\,\rho_{00}(r)\,.
\end{eqnarray}
Then the result can be represented as
\begin{eqnarray}
\tilde{Z}_{00} &=& \exp \beta\biggl(\biggr.
-\frac{1}{2}\int_{0}^{R} x^{2}dx\,y^{2}dy\,\frac{\rho_{00}(x)\,
\rho_{00}(y)}{\max(x,y)}
-\frac{3}{2\epsilon^{2}R^{3}}
({\cal Q}_{00}+\chi_{00})^{2} \nonumber\\
&+& \frac{1}{R}(\frac{9}{10}{\cal Q}_{00}^{2}
-\frac{1}{10}\chi_{00}^{2} + \frac{3}{10}\chi_{00}{\cal Q}_{00})
-\frac{1}{2R^{3}}({\cal Q}_{00}+\chi_{00}){\cal G}_{00} \biggl.\biggr)\,.
\end{eqnarray}
In the limit $\epsilon \rightarrow 0$, this expression will contain 
the delta function of the condition ${\cal Q}_{00} + \chi_{00} = 0$,
and we arrive at the following correction to the standard answer
\begin{equation}
\tilde{Z}_{00} = \exp\left(\frac{\beta}{2R}{\cal Q}_{00}^{2}\right)\,
\delta({\cal  Q}_{00} + \chi_{00})\,.
\end{equation}

As a simple illustration of this result, consider the charge density 
of charge $\rho_{00} = - \kappa/(\sqrt{\pi}\,r)$
corresponding to the linearly increasing potential $\varphi = \kappa\,r$.
For this exotic charge distribution, modelling the confinement potential,
we obtain an additional constant contribution to the free energy density  
\begin{equation}
\Delta {\cal F} = - \frac{\log Z}{\beta V} = -\frac{3}{32\pi^{2}}
\kappa^{2}\,.
\end{equation}
It is interesting to emphasise that its occurrence is entirely due to the boundary
terms, and this addition reduces the free energy. This example suggests
a possible connection between the confinement phenomenon in the non--Abelian gauge theory
and boundary effects.

\subsection{Formulation in terms of the collective variables}\label{P-C}

Here I am going to consider a slightly different formulation of this problem by means of
introducing a collective variable, $\sigma$, conjugate to the Gaussian constraint.
Both formulations are equivalent in the Abelian theory, where the transition from one to the other is carried out by a trivial change of the integration variables.
The new formulation turns out to be more fruitful in the non--Abelian theory.
We can rewrite the formula \Ref{Z} in terms of the collective variable $\sigma$,
introduced by means of the definition
\begin{equation}
\exp\left(-\frac{1}{2\epsilon^{2}}\int_{\Lambda}d^{4}x\,(\partial {\bfM E}
-\rho)^{2} \right) = \int {\cal D}\sigma\,\exp\int_{\Lambda}d^{4}x\,
\left(-\frac{\epsilon^{2}}{2}\sigma^{2} + i\sigma(\partial {\bfM E} -\rho)\right).
\end{equation}
We calculate the integral over ${\bfM E}$
\begin{eqnarray}
&& \int {\cal D}{\bfM E}\, \exp\int_{\Lambda}d^{4}x\, \left(
-\frac{1}{2}{\bfM E}^{2} +
i{\bfM E}\dot{{\bfM A}}+ i\sigma\partial{\bfM E}\right)\,\delta(R^{2}
E_{\Vert}(R\hx)+\chi(\hx)) \nonumber \\
&& = \exp\left(-\frac{1}{2}\int_{\Lambda} d^{4} x(\dot{{\bfM A}}^{2} 
+(\partial\sigma)^{2}) - i\int_{\partial \Lambda} dt\, d\hx\,\chi \sigma\right)\,.
\end{eqnarray}
After integration by parts in the summand $\int d{\bfM x}\,
\sigma\partial{\bfM E}$
we have applied the condition that follows from the delta function in the surface term.
In addition, we have performed integration by parts in the summand 
$\int d{\bfM x}\,\dot{{\bfM A}}\partial \sigma$,
where the corresponding surface term vanishes due to the gauge condition.
The dependence on the variable $\chi$ is contained only in the integral
\begin{equation}
\tilde{Z} = \int{\cal D}\sigma\,\exp\left(-\int_{\Lambda} d^{4}x\,\bigl(
\frac{1}{2}(\partial \sigma)^{2}+\frac{\epsilon^{2}}{2}\sigma^{2} +i\sigma\rho \bigr)
- i\int_{\partial\Lambda} dt\, d\hx\,\chi \sigma\right)\,.
\end{equation}
It can be seen that the boundary condition on $E_{\Vert}$ is equivalent to an 
additional surface term in this formula.  The integral is calculated
directly by shifting the integration variable $\sigma =\sigma_{1}+\varsigma$.
The new variable satisfies the boundary condition $R^{2}\sigma_{1}'(R\hx) = 0$,
and $\varsigma$ is found from the condition of
of absence of the linear term in $\sigma_{1}$ after this shifting. This gives the equation
on $\varsigma$
\begin{eqnarray}
&&(\Delta-\epsilon^{2})\varsigma = i\rho\,,
\end{eqnarray}
from which we obtain
$\tilde{Z}=\tilde{Z}_{1}\tilde{Z}_{2}$,
\begin{eqnarray}
\tilde{Z}_{1} &=& \int{\cal D}\sigma_{1}\,\exp\biggl(\biggr.
-\int_{\Lambda}d^{4}x\, \bigl(\frac{1}{2}(\partial \sigma_{1})^{2} +\frac{1}{2}
\epsilon^{2}\sigma_{1}^{2}\bigr) \nonumber\\
 &-& \int_{\partial\Lambda} dt\, d\hx\,\sigma_{1}(R^{2}\varsigma'+i\chi) 
\biggl.\biggr)\,, \\
\tilde{Z}_{2} &=&
\exp\left(-\frac{i}{2}\int_{\Lambda}d^{4}x\, \rho\,\varsigma -
\int_{\partial\Lambda} dt\,d\hx\,\varsigma(\frac{1}{2}R^{2}\varsigma'+i\chi)
\right)\,. 
\end{eqnarray}
Note that the boundary condition on $\sigma_{1}$ allows a certain
arbitrariness, namely
$\sigma_{1} =\sigma_{2}+\sigma_{R}(\hx),
\ \sigma_{2}(R\hx)=0$.
Integration over {\it variables at the boundary} $\sigma_{R}(\Hx)$ gives a surface delta function
\begin{equation}
\int{\cal D}\sigma_{R}(\hx)\exp\left(-\int_{\partial\Lambda} dt\,d\hx\,
\sigma_{R}(\hx) (R^{2}\varsigma'+i\chi)\right) =
\delta(R^{2}\varsigma'+i\chi)\,.
\end{equation}
Applying the boundary condition 
$\varsigma$, we finally obtain
\begin{equation}
\tilde{Z} =\exp\left( 
-\frac{i}{2} \beta \int_{V} d{\bfM x}\, \varsigma\rho
-\frac{i}{2}\beta\int_{V} d\hx\,\varsigma\,\chi
\right)\,.
\end{equation}
By renaming $\varsigma = -i\varphi$, we completely reproduce the answer 
(\ref{eqFi},\ref{boFi},\ref{Zt}) obtained earlier.
Thus, the formulation in terms of the collective variable is equivalent to the
traditional if we take into account the contribution of surface terms and the
integration over the variable at the boundary  $\sigma_{R}(\Hx)$.


\subsection{Effective action of gluodynamics
}\label{P-N}

Let us consider the Faddeev--Popov functional integral for the generating 
functional of the Green's functions 
 \cite{SveTim92PL,SveTim91Pr}
\begin{eqnarray}
\backspace\backspace
Z[\chi;\zeta,\eta]&=&\int\DD\Ap\,\DD\Ep\,\DD A_{0}(R\Hx)\,
\exp\biggl(\biggr.-\int_{\Lambda}d^{4}x\,
[{\bf E}\dot{{\bf A}} -\frac{1}{2}{\bf E}^{2}+\frac{1}{2}{\bf B}^{2} 
\nonumber \\ \backspace\backspace
&+&i\zeta \Ev-\eta B_{\Vert}]
+\int_{\partial \Lambda}dt\,d\Hx\,A_{0}(R\Hx)\,(R^{2} \Ev(R\Hx) +\chi(\Hx))
\biggl.\biggr). \label{FPi}
\end{eqnarray}
In this formula, it is assumed that $\Ev$ is expressed in terms of the transverse 
components using the relation \Ref{G-6}.
We shall use the Euclidean notation here, in which the electric field is purely imaginary. 
The fields satisfy the periodic boundary conditions in time
on the interval $[0,\beta]$ and the boundary conditions (\ref{H-20}-\ref{H-25})
at spatial infinity. We have introduced sources for the longitudinal components of the fields and the surface term with $\chi$.
For the sake of brevity, the source terms below will generally not be written out explicitly if not necessary. 

We can use the following simple observation. In the expressions  
\begin{displaymath}
{\bf E}^{2} = {\bf E}^{2}_{\perp }+E^{2}_{\Vert }\,, \qquad {\bf B}^{2} =
{\bf B}^{2}_{\perp }+B^{2}_{\Vert }\,, \qquad {\bf B}_{\perp } = P\,{\bf B} =
P\,\mbox{rot}\,\Ap
\end{displaymath}
it can be seen that all nonlocal and non--Abelian constructions are contained
only in the longitudinal components. It is useful to introduce two auxiliary periodic in
time real variables, $\lambda$ and $\nu$ by means of the integral representations
\begin{eqnarray}                     \label{li}
\exp\,(\frac{1}{2}\int _{\Lambda }dx\,E_{\Vert }^{2}\,) &=& \int
{\cal D}\lambda \
\exp\,(\,\int _{\Lambda }dx\,[-\frac{1}{2}\lambda ^{2}
+\lambda E_{\Vert }\,]\,)\,,
\\ 
\exp\,(-\frac{1}{2}\int _{\Lambda }dx\,B_{\Vert }^{2}\,) &=& \int
{\cal D}\,\nu \
\exp\,(\,\int _{\Lambda }dx\,[-\frac{1}{2}\nu ^{2}+i\,\nu B_{\Vert }\,]\,)\,.
\label{lili}
\end{eqnarray}
We perform a transformation 
\begin{eqnarray}
&&\int _{V_{R}}d{\bf x}\,\lambda (x\hat{{\bf
x}})\frac{1}{x^{2}}\int _{0}^{x}y^{2}\,dy\,\Phi_{\perp} 
(y\hat{{\bf x}}) = \int
_{V_{R}}d{\bf x}\,\Phi_{\perp} ({\bf x})\,\tilde{\sigma} _{R}({\bf x})\,,
\\&&
\tilde{\sigma} _{R}({\bf x}) \equiv \int _{x}^{R}dy\,\lambda (y\hat{{\bf x}})
\end{eqnarray}
and then integrate by parts
\begin{eqnarray}
&&\backspace\backspace\backspace
\int _{V_{R}}d{\bf x}\ \tilde{\sigma}_{R}({\bf x})\,\partial {\bf E}_{\perp }
= -\int _{V_{R}}d{\bf
x}\ {\bf E}_{\perp }\,P\partial \,\tilde{\sigma} _{R}
+\int _{\partial V_{R}}d\hat{{\bf x}}\ R^{2}{\bf
E}_{\perp }(R\hat{{\bf x}})\,\tilde{\sigma} _{R}(R\hat{{\bf x}}),
\nonumber \\&&\backspace\backspace\backspace
\int _{V_{R}}d{\bf x}\ \nu \,\Hx\mbox{rot}\,{\bf A}_{\perp } 
= \int _{V_{R}}d{\bf x}\ {\bf
A}_{\perp }\ [\,\partial \,\nu \ ,\ \hat{{\bf x}}\,]
+\int _{\partial V_{R}}d\hat{{\bf
x}}\ R^{2}\nu \ [\,\hat{{\bf x}}\ ,\ {\bf A}_{\perp }\,]\,.
\end{eqnarray}

The surface terms are identically zero due to the definition of the
of transversality.
Note that the integrals over the transverse components become Gaussian after introduction of the collective variables and can be explicitly computed.
For a more compact notation it is convenient to use
a new variable similar to the one we introduced in the formula \Ref{H-36},
and taking non--trivial values at the boundary,
\begin{equation}
\sigma(\Bx) \equiv A_{0}(R\Hx) + \tilde{\sigma}_{R}(\Bx),
\end{equation}
and by integration over this variable we shall understand the integrations 
over the original variables 
$\DD\sigma \equiv \DD\lambda\,\DD A_{0}(R\Hx)$.
Next, we integrate over $\Ep$
\begin{eqnarray}
I &=& \int {\cal D}{\bf E}_{\perp }\,\exp\,(\,\int
_{\Lambda }dx\,[\,\frac{1}{2}({\bf E}^{a})^{2}-{\bf E}^{a}\,(\,{\bf
\dot{A}}^{a}+P\partial\,\sigma^{a}-gt^{abc}\,{\bf A}^{b}\,\sigma^{c}\,)\,]\,)
\nonumber \\
& =& \exp\,(\,-\frac{1}{2}\int _{\Lambda }dx\,[\,\dot{{\bf
A}}^{a}+P\partial\,\sigma^{a}-gt^{abc}\,{\bf A}^{b}\,\sigma ^{c}\,]^{2}\,)\,.
\end{eqnarray}
In the integral over $\Ap$, which arises after integration over the electric field, we make a rotation
\begin{displaymath}
A_{\perp \,i} = \varepsilon _{ijk}\,{\cal A}_{\perp \,j}\hat{x}_{k},
\end{displaymath}
preserving the integration region and having as the Jacobian a non--zero normalisation constant.
In the integral\\ $\int_{V_{R}}d\Bx\,\left(\frac{1}{x}\Hx\partial 
(x{\cal A}_{\perp})\right)^{2}$ 
it is possible to integrate by parts,
and the surface term is zero due to the boundary condition \Ref{H-25}.
After these calculations, the partition function is written as
\begin{eqnarray}              \label{res}
\backspace
Z &=& \int{\cal D}\,{\bf {\cal A}}_{\perp }\,{\cal D}\,\sigma \,{\cal D}\,\nu\
\exp\,(\
-\frac{1}{2}\nu\bullet\nu-\frac{1}{2}\partial\sigma
\bullet\partial\sigma \nonumber\\
\backspace
& - &\frac{1}{2}{\cal A}\bullet M \bullet {\cal A}
+{\cal A}\bullet N\ )\,,
\\ \backspace 
M^{ab} &=& (-\Delta _{x}\,\delta ^{ab}-
(\nabla _{t}^{2})^{ab}\,)\,P+i\,g t^{abc}\,\nu ^{c}\,L\,,
\\ \backspace
N^{a}_{i} &=& i\,P_{ij}\,\partial _{j}\,\nu ^{a}-L_{ij}\,
\nabla _{t}^{ab}\,\partial _{j}\,\sigma ^{b}\,, \qquad
\nabla _{t}^{ab} = \delta ^{ab}\,\partial _{t}-g t^{abc}\,\sigma ^{c}    \,.
\end{eqnarray}
Here the bullet $\bullet$ means integration over the domain $\Lambda$, and operators 
 $P$ and $L$ are defined as
\begin{equation}
L_{ij} = \varepsilon_{ijk}\, \hat{x}_{k}, \qquad
P_{ij} = \delta_{ij} - \hat{x}_{i}\,\hat{x}_{j}.
\end{equation}
It is trivially verified that they satisfy a simple algebra
\begin{equation}\label{PL-Alg}
P^{2}=P, \qquad L^{2}= - P, \qquad L P = P L = L.
\end{equation}
Let $\alpha$, $\beta$ and  $\gamma$ be two arbitrary operators, which commute with 
$P$ and $L$, but not necessarily with each other. 
We denote the full trace with a capital letter, and the trace over the degrees of
of freedom except those corresponding to  $P$ and $L$ with a lowercase one. 
The algebra \Ref{PL-Alg} allows one to derive
\begin{eqnarray}
&&(\,\alpha \,P+\beta \,L\,)^{n} = \alpha _{n}\,P+\beta _{n}\,L\,,\nonumber\\
\alpha _{n} &=& \mbox{Re}\,(\alpha +i\,\beta )^{n}\equiv \frac{1}{2}[\,
(\alpha +i\,\beta )^{n}+(\alpha -i\,\beta )^{n}\,]\,,
\nonumber\\
\beta _{n}& =& \mbox{Im}\,(\alpha +i\,\beta )^{n}\equiv 
\frac{1}{2\,i}[\,(\alpha +i\,\beta )^{n}-(\alpha -i\,\beta )^{n}\,]\,,
\nonumber
\end{eqnarray}
which gives 
{\hfuzz 18pt
\begin{eqnarray}
\mbox{\mbox{Tr}\,log}\,[\,\gamma \,{\bf 1}+\alpha \,P+\beta \,L\,] 
&=&\mbox{tr}\log \gamma +\mbox{tr}\log[\,(\alpha +\gamma +i\,\beta )
\,(\alpha +\gamma -i\,\beta )\,]
\,,\nonumber\\
(\,\gamma \,{\bf 1}+\alpha \,P+\beta \,L\,)^{-1}
&=&\gamma ^{-1}({\bf 1}-P)+\mbox{Re}\,(\,\alpha +\gamma +i\,\beta )^{-1}\,
P+  \nonumber\\
&+&\mbox{Im}\,(\,\alpha +\gamma +i\,\beta )^{-1}\,L\,.   \label{TrLog}
\end{eqnarray}
}

The integral over ${\cal A}_{\perp}$ is calculated directly and with the use of 
the above formulae one can exclude the spatial
vector structure from the result. A slightly different and more elegant way to treat this is to 
introduce the projection operators and the corresponding components of the 
transverse vectors
\begin{equation}
\Pi_{\pm}=\frac{1}{2}(P \pm i L) \,,\quad
{\bf a}_{\pm}=\Pi_{\pm}\,{\bf a}.
\end{equation}
They also possess a simple algebra
\begin{equation}
\Pi _{\pm }= \Pi _{\pm }^{2}\,,\quad \Pi _{+}\Pi _{-}  =  0\,,\quad
\Pi _{+}+\Pi _{-} = P\,,
\end{equation}
We can rewrite \Ref{res} via ${\cal A}_{\pm}$ projections 
\begin{equation}  \label{Res}
Z = \int{\cal D}\,{\cal A}_{+}\,{\cal D}\,{\cal A}_{-}\,
{\cal D}\,\sigma \,{\cal D}\,\nu \
\exp\,(\,-W[\sigma ,\nu ,{\cal A}_{\pm }\,]\,)\,,
\end{equation}
\begin{displaymath}
W[\sigma ,\nu ,{\cal A}_{\pm }] = \frac{1}{2}\nu\bullet\nu
+\frac{1}{2}\partial\sigma \bullet\partial\sigma  + \frac{1}{2}{\cal A}\bullet
{\cal C}\bullet {\cal A} + i\,{\cal A}\bullet {\cal K}\,,
\end{displaymath}
\begin{displaymath}
{\cal A} = (\,{\cal A}_{-}\,,\,{\cal A}_{+}\,)\,, \qquad  \quad {\cal K} 
= (\,K_{+}\,,K_{-}\,)\,,
\end{displaymath}
\begin{displaymath}
{\cal C} =
\left(
\begin{array}{cc}
0 & C_{+} \\
C_{-} & 0
\end{array}              \right)\,,  \quad \qquad   
D^{ab} = g t^{abc}\,\nu ^{c}\,,
\end{displaymath}
\begin{displaymath}
C_{\pm } = -\Delta _{x}-\nabla _{t}^{2}\pm D\,, \qquad  \quad 
K_{\pm } = \partial _{\pm }\,\nu \pm \nabla _{t}\,\partial _{\pm }\,\sigma\,.
\end{displaymath}
The integrals over ${\cal A}_{\pm }$ are now elementary. We have
\begin{eqnarray}
Z[\chi] &=& \int{\cal D}\,\sigma \,{\cal D}\,\nu \ 
\exp\,(\,-W[\sigma ,\nu ]\,
+\int_{\partial\Lambda}dt\,d\Hx\,\sigma\chi
)\,,  \label{Result} \\
 W[\sigma ,\nu ] &=& \frac{1}{2}\nu\bullet\nu
+ \frac{1}{2}\partial\sigma\bullet\partial\sigma  + \frac{1}{2}K_{-}\bullet
C_{+}^{-1}\bullet K_{+} \nonumber\\
&+& \frac{1}{2}K_{+}\bullet C_{-}^{-1}\bullet K_{-} 
+ \frac{1}{2}\mbox{tr}\,\mbox{log}\,C_{+}C_{-}\,,    \nonumber
\end{eqnarray}
where we have explicitly written out the dependence on the parameter $\chi$.
Allow me to discuss the issue of accounting for the sources. Firstly, in order to include
the source $J$ of the gauge field $\Ap$ one must add to the action the term
$\Delta W = gJ_{0}\bullet\sigma$
and change
$K_{\pm}\rightarrow
K_{\pm}\pm g J_{\pm}$
in the previous formula.

We can relate the generating functional of the longitudinal components of the strengths
\Ref{FPi} with the generating functional
\begin{equation}
{\cal Z}[\zeta ,\eta ] = \int {\cal D}\,
\lambda \ {\cal D}\,\nu \
\exp\ [\,-W[\lambda ,\eta ]-i\,(\,\lambda\bullet\zeta +\eta\bullet\nu \,)\,]\,.
\end{equation}
For this in Eq.  (\ref{FPi}) it is sufficient to represent this as
\begin{displaymath}
e^{\int _{\Lambda }dx\,(\,\eta B_{\Vert }-\frac{1}{2}\nu ^{2}
+i\,\nu B_{\Vert }\,)} = e^{-\int _{\Lambda }dx
\,\frac{\nu ^{2}}{2}}e^{-\int _{\Lambda }dx\,i\,\eta 
\frac{\delta }{\delta \,\nu }}e^{\int _{\Lambda }dx\,i\,
\nu B_{\Vert }},
\end{displaymath}
and then functionally integrate by parts 
\begin{displaymath}
\int {\cal D}\nu \,\mbox{(r.h.s.\ of \ prev.\ f.)} = \int
{\cal D}\nu \,e^{-\int
_{\Lambda }dx\,(\,\frac{1}{2}[\nu +i\,\eta ]^{2}-i\,\nu B_{\Vert }\,)},
\end{displaymath}
and completely analogously for $\lambda$, which gives
\begin{equation}       \label{gfc}
Z[\zeta ,\eta ] = \exp\,[\,\frac{1}{2}\int _{\Lambda }dx\ (\,\zeta ^{2}
+\eta ^{2}\,)\,]\ {\cal Z}[\zeta ,\eta ] \,.
\end{equation}

A connection formula for the averages can be deduced after performing some transformations with the shift operators in the functional space and utilising their functional integral representations.
Let $F(E_{\Vert },B_{\Vert })$
be an arbitrary functional of the longitudinal strengths. Then the mean value 
\begin{equation}    \left .
\langle \,F(E_{\Vert },B_{\Vert })\,\rangle  =
Z^{-1}[\zeta ,\eta ]\,F(\,i\,\frac{\delta }{\delta \,\zeta }\,,\,
\frac{\delta }{\delta \,\eta }\,)\
Z[\zeta ,\eta ]       \right |_{\zeta  = \eta  =0}
\end{equation}
is expressed via the mean $\langle F(\lambda ,\nu )\rangle $, 
defining ${\cal Z}[\zeta ,\eta ]\ $, by means of the formula
\begin{equation}     \label{Conn}
\begin{array}{c}
\langle \,F(\,E_{\Vert }\,,\,B_{\Vert }\,)\,\rangle  = 
\int{\cal D}\,
\xi \ {\cal D}\,\theta \
\exp\,(\,-\frac{1}{2}\int
_{\Lambda }dx\,[\,\xi ^{2}+\theta ^{2}\,]\,)\,   \\
\qquad\langle \,F(\,\lambda +i\,\xi \,,\,-i\,\nu +\theta \,)\,\rangle \,.
\end{array}
\end{equation}

The latter relationship is the basis for the interpretation of the collective
variables. They can be thought of as longitudinal (chromo) electric
and magnetic fields smoothed out by a Gaussian noise. It is interesting to note, in addition,
that although $\lambda$ and $\nu$ are real in the integral expressions, their averages are 
purely imaginary
\begin{equation}  \label{BE}
\begin{array}{lcrclcr}
\langle \,E_{\Vert }\,\rangle  &=& \langle \,\lambda \,\rangle \,,
& \qquad  \qquad &\langle \,B_{\Vert }\,\rangle  &=& -i\,\langle 
\,\nu \,\rangle    \,, \\
\langle \,E_{\Vert }\,\rangle ^{*} &=& -\langle \,E_{\Vert }\,\rangle \,, 
&\qquad  \qquad &\langle \,B_{\Vert }\,\rangle ^{*}
 &=& \langle \,B_{\Vert }\,\rangle \,.
\end{array}
\end{equation}

Let us apply the stationary phase method to the action of \Ref{Result}.
In the ``classical'' approximation on the variables $\sigma$ and $\nu$,
which is a quasi--classical in the initial gauge fields,
we need to find the minima of the effective action. We can use the natural 
Ansatz of constant collective variables.
According to Eq.  \Ref{BE}, we need to take the purely imaginary magnetic field
at the saddle point, so we may change the notation
$\nu\rightarrow i\nu$.
In terms of spectral expansions for the operators $\partial_{t}$
(periodic boundary conditions) and $\Delta_{x}$ (the Neumann condition on the boundary),
we can write 
\begin{eqnarray}
&& W[\sigma,\nu] = -\frac{\beta V_{R}}{2}\nu^{2} 
+2\pi\delta(\hat{{\bf 0}})\sum_{m=0}^{\infty}\sum_{n=-\infty}^{\infty}
\tr\log C_{+} C_{-}, \\
&& C_{\pm}^{ab} = (\omega_{n}^{2}+\Omega_{m}^{2})\delta^{ab}
+ i(2\omega_{n}\sigma^{c}\pm \nu^{c})gt^{abc} -(gt^{abc}\sigma^{c})^{2},
\end{eqnarray}
where $\omega_{n}= 2\pi \beta n$ and  $\Omega_{m} = \pi(m+1/2)/R$.
It is convenient to introduce the notations for the free energy density,
${\cal F}_{R}$,
\begin{equation}
W_{R} = \beta V_{R}\, {\cal F}_{R}\,, \quad 
{\cal F}_{R} = \gamma_{R}\,F_{R}\,, 
\quad \gamma_{R} = \frac{8\pi^{2} R \delta({\hat{0})}}
{\beta^{2}\,V_{R}}\,,
\end{equation}
where $V_{R} = 4\pi R^{3}/3$ is the domain volume.
Here $\delta(\hat{0})=\frac{1}{4\pi}\sum_{l}(2 l +1)$ 
denotes 
the angular delta function with coinciding arguments,
which is ultraviolet divergent.
It is more convenient to rewrite the result in terms of the variables
$s =(g\beta/2\pi)\sigma$, $v=(g\beta/2\pi)^{2}\nu$.

We first restrict ourselves to the simpler case of the gauge group SU(2),
and then generalise the result to SU(3) group. We denote $v=u^{2}$ and use the
formulae
\begin{equation}
\prod_{m=0}^{\infty}\left(1+\frac{a^{2}}{(m+1/2)^{2}}\right) = \cosh \pi a,
\quad 
\prod_{m=1}^{\infty}\left(1+\frac{a^{2}}{m^{2}}\right) = \frac{\sinh \pi a}
{\pi a}.
\end{equation}
It can be shown (see Appendix B), 
that the presence of a part of the colour vector $v$ transverse to
$s$ leads to the imaginary contribution to the effective action.
Therefore, we restrict ourselves to the Ansatz in which these vectors are collinear.
Note that for the group $SU(2)$ the operator ${\cal L}^{ab} = t^{abc}\hat{s}^{c}$
has the same algebra as the operator $L$ considered earlier in
${\bf x}$-space, and a determinant
of the linear combination of its powers is calculated using Eq.
\Ref{TrLog} formulae.
After computing the colour trace we arrive at the answer 
(up to an irrelevant constant)
\begin{eqnarray}
&&\backspace \backspace
F_{R}[u,s] = -a\,u^4 + {\cal U}_{R}[u,s]\,, \\
&&\backspace\backspace
 {\cal U}_{R}[u,s] = {\cal U}_{R}[s] + {\cal V}_{R}[u,s]\,, \\
&&\backspace\backspace
 {\cal V}_{R}[u,s] = \frac{\beta}{2\pi R}\sum_{n=-\infty}^{\infty}
\log \frac{L_{R}((n+s)^{2}+u^{2})\,L_{R}((n+s)^{2}-u^{2}) }
{L_{R}^{2}((n+s)^{2})}\,, \\
&&\backspace\backspace
 {\cal U}_{R}[s] = \frac{\beta}{\pi R}\sum_{m=0}^{\infty}\log \left( 1 -
\frac{\cos 2\pi s}{\cosh (\pi(m+1/2)\beta/R)}\right)\,,
\end{eqnarray}
where $L_{R}(x) = \cosh (2\pi R\sqrt{x}/\beta)$ and
$a=(2\pi)^{4}/(2g^{2}\beta^{4}\gamma_{R})$.
The expression is periodic in $s$ and we can limit ourselves to the period
$0 \le s \le 1$, or a half--period, due to parity with respect to 
$s=1/2$.
At finite $R$, calculating the derivative of the effective action on $u^{2}$ 
and equating it to zero, we can see that the resulting equation 
admits only a trivial solution $u=0$. Further, by assuming $u$ to be zero,
it is easy to see that the minimum of the action is reached at $s=0$.
Thus, there is no phase transition in a finite size system, which
can serve as an illustration of the Yang and Lee's theorem \cite{Glim81}.
The behaviour of the effective action changes in the thermodynamic limit
$R\rightarrow\infty$:
\begin{eqnarray}
{\cal U}[s,u] & = & \frac{1}{2\pi ^{2}}\int
_{0}^{\infty }dv\ \log\,\biggl[\,(\,\cosh\,\sqrt{v^{2}
+(2\pi u)^{2}}-\cos\,[2\pi s]\,)
\biggr. \nonumber\\
& &\biggl. (\,\cosh\,\sqrt{v^{2}-(2\pi u)^{2}}
-\cos\,[2\pi s]\,)\, \biggr/ \biggl.
\cosh^{2}v \biggr]\,, \\ 
{\cal U}[s,0] &=& -\frac{1}{2}(1-2s)^{2}+\frac{1}{8}\,,
\qquad  \quad (0 \leq  s \leq  1)\,,
\end{eqnarray}
where we have used the formulae for calculating the integral
$$I(a,b) = \int_{0}^{\infty}dx\frac{\cosh bx}{\cosh x +a}
= \frac{\pi}{\sqrt{1-a^{2}}}\frac{\sin(b\arccos a)}{\sin\pi b},$$
since
$${\cal U}[0,s] =-2\frac{\partial}{\partial b}\biggl.\biggr|_{b=1}
\left(I(a,b) - I(0,b) \right).$$
The integral can be rewritten as the series
\begin{eqnarray}
\mbox{Re}\,{\cal V}[s,u]&=&\sum_{n = -\infty }^{\infty }\biggl(
\,[\,\sqrt{(n+s)^{2}+u^{2}}+\sqrt{(n+s)^{2}-u^{2}}_{+}-\biggr. \nonumber\\
&-&\ \ 2\mid n+s \mid \biggl. \biggr)\,,\nonumber \\
\mbox{Im}\,{\cal V}[s,u]&=&\sum_{n = -\infty }^{\infty
}\sqrt{u^{2}-(n+s)^{2}}_{+}\ .
\end{eqnarray}
The graphs of the real and imaginary parts are presented in Fig.\ 1 and 2.
Applying to these series the Poisson summation formula from Appendix B we
find the expansions via the cylindrical functions
\begin{eqnarray}
\mbox{Re}\,{\cal V}[s,u] &= &2s(s-1)+u\sum_{n =
1}^{\infty }\frac{\cos[2\pi ns]}{n}\biggl
(Y_{1}(2\pi nu)-\frac{2}{\pi }K_{1}(2\pi nu)\biggr)\,,
\nonumber    \\
\mbox{Im}\,{\cal V}[s,u]&=&\frac{\pi u^{2}}{2}
+u\sum_{n = 1}^{\infty }\frac{\cos[2\pi ns]}{n}J_{1}(2\pi nu)\,.
\end{eqnarray}
The asymptotic formulae can be derived from the asymptotics of the Bessel and Neumann functions and are expressed through the periodic continuation by the 
Hurwitz formula from Appendix B of the generalised zeta function
\begin{eqnarray}
\mbox{Re}\,{\cal U}[s,u]&=&\sqrt{2u}\,[\,\zeta _{H}(-\frac{1}{2},u+s)
+\zeta _{H}(-\frac{1}{2},u-s)
\,] \,,\nonumber \\
\mbox{Im}\,{\cal U}[s,u]&=&\frac{\pi u^{2}}{2}
+\sqrt{2u}\,[\,\zeta _{H}(-\frac{1}{2},u+s)+\zeta _{H}(-
\frac{1}{2},u-s)\,]\,.
\end{eqnarray}
Given the periodicity and the values at the endpoints of the zeta function,
it is easy to come to the following conclusion below, which is also valid beyond the limits of
the applicability of the asymptotic formulae.

We calculate the derivative 
 $\partial F[u,s]/\partial u^{2}$
 and equate it to zero at the point $u=s=1/2$.
 Hence we determine the critical 
the value of the parameter $a$
\begin{equation}
\backspace
a_{c} = 2\sqrt{2}+4\sum_{n=1}^{\infty}\left(
\left((2n+1)^{2}+1 \right)^{-1/2} - \left((2n+1)^{2}-1 \right)^{-1/2}
\right).
\end{equation}
By using a different summation 
\begin{equation}
(m^{2} \pm 1)^{-1/2} = \frac{1}{m}+\sum_{k=0}^{\infty}\frac{(\mp 1)^{k}
(2k-1)!!}{2^{k}k!}\frac{1}{m^{2k+1}}
\end{equation}
and connection to the Riemann $\zeta$-function  $\sum_{k=0}^{\infty}(2k+1)^{-s}
=(1-2^{-s})\,\zeta(s)$ we arrive at the rapidly converging series form
\begin{eqnarray} 
&& \backspace\backspace a_{c} = 2\sqrt{2} - 8
\sum_{k = 0}^{\infty}\frac{(4 k +1)!!}{(2k + 1)!\,2^{2 k + 1}}
\left((1 - \frac{1}{2^{4 k+3}})\,\zeta(4 k + 3) - 1 \right),
\label{PL23}
\end{eqnarray}
and this can be computed numerically as approximely  $a_{c} \simeq 2.61882$.

If the parameter ``$a$'', which has the meaning of temperature squared, 
is greater than the value of $a_{c}$, then
there is only a trivial phase with minima at integer 
values of $s$ and $u$. At smaller $a$, there are
the free energy minima at half--integer values of 
$u$ and $s$ (see Fig. 3).
The stability condition $\mbox{Im}\ F =0$ (Fig. 2) gives
$u \le s$ and excludes all minima except the trivial $u=s=0$ and
non--trivial $u=s=1/2$, the second of which is deeper.
Note that the minima are also achieved at the results of the action on the
$s=1/2$ of transformations from the centre $Z_{2}$ by means of shifts
$s\rightarrow s +1$.

It is easy to see that in the thermodynamic limit, the points of the stationary phase
will be not only constant fields. Indeed, the operators
under the trace are diagonal over $\Hx$, and therefore 
the existence of the dependence $s(\Hx)$ leads to the replacement of the value $4\pi$ by the
integral over the sphere.
The kinetic terms thus become nontrivial, but they are proportional to
$R$ and can be discarded in the thermodynamic limit.
As we have seen, the unit vector in the colour space,
to which the stationary phase solution is proportional,
has been an arbitrary constant, but now we can see
that it can be even an arbitrary function of 
 $\hat{s}(\Hx)$.  

Our analysis of the formulae for SU(2) group suggests, 
that we can restrict ourselves to an Ansatz in which the colour vectors 
can be simultaneously 
transformed to a basis with only their Cartan components
($a=3,8$) are different from zero,
since the presence of the $v$ part of the transverse $s$ leads to an imaginary part of the
are effective, and hence are not of
interests. We introduce the notations 
\begin{equation}
s_{\pm} \equiv \frac{1}{2}(s_{3} \pm \sqrt{3}\,s_{8}), \qquad
s_{0} \equiv s_{3} = s_{+}+s_{-}
\end{equation}
and analogously for $v$. Then after calculation of the colour trace one can obtain
 \cite{SveTim92Pr}
\begin{eqnarray}
F_{SU(3)}[v,s] = \sum_{A=0,\pm}\left( 
-\frac{2a}{3}v_{A}^{2}+ {\cal U}_{SU(2)}[v_{A},s_{A}] \right).
\end{eqnarray}

Thus, the free energy density of SU(3) group is arranged as the
sum of three independent expressions of SU(2) group.
By virtue of the behaviour of this function for the group $SU(2)$ one can see,
then a nontrivial minimum is reached when one of the values of 
$s_{+}$, $s_{-}$, $s_{+}+s_{-}$, and the other two -- $1/2$,
and the corresponding components $v$ are similarly zero and $1/4$.
Thus the minima of the effective action 
(see Fig.~4)
are the colour vectors with components $3,\ 8$:
$\vec{s}_{\pm}=(1/2,\pm 1/(2\sqrt{3}))$ and
$\vec{s}_{0}=(0,1/\sqrt{3})$,
as well as the ones obtained from them
by the centre transformation actions $Z_{3}$:
$s_{3}\rightarrow s_{3}+1$, $s_{8}\rightarrow s_{8}+1/\sqrt{3}$ 
or $s_{3}\rightarrow s_{3}$, $s_{8}\rightarrow s_{8}+2/\sqrt{3}$.

All these minima have the same depth, which at temperatures
below critical, is greater than the depth of the
trivial minima obtained by centre transformations from zero.
The phase transition temperature at which the non--zero values occur, is equal to 
\begin{equation}
T_{SU(3)}= \sqrt{\frac{3}{2}}\,T_{SU(2)}.
\end{equation}

Allow me to return to the consideration of the dependence on the boundary variable $\chi$.
Without loss of generality, we can restrict ourselves to the formulation for SU(2) group again.
Thus, the formula \Ref{Result} gives us the desired representation of the form
\begin{equation}
Z[\chi] = \int {\cal D}\sigma\,\exp \left( -W[\sigma] 
- i\int_{\partial\Lambda}dt\,d\hx\,\sigma\chi \right)\,.
\end{equation}
In the zero stationary phase approximation, the partition function is determined by the classical
equation of motion for the collective variable.
We decompose the action in the neighbourhood of the saddle point
\begin{eqnarray}
W[\varsigma + \sigma_{1}] &=& W[\varsigma]+ \int_{\Lambda} d^{4}x\,
\frac{\delta W}{\delta \varsigma({\bf x})} \sigma_{1}({\bf x}) \nonumber \\
&& + \int_{\partial\Lambda} dt\,R^{2}d\hx\,
{\cal E}^{(1)}[\varsigma]\,\sigma_{1}(R\hx)+
\ldots,
\end{eqnarray}
where ${\cal E}^{(1)}[\varsigma]$ is the first Euler derivative of the action. 
In the zero approximation, after integration over the
variable at the boundary $\sigma_{R}(\Hx)$ we get
\begin{eqnarray}
&& \backspace\backspace
Z[\chi] \propto \exp\left(-W[\varsigma]-i\int_{\partial\Lambda}
dt\,d\hx\, \varsigma \chi\right)\,
\delta\left(R^{2}{\cal E}^{(1)}[\varsigma](R\hx)+ 
i\chi(\hx)\right), \\
&& \backspace\backspace
\frac{\delta\,W}{\delta \varsigma} = 0\,.
\end{eqnarray}
The equation and the related boundary condition 
\begin{equation}
\frac{\delta W}{\delta \varsigma}=0,\qquad R^{2}\varsigma'(R\hx) = -i\chi(\hx)
\label{NonAbSys}
\end{equation} 
are rather complicated in the non--Abelian case.
For small $\chi(\Hx)$ we can 
decompose $\varsigma = \varsigma^{(0)}+ \varsigma^{(1)}$
solution near the solution with the trivial boundary condition
$\varsigma^{(0)\,'}(R\hx) =0$. The modulus of the solution is defined by
the stationary phase value, and the unit vectors $\hat{\chi}$ and
$\hat{\varsigma}$ must coincide as can be seen from the limiting
behaviour at small $\chi$.

As we have demonstrated, the Abelian theory admits only a trivial solution
$\varsigma^{(0)}=0$.
In the non--Abelian theory, the same situation is realised for $T > T_{c}$.
In this case, $\varsigma$ is determined from the linearised equation
and the dependence $Z_{R}[\chi]$ has a behaviour similar 
to the Abelian theory. Namely, it contains a delta function expressing the
the global charge conservation law, but besides this, it is trivial in the
sense that $Z[\chi_{lm}]\rightarrow 1$ as $R\rightarrow\infty$.

However, at temperatures below the critical point, there is a constant
non--zero solution $|\varsigma^{(0)}|=\pi/g\beta$ 
and the relation takes the form
\begin{equation}\label{PROd}
Z[\chi] = \prod_{\hx}\cos \beta\varsigma^{(0)} \chi(\hx) \,.
\end{equation}
In this formula, we used the fact that $\varsigma^{(0)}$
can have both signs independently at every point on the sphere $\Hx$.
The product of \Ref{PROd} is zero if $\chi(\Hx)$ is different 
from zero and equal to one otherwise. Thus,
only the value $\chi(\hx) = 0$ is statistically possible in the phase of
confinement.

When calculating averages, we need to consider the measure of the phase space
of the variables $\hat{\chi}$, the orbits of the adjoint transformation of the group,
and either sum over all admissible values of $|\chi|$,
or consider distinct ``domains'' with arbitrary values of this quantity.
Hence, the temperature average of some observable $A$ 
is given by the following formula
\begin{equation} \label{AVv}
<A> = \left\{ \begin{array}{ll}
\langle A \rangle_{0}, & T < T_{c},\\
\int d\chi(\hx)\, \langle A \rangle_{\chi}, & T>T_{c}, \\ 
\end{array}\right. 
\end{equation}
where
\begin{equation}
\langle A \rangle_{\chi} = \frac{1}{Z[\chi]}
\int {\cal D}\sigma\,{\cal D}\nu\, e^{-W[\sigma,\nu]
- i\int_{\partial\Lambda} dt\,d\hx\,\sigma\chi}\,A[\sigma,\nu]\,.
\end{equation}

We shall not discuss here the proper procedure for calculating averages
in the deconfinement phase, although the situation seems to be analogous to the
known physical systems such as spin glasses \cite{Meza87}.
If we compute the integral over all $\chi$ at high temperature,
it will yield $\delta(\sigma_{\infty}(\hx))$, and hence effectively remove the 
integration over the variable at the boundary. In contrast, at low temperatures
only the value $\chi=0$ is realised and the
the integral over the variable at the boundary in the mean $\langle A \rangle_{0}$
gives the singlet projector $P_{s}$ of the group of the large gauge transformations
\begin{eqnarray}
\langle A \rangle _{0} &=& \int \DD\sigma_{\infty}(\hx)\,
\lim_{R\rightarrow\infty}\frac{1}{Z_{R}[0]} \mbox{Tr}\, 
(e^{-\beta H_{R} - iQ(\sigma_{\infty})}\, A) 
\nonumber\\
&=& \lim_{R\rightarrow\infty} \frac{1}{Z_{R}[0]}\mbox{Tr}\, 
(e^{-\beta H_{R}}
\,P_{s}\,A)\,,
\end{eqnarray}
where $Q(\sigma_{\infty})$ are generators of these transformations.

Finally, let us discuss the symmetry with respect to the centre group
$Z_{N}$ (this we shall consider in more detail below), which
is realised as shifts $g\beta A_{0}$ of a special kind.
For example, for the group $SU(2)$ in the adjoint representation, only
shifts to integers are allowed. Recall that the centre of the fundamental representation of
$SU(N)$ is generated by the unit roots of $\exp( i2\pi n/N)$, but the adjoint
representation of $SU(2)$ coincides with the adjoint representation of 
$SO(3)$ and so the centre elements are trivially represented in it.

From the above analysis it is obvious that there is a spontaneous breaking of $Z_{N}$ symmetry at high temperature, in the deconfinement phase,
by a surface term in any state
with $\chi$ different from zero. In the confinement phase this symmetry is not broken,
since $\chi=0$ and the partition function contains the integration over the surface Lagrange multiplier, by virtue of which a shift of this integration variable 
by the value of the mean $\langle A_{0} \rangle$ is trivial on the surface of the 
delta function and does not break the symmetry. 
Moreover, by the action of $Z_{N}$ all minima (including non--trivial ones)
are transformed into each other.


\section{Confinement criteria. The Wilson loop}
\label{W}

In the considered formalism the notion of confinement has been introduced
as the singletness with respect to the group of the large gauge transformations
of all states contributing to the temperature averages of the observables,
on which the generators of transformations are nullified,
\begin{equation} \label{W-0}
\int_{0}^{R}y^{2}\,dy\,\Phi_{\perp}(y\Hx) =0.
\end{equation}
Using the Gauss's law, we rewrite
$\Phi_{\perp}=\partial 
{\bf E}_{\Vert}$  and integrate \Ref{W-0} over an arbitrary area
$\Delta S$ on the boundary by converting the volume integral into a surface integral
over the encompassing conic surface plus the integral over the area itself.
Obviously, the normal to the conic surface is orthogonal to $\Hx$, 
and therefore the integral
over it is zero. Thus, the condition \Ref{W-0} is equivalent to
{\it the equality to zero of the chromo--electric field flux through any element of the boundary}. 
Note that the condition of the zero chromo--magnetic field flux
through an arbitrary element of the boundary also naturally appeared in our approach.
This formulation of confinement essentially coincides with intuitive
physical notions and it leads to the unobservability of coloured states at infinity.

It would be natural to compute another physical characteristic of the colour no--escape, namely the energy $E(L)$ of a state with two static coloured sources
quark -- antiquark located at distance $L$ from each other.
This quantity is a function of the response of the gauge field
to the introduction of sample sources and is expressed in terms of a geometric
field characteristic called the Wilson loop \cite{Wils74,Band81}
\begin{equation}    \label{W-1}
{\cal W}[\Gamma ] = \langle \mbox{Tr}\,\mbox{P}\,
\exp(g \oint_{\Gamma }A_{\mu }dx^{\mu })\rangle ,\quad A_{\mu } =
A_{\mu }^{a}t^{a},\quad (t^{a})^{\dagger} = -t^{a}\,,
\end{equation}
where $t^{a}$ are generators of the fundamental representation of the group $SU(N)$, and
$\Gamma$ is a closed contour in space--time (for definiteness
we shall consider a rectangular contour of length $T$ in the temporal 
direction and length $L$ in space). 
The asymptotic of the Wilson loop at large $T$ 
is related to the interaction energy of the charges
${\cal W}[\Gamma] \propto \exp(-T\,E(L))$. According to the confinement criterion by
Wilson \cite{Wils74}, the confinement phase follows the area law 
${\cal W}[\Gamma] \propto \exp(-\chi S(\Gamma))$,
and the Coulomb phase 
obeys the perimeter law
${\cal W}[\Gamma] \propto  \exp(-\mu P(\Gamma))$, where
$S(\Gamma) = LT$ and $P(\Gamma) = 2(T+L)$
and the area and the perimeter of the contour respectively. The proportionality parameter in the linearly rising potential energy, $E(L) = \chi\,L$, is called the string tension coefficient.
Such a name has a clear visual interpretation, since the 
the flux of the coloured chromosome--electric field between the charges is compressed into tubes (strings) of a fixed thickness \cite{Band81}.
The Wilson loop is a gauge--invariant object by definition. 

In lattice calculations, a different confinement  criterion is more popular. It is associated with the Polyakov line \cite{Poly87,Svet86}
\begin{equation}\label{W-1p}
{\cal P}(\Bx)= \mbox{T} \exp \int_{0}^{\beta}dt\,A_{0}(t,\Bx).
\end{equation}
The temperature average of the trace of products of this operator is related to the
the free energy of sample point charges introduced into the system at the points which are its arguments.
The average trace of the Polyakov line, which is proportional to the 
partition function of the solitary charge in the field, invariant only with respect to the 
gauge transformations satisfying periodic boundary conditions
$U(0,\Bx)=U(\beta,\Bx)$. 
According to another confinement criterion \cite{Poly87},
the mean of the Polyakov line is zero in the confinement phase, i.e. the free energy of the
solitary charge is infinite, and in the Coulomb phase it is different from zero.
Thus, this quantity is the order parameter of the confinement-deconfinement phase transition and allows us to distinguish  
the global symmetry breaking with respect to the centre $Z_{N}$ of the gauge group $SU(N)$. Indeed, the gauge transformations satisfying the condition 
$U(t+\beta,\Bx) = z U(t,\Bx)$, where $z=\exp(i 2 \pi n/N) \in
Z_{N}$
do not change the periodic boundary conditions on the gauge fields in the 
functional integral and, of course, any $SU(N)$ gauge invariant objects. 
In this case, they non--trivially transform the Polyakov line 
${\cal P}\rightarrow z\,{\cal P}$.
Therefore, in the absence of $Z_{N}$ symmetry breaking (the 
confinement phase) $\langle {\cal P} \rangle =0$, and in the deconfinement phase
the mean is non--zero and  the symmetry is spontaneously broken. The universality inherent to all critical \cite{Binn92,Itzy89,Amit78} phenomena would imply that the properties of the the confinement--deconfinement phase transition
for the gauge fields on a lattice should be analogous to those of any 
$Z_{N}$ symmetric lattice system \cite{Svet86}.
However, the continuous limit of the theory gives rise to ultraviolet and
infrared singularities requiring a non--trivial renormalisation procedure.
Therefore, the question of establishing a firm connection between the system on a lattice and the limiting continuous system remains a serious theoretical challenge.

The calculation of the Polyakov line by analytical methods in the continuous theory is
rather difficult due to the lack of its full gauge invariance,
although it is easily realisable numerically on the lattice.
We shall therefore restrict ourselves to the consideration of the Wilson confinement criterion.

Let us proceed to the direct calculation of the Wilson loop in the confinement phase \cite{rusTim93,Tim92NPI}. In the FS gauge, the zero component of the 
of the gauge field is expressed in the form \Ref{H-26}. For an arbitrary
of the functional $F$ it is easy to demonstrate that 
\begin{equation}   \label{W-2}
\Av{F[A_{0}(\Bx)]} = \Av{\int_{x}^{R}dy\,E_{\Vert}(y\Hx)}\,.
\end{equation}
To obtain this ratio, we can write the mean in the form 
$$F[\frac{\delta}{\delta J_{0}}]
\biggl|_{0}\biggr.
e^{i\int_{\Lambda} d^{4}x\,A_{0}(\Phi - J_{0})}.$$ 
Integrating over $A_{0}$ and expressing $E_{\Vert}$ using the $\delta$-function, we indeed have
\begin{eqnarray}
&& F[\frac{\delta}{\delta J_{0}}]\biggl|_{0}\biggr.\,
\exp\left(\frac{1}{2}\int_{\Lambda} d^{4}x\, (E_{\Vert} - \frac{1}{x^{2}}
\int_{0}^{x} y^{2}dy\,
J_{0}(y\Hx))^{2}\right)  \nonumber\\
&& = F[\int_{x}^{R}dy\,
E_{\Vert}(y\Hx)]\,\exp\left(\frac{1}{2}\int_{\Lambda} d^{4}x\,
E_{\Vert}^{2}\right)\,. \nonumber
\end{eqnarray}
The most convenient $\Gamma$ contour consists of two straight lines
along the radius in some fixed direction $\Hx_{0}$ 
at times $t=0,T$,
which give zero contributions due to the gauge condition, 
and two lines parallel to the time axis at radii $x=R',R''$,
for which we can write
\begin{eqnarray}
{\cal W}[\Gamma] &=& \left\langle
\Tr\left[
\left(\mbox{T}\exp\int_{0}^{T}dt\,A_{0}(R''\Hx_{0})\right)^{\dagger} 
\right.\right.\nonumber\\ &&
\left.\left.\mbox{T} \exp \int_{0}^{T}dt\,
(A_{0}(R''\Hx_{0})+\int_{R'}^{R''}dy\,E_{\Vert}(y\Hx_{0}))
\right]\right\rangle . \label{W-3}
\end{eqnarray}
For the sake of brevity, we shall further omit the angular coordinates unless this
complicates the understanding of the formulae.
Taking advantage of the gauge invariance of the expression under the trace,
we can transform the variables as follows
\begin{eqnarray}
A_{0} &\rightarrow & A_{0}' = U^{-1}(A_{0}+
g^{-1}\partial_{t})U(t), \nonumber \\
E_{\Vert} &\rightarrow & E_{\Vert}' = U^{-1} E_{\Vert} U(t).
\label{W-4}
\end{eqnarray}
From the definition of T-exponent 
${\cal P}[A_{0}](t) =\mbox{T} \exp\int_{0}^{t}dt'\,A_{0}(t')$,
as the solution of the following differential equation,
\begin{equation} \label{W-5}
\dot{{\cal P}}[A_{0}](t)  = gA_{0}(t)\,{\cal P}[A_{0}](t), 
\quad {\cal P}[A_{0}](0)={\bf 1}
\end{equation}
we obtain 
\begin{eqnarray} 
&&\backspace {\cal P}\left[A_{0}(R'')+\int_{R'}^{R''}dy\,E_{\Vert}(y)\right](t) 
\nonumber\\ 
&& \backspace = {\cal P}[A_{0}(R'')](t)\,
{\cal P}\left[{\cal P}[A_{0}(R'')]^{-1} \int_{R'}^{R''}dy\,
E_{\Vert}(y)\ {\cal P}[A_{0}(R'')] \right].
\label{W-6}
\end{eqnarray}
Under gauge transformations \Ref{H-4} T-exponent changes as
\begin{equation} \label{W-6p}
{\cal P}(t)\rightarrow {\cal P}'(t) = U^{-1}(0) {\cal P}(t) U(t).
\end{equation}
To perform the Gibbs averaging, we use the formula \Ref{Conn},
which allows us to replace the average of an arbitrary functional from $E_{\Vert}$ with the average of the functional from the conjugate variable $\lambda$
with the help of an additional
Gaussian integration over the auxiliary variable $\xi$.
We can try to simplify the contribution of the surface term, which,
as we have shown, should be taken with $\chi=0$ in the confinement phase. The integral
over $A_{0}(R\Hx)$ gives the surface $\delta$-function and we can write the 
partition function in the form
\begin{eqnarray}
Z[0] &=& \int\DD\Ap\,\DD\Ep\,e^{-W[\Ap,\Ep]}\,\delta\left(R^{2}\Ev(R\Hx)\right)
=\int\DD\lambda\,\DD\Ap\,\DD\Ep\,
\nonumber\\
&& e^{\int_{\Lambda}d^{4}x\,\left[ 
\frac{1}{2}(\Ep^{2}-{\bf B}^{2})
-\frac{1}{2}\lambda^{2}+\lambda\Ev\right]}\,
\delta\left(R^{2}\Ev(R\Hx)\right)\,\delta\left(R^{2}\lambda(R\Hx)\right).
\label{W-7}
\end{eqnarray}
Here we have used the fact that the addition of a second surface 
$\delta$-function is equivalent to changing the normalisation constant 
due to the presence of the first $\delta$-function. After we have taken into account the 
constraint on $\lambda$, we again represent the first $\delta$-function as an
integral over $A_{0}(R\Hx)$, integrate over the transverse physical 
variables and so we obtain a representation in terms of the effective action of the 
of collective variables of the form (\ref{li},\ref{lili}). 
Thus, we have shown that actually 
the integration is performed only over the fields with an additional boundary condition 
on $\lambda(R\Hx)$.
Hence, the integration space of the conjugate variable 
satisfies the same boundary condition as the space $\Ep$.
The Wilson loop will be expressed through the following functional 
integral
\begin{eqnarray}
&& {\cal W}[\Gamma] = \int\DD\xi\,\DD\sigma\,\DD\nu\,e^{-W[\sigma,\nu]
-\frac{1}{2}\int_{\Lambda}d^{4}x\,\xi^{2}}\,
\delta\left( R^{2}\lambda(R\Hx)\right)
\nonumber\\
&& \tr\, {\cal P}\left[ 
{\cal P}[\sigma(R'')]^{-1} \int_{R'}^{R''}dy\,
(\lambda+i\xi)(y)\ {\cal P}[\sigma(R'')]
\right].
\label{W-8}
\end{eqnarray}
We have converted this formula to the Euclidean time by substituting
$t\rightarrow i\,t$, and ``tr'' denotes the
trace in the fundamental representation $SU(N)$, in contrast to ``Tr'' 
which denotes a trace in Hilbert space.
Note that the coefficient functions of the fields are time dependent and hence,
the T-exponent is not reducible to the ordinary exponent.

The expression \Ref{W-8} at finite $R,R',R''$ is extremely 
computationally difficult. One can, however, expect some noticeable
simplifications in the limit of the infinite $R\rightarrow\infty$ system, when the pair
quark--antiquark is located infinitely far from the centre of coordinates.
In other words, we are only interested in the asymptotic interaction energy
in the limit
$R',R''\rightarrow\infty,\ \Delta R=R''-R'=const$.
We introduce the variable $\varphi$ using the relation
$$\lambda(\Bx)=\frac{1}{x^{2}}\int_{0}^{x}y^{2}dy\,\varphi(y\Hx).$$
By virtue of the confinement condition \Ref{W-0}, 
the electric field flux through an infinitely small solid angle at spatial infinity in arbitrary 
direction is zero.
In terms of the Fourier components of the basis of the form
$$f_{lm\,n}(k)[\Bx,t]=
\left(\frac{2}{\pi}\right)^{1/2}\frac{\sin k x}{kx}\,Y_{lm}(\Hx)\,
\exp (i 2\pi n\,t/\beta)$$
the $\delta$-functional gives in the considered limit the condition 
\begin{equation}
\tilde{\varphi}_{lm\,n}(k=0)=0.
\label{W-st}
\end{equation}
The expression in the T-exponent tends to 
\begin{equation}
\int_{R'}^{R''}dy\,\lambda_{lm\,n}(y) \rightarrow -(8\pi)^{1/2}
\frac{\Delta R}{R'R''}\,\tilde{\varphi}_{lm\,n}(k=0),
\label{W-stst}
\end{equation}
and therefore can be neglected. 
It is easy to realise that the expression in this formula oscillates strongly 
and the convergence is very slow. Using such an approximation
in terms of the original integral over the gauge fields would give only a trivial 
result, nevertheless, it is natural in the integral over the collective 
variables. This circumstance has the same reason as the fact of the 
non--triviality of the stationary phase approximation in the last representation of the 
of the functional integral compared to the original integral. The collective variable
includes in its definition the averaging over  a Gaussian noise, and therefore,
in some sense its quantum fluctuations give a less significant contribution.

Once the contribution of \Ref{W-stst} is neglected, the 
expression can be significantly simplified by 
by choosing a special gauge transformation.
Let us perform some gauge transformation without 
changing the magnitude of the trace \Ref{W-8}. 
The auxiliary variable is transformed homogeneously 
$\xi\rightarrow\xi'=U^{-1}\xi U(t)$. 
The action for $\xi$ does not contain time derivatives, so we can
replace the variables in the integral
$\xi\rightarrow\xi'$ (whose Jacobian is equal to unity), without violating the periodic boundary conditions in time.
Removing the prime in the notation of
the integration variable, we obtain the original expression where
${\cal P}[A_{0}(R'')]$ is transformed according to the formula \Ref{W-6p}.
Note also that the constant homogeneous transformation is taken out of the T-exponent and vanishes due to the cyclic nature of the trace.
Thus, we have shown that the Wilson loop does not change under the 
transformation affecting only the construction
${\cal P}[A_{0}(R'')]\rightarrow {\cal P}' 
\equiv U^{-1}(t)\,{\cal P}[A_{0}(R'')]$.

Since $U(t)$ is arbitrary, we can get ${\cal P}'$ to be an arbitrary time--dependent 
transformation. We introduce the notation 
$\Xi =\int^{R'}_{R''}dy\, \xi(y)$.
There holds the following 
\begin{quote}
Lemma. {\it
For an arbitrary function of the form $\hat{\Xi}^{a}(t)= \Xi^{a}/|\Xi|$, there exists such a
gauge transformation $U(t)$, which makes the corresponding element of the gauge algebra
$\hat{\Xi} =\hat{\Xi}^{a}t^{a}$ independent of time.
}\end{quote}
For prove this it suffices to show that for an arbitrary normalised
vector in the colour space of the group $SU(N)$ there always exists a
(in general, not the unique) solution of equation 
\begin{equation} \label{W-9}
\frac{d}{dt}\left(\,U^{- 1}(t)\,\hat{\Xi}(t)\,U(t)\,\right) = 0.
\end{equation}
This equation can be rewritten as
\begin{equation} \label{W-10}
[V(t)\,,\,\hat{\Xi}(t)] = \partial_{t}\,\hat{\Xi}(t)\,, \qquad \quad V(t) =
\partial_{t}\,U(t)\ U^{- 1}(t)\,.
\end{equation}
This, obviously, can be resolved in
$V(t)$, and then $U(t)$ is equal to
\begin{equation} \label{W-11}
U(t) = \mbox{T}\,\exp \int^{t}dt'\,V(t')\,.
\end{equation}

After choosing such a transformation, the T-exponent is reduced to 
the normal exponent. To get rid of the vector structure we apply 
an additional global transformation, so that the vector $\hat{\Xi}$
belongs to the Cartan subalgebra. Due to the diagonality of the Cartan 
generators, the trace, and then the Gaussian integral, are easily computed
\begin{eqnarray}
&& \backspace {\cal W}[\Gamma] = 
\int\prod_{\alpha = 1}^{N - 1} {\cal D}\xi^{\alpha}\,e^{- \frac{1}{2}\int
_{\Lambda}
d^{4}x\,(\xi^{\alpha})^{2}}\,\sum_{i = 1}^{N} \exp \left[\sum_{\alpha = 1}^{N
- 1} (j \bullet \xi^{\alpha})\,\lambda_{i}^{(\alpha)}\right]\,,
\label{W-12}\\  \backspace
&& j(x) = \frac{i\,g}{2}\,\theta(0 \leq t \leq T)\,\frac{\theta(R' \leq x \leq
R'')}{x^{2}}\,\delta(\hat{\bf x} - \hat{\bf x}_{0}),\,
\label{W-13}
\end{eqnarray}
leading to the result 
\begin{equation}
{\cal W}[\Gamma] = \sum_{i = 1}^{N}\exp [ - \sum_{\alpha = 1}^{N - 1}
(\lambda_{i}^{(\alpha)})^{2} \,\chi_{0}\, S]\,,
\label{W-14}
\end{equation}
where $\left\{\lambda_{i}^{(\alpha)}\,\equiv\, (\lambda^{\alpha})_{i\, i}\,,\ 
(\alpha = 1,\,\ldots ,\  N - 1 )\right\}$
are the elements of the diagonal generators in the fundamental representation of $SU(N)$.
The generator matrices are normalised as follows
\begin{equation}
\mbox{tr}(\lambda^{a}\lambda^{b}) = 2\delta^{ab}\,.
\label{W-15}
\end{equation}
The condition \Ref{W-15} can be formally represented in terms of N-dimensional vectors
${\bf \lambda}^{(\alpha)}$, if we introduce an additional vector with the components
\begin{equation}\label{W-16}
\lambda^{(0)} = \frac{1}{\sqrt{N}}\,\bigl(\, \underbrace{1, \ldots,
1}_{N}\, \bigr)\,,
\end{equation}
as a condition of orthonormalisation of the N-dimensional basis
$\biggl\{\lambda^{(0)},\lambda^{(\alpha)}/\sqrt{2}\biggr\}$.
The completeness condition of such a basis gives
\begin{equation} \label{W-17}
\sum_{\alpha = 1}^{N - 1}\lambda_{i}^{(\alpha)}\,\lambda_{j}^{(\alpha)} =
2\,\biggl(\,\delta_{ij} - \lambda_{i}^{(0)}\,\lambda_{j}^{(0)}\biggr)\,.
\end{equation}
The sum in \Ref{W-14} is easily calculated using the last relation
\begin{equation}
\sum_{\alpha = 1}^{N - 1}(\lambda_{i}^{(\alpha)})^{2} = 2\,\biggl( 
1 - \frac{1}{N}\biggr).
\label{W-18}
\end{equation}

Combining the previous results, we obtain the area law
\begin{equation}
{\cal W}_{_{SU(N)}}[\Gamma] = N\, e^{- \chi_{_{SU(N)}}\, S}\,,
\label{W-19}
\end{equation}
with the string tension coefficient equal to
\begin{equation}
\chi_{_{SU(N)}} = 2\,\biggl( 1 - \frac{1}{N}\biggr)\,\chi_{0}\,.
\label{W-20}
\end{equation}
The constant $\chi_{0}$ contains ultraviolet and infrared divergences:
\begin{equation}
\chi_{0} = \frac{ g^{2}\,\delta (\hat 0)}{8\, R'\, R'' }\,,\qquad
\delta (\hat 0 ) = \sum_{l} \frac{(2\, l + 1)}{ 4\,\pi}\,,
\label{W-21}
\end{equation}
but it has the same structure as the square of the temperature $T_{c}$ of the phase transition 
of the confinement-deconfinement transition.
The formula \Ref{W-21} is expressed through the bare parameters of the theory and 
requires the application of a renormalisation procedure, which is however
difficult in non--covariant gauges.

For the group $SU(2)$ for the dimensionless combination
$\xi \equiv T_{c}/\sqrt{\chi}$
we obtain the following formula 
\begin{equation} \label{XI}
\xi^{2} = \frac{a_{c} 8 R^{3}}{\pi^{2} V_{R}} =\frac{6}{\pi^{2}}a_{c}.
\end{equation}
Given the formula \Ref{PL23}, the numerical value is
$\xi \simeq 0.71$ and is in qualitative agreement with the
the result obtained from Monte Carlo simulations
in the lattice theory, which equals 
$0.69 \pm 0.02$. The free energy density can now be
rewritten in terms of the string tension as
\begin{equation}
{\cal F}_{R}[a_{R}] = \frac{54\,\pi}{g^{2}\,\beta^{2}}\,\chi_{0}\,F_{R}\biggl(
\frac{\pi^{3}}{6\,\beta^{2}\,\chi_{0}}\biggr)\,,
\label{Fr}
\end{equation}
which emphasises the clearly non--perturbative character
of the approximation we have considered.
The computation of the Wilson loop in the deconfinement phase is much more complicated due to the absence of the $\delta$-function, which actually has allowed us to get rid of
of the most complicated functional integration above.

\newpage
\section{Conclusion}

In conclusion, let us formulate the main results obtained in the thesis:

1. The formulation of the non--Abelian gauge theory in the 3-dimensional Fock--Schwinger gauge is constructed. A generalisation of this gauge to a class of gauges, which we call the generalised Fock--Schwinger gauges, for which the Gauss law constraint is still exactly solvable and the gauge fields are expressed via the the strength fields by a linear differential relation, is proposed.

2. The problem of the group of residual gauge transformations 
allowed by these gauges is studied and the choice of boundary conditions fixing
such arbitrariness is discussed. Explicit formulae for the connection of gauge fields in the given gauge with those in an arbitrary gauge are obtained.  These formulae are analysed in detail 
on the example of the Coulomb gauge in the Abelian theory, and allow one, for instance, 
to obtain a simple representation for the free propagator in the Fock--Schwinger gauge.

3. It is shown that among the gauges of the class considered here, 
there is a preferred one for the consideration of the dynamics in a given compact 
region, the form of which is consistent with the field of normal vectors to the boundary of the region. Such a gauge leads to a substantial simplification of the surface terms in the theory. 
The Hamiltonian formalism for the Yang--Mills theory in a finite domain, which allows one to take into account the variables at the boundary as true Hamiltonian variables and the 
the contribution of these surface terms is quite essential, whereas the formal 
neglect of these terms in the standard approach is invalid and leads to the loss of essential physical effects.
Any boundary conditions in this formalism are thus treated as additional constraints. Admissibility of this or that boundary condition is determined by the conditions of its preservation by the dynamics of the system in the finite domain.

4. The formulation of the Yang--Mills theory in the generalised
Fock--Schwinger gauge is constructed in the formalism of the ``variables at infinity'' proposed by Morcio and Strocchi in the framework of the algebraic Quantum Field Theory \cite{Morc85,Morc87,Morc87jmp}.
This formalism allows us to construct the correct limit of the dynamics of the system
in infinite volume as a group of automorphisms of the extended algebra of observables 
with a nontrivial centre. The latter is generated by weak bounds 
of delocalised variables, called variables at infinity.
The non--trivial dynamics of variables at infinity in the  
Yang--Mills theory is deduced, which is essential for the confinement mechanism.
In the limit of a system of infinite size, the formalism of point 3 turns into the
the present one, and thus their equivalence is established.
Note that the former version is useful for practical  calculations, for example, of thermodynamic quantities at finite temperature and is used further in this thesis,
while the latter is of great value for analysing the phenomena of 
spontaneous symmetry breaking, the global structure of the state space of the
limiting system, and is also more convenient in the axiomatic
field theory for a rigorous proof of general results.
The representation of the Poincar\'e algebra in the Fock--Schwinger gauge is briefly discussed.

5. The main achievement of the thesis is the analysis of the dependence of the
of the partition function of the non--Abelian gauge theory on the boundary value of the longitudinal component of the electric field, which by virtue of the Gauss law in the given gauge coincides with the electric field flux through an infinitesimal element of the 
of the boundary, and the establishment of the connection between such dependence and the mechanism of the confinement--deconfinement phase transition.

6. The analysis of the non--Abelian theory is preceded by consideration of a simpler
quantum electrodynamics problem with an external charge density 
in a spherical region at finite temperature,  where the dependence of the partition function on the boundary variable is calculated exactly.
The result suggests that in the case of a charge density corresponding to a
linearly increasing potential, which models the confinement situation, the 
surface effects lead to a finite correction that decreases the free energy density.
Therefore the tendency towards confinement is a statistically 
favoured effect.

7. A formulation of quantum gluodynamics in terms of the
functional integral of the collective variables, conjugate to the longitudinal components 
of the chromo--electric and --magnetic fields is developed. Remarkably, the integrals over the gauge fields themselves are exactly calculated in the Fock--Schwinger gauge. 
Thus, the generating functional of the theory is represented in terms of the 
variables transformable homogeneously under the action of gauge
transformations.
The effective action (free energy density) is calculated in cases of 
$SU(2)$ and $SU(3)$ gluodynamics in the mean--field approximation for the
collective variables. The analysis of the free energy minima 
revealed a phase transition at a certain temperature, 
below which the mean value of the collective variables is different from zero.

8. It is found that the latter transition can be interpreted 
as a confinement--deconfinement phase transition. 
In the confinement phase the flux of the chromo--electric
field through an arbitrary element of the boundary is strictly zero, which is the condition 
of singletness with respect to the group of the residual gauge transformations,
and this physically implies the impossibility of observing coloured objects at
spatial infinity (in asymptotic states).

9. It is shown that our confinement condition satisfies the traditional
confinement criteria. Firstly, it is shown that in the confinement phase the 
Wilson loop for the $SU(N)$ theory satisfies the area law. The ratio of the phase transition temperature to the square root of the string tension coefficient does not contain any
divergences and, despite the approximations used in the calculation of the phase transition temperature (the mean--field contribution and the lack of a 
full renormalisation procedure), it is in a qualitative
agreement with the result obtained using the Monte Carlo simulations
of the lattice gauge theory.

10. It is shown that in the deconfinement phase the global $Z_{N}$-symmetry 
with respect to the centre of the group $SU(N)$ is spontaneously broken by the surface 
terms. The confinement phase is characterised by the unbroken symmetry.
In the latter phase all nontrivial minima of the effective action have the same
depth and are transformed into each other by the action of $Z_{N}$ transformations,
which allows us to expect the structure of the ground state of the theory 
similar to the $\theta$-vacuum.
\newpage


\centerline{{\Large\bf Acknowledgements}}

\vspace{2cm}

The author would like to warmly thank his scientific supervisor
N.A.~Sveshnikov for the choice of an interesting problem, constant attention to the work,
useful creative advice, critical comments and fruitful collaboration.

I would also like to express my deep gratitude to
V.O.~Soloviev for his active interest in this problem
and many important discussions, as well as to B.A.~Arbuzov, E.E.~Boos, A.S.~Vshivtsev,
V.A.~Ilyin, G.S.~Iroshnikov, D.D.~Kazakov, Yu.A.~Kuznetsov, 
V.A.~Petrov, V.E.~Rochev, A.P.~Samokhin, V.V.~Skalozub, A.A.~Slavnov,
F.V.~Tkachev and  A.M.~Shirokov
for useful discussions and important remarks. 

The author would like to thank Professor K.A.~Dawson (UCD, Dublin) for his support and helpful discussions during the writing of the thesis.

The author is sincerely grateful to V.I.~Savrin
and other members of the Division of Theoretical High Energy Physics of Nuclear Physics Institute of 
Moscow State University for valuable advice
and constant support.


\newpage

{
\renewcommand{\a}{\alpha}\renewcommand{\b}{\beta}\newcommand{\g}{\gamma}
\renewcommand{\d}{\delta}\newcommand{\e}{\epsilon}\newcommand{\ve}{\varepsilon}
\newcommand{\z}{\zeta}\newcommand{\h}{\eta}\newcommand{\q}{\theta}
\newcommand{\vq}{\vartheta}\renewcommand{\i}{\iota}\renewcommand{\k}{\kappa}
\renewcommand{\l}{\lambda}\newcommand{\m}{\mu}\newcommand{\n}{\nu}
\newcommand{\x}{\xi}\renewcommand{\o}{o}\newcommand{\p}{\pi}
\renewcommand{\r}{\rho}\newcommand{\vr}{\varrho}\newcommand{\s}{\sigma}
\newcommand{\vs}{\varsigma}\renewcommand{\t}{\tau}\renewcommand{\u}{\upsilon}
\newcommand{\f}{\phi}\newcommand{\vf}{\varphi}\renewcommand{\c}{\chi}
\newcommand{\y}{\psi}\newcommand{\w}{\omega}\newcommand{\G}{\Gamma}
\newcommand{\D}{\Delta}\newcommand{\Q}{\Theta}\renewcommand{\L}{\Lambda}
\newcommand{\X}{\Xi}\renewcommand{\P}{\Pi}\renewcommand{\S}{\Sigma}
\newcommand{\U}{\Upsilon}\newcommand{\F}{\Phi}\newcommand{\Y}{\Psi}
\newcommand{\W}{\Omega}                                    

\newcommand{\cedilla}[1]{\c{#1}}
\newcommand{\be}[1]{\begin{#1}}
\newcommand{\ee}[1]{\end{#1}}
\newcommand{\eq}{equation}
\newcommand{\en}{eqnarray}
\newcommand{\si}{\vec{\sigma}_{\infty}(\hat{{\bf x}})}
\renewcommand{\Tr}[1]{\mbox{Tr}_{#1}}
\renewcommand{\tr}[1]{\mbox{tr}_{#1}}
\newcommand{\iy}{\infty}
\newcommand{\dd}{{\cal D}}
\newcommand{\la}[1]{\label{#1}}
\newcommand{\nn}{\nonumber}
\renewcommand{\v}[1]{\hat{#1}}
\newcommand{\rar}{\rightarrow}
\newcommand{\Rar}{\Rightarrow}
\newcommand{\lar}{\leftarrow}
\newcommand{\Lar}{\Leftarrow}
\newcommand{\lra}{\leftrightarrow}
\newcommand{\av}[1]{\langle #1 \rangle}
\newcommand{\ex}[1]{\mbox{#1-exp}}
\renewcommand{\j}[1]{{\cal #1}}
\newcommand{\pp}{_{\perp}}
\newcommand{\fr}[2]{\frac{#1}{#2}}
\newcommand{\ti}[1]{\tilde{#1}}
\newcommand{\mb}[1]{\mbox{#1}}
\newcommand{\rf}[1]{(\ref{#1})}
\newcommand{\pa}[1]{\partial_{#1}}
\newcommand{\na}[1]{\nabla_{#1}}
\newcommand{\mo}[1]{\mid\,#1\,\mid}
\newcommand{\bu}{\bullet}


\section{Appendices}
\subsection*{Appendix A.$\quad$ Gluon propagator in the FS gauge}

Let us consider the propagator in the Fock--Schwinger gauge
\begin{displaymath}
D_{FS}^{ij}(x,y) = -i\,\langle 0\mid T_{D}\,
(\,A_{\perp }^{i}(x)\,A_{\perp }^{j}(y)\,)\,
\mid 0\rangle\,.
\end{displaymath}
After rewriting these expressions via the Coulomb fields with the use of the algebraic properties 
 (\ref{G-26},\ref{G-27}) one can derive the following representation 
\begin{eqnarray}
\backspace
D_{FS}^{ij}({\bf x},{\bf y};x_{0},y_{0}) &=& -\int
\frac{d^{4}k}{(2\pi )^{4}}\,\frac{e^{-i\,k_{0}\,(x_{0}-y_{0})}}{(k^{2}+i\,0)}
\biggl[ \delta ^{ij}\,e^{i\,{\bf k}({\bf x}-{\bf y})}-
\frac{\partial }{\partial \,x_{i}}x^{j}\frac{e^{i\,{\bf k}{\bf
x}}-1}{i\,{\bf k}{\bf
x}}- \nonumber\\
\backspace
&-&\frac{\partial }{\partial \,y_{j}}y^{i}\frac{e^{-i\,{\bf k}{\bf
y}}-1}{-i\,{\bf k}{\bf y}}+
\frac{\partial }{\partial \,x_{i}}x^{l}\frac{\partial }{\partial 
\,y_{j}}y^{l}\frac{e^{i\,{\bf k}{\bf
x}}-1}{i\,{\bf k}{\bf x}}\frac{e^{-i\,{\bf k}{\bf y}}-1}{-i\,{\bf k}{\bf
y}}\biggl. \biggr]\,.
\nonumber
\end{eqnarray}
Due to the boundary condition for the gauge field at zero this propagator is well--defined, whereas the propagator in the axial gauge has an unpleasant double pole. 
Despite many similarities of our gauge with the axial one, the latter posses serious problems with the boundary conditions  \cite{Bass84}.

{
\newcommand{\ARG}{[\vec{s},\vec{\n}]}
\newcommand{\ARg}{[s,\vec{\n}]}
\newcommand{\V}[1]{\vec{#1}}
\newcommand{\vv}{_{\Vert}}
\renewcommand{\n}{v}

\subsection*{Appendix B. $\quad$SU(2) Group: the generic Ansatz}

The general formula for the free energy density in terms of the dimensionsless parameters has the form
\be{\en}
F\ARG &=& - a\,\V{\n}^{\,2} + \j{U}\ARG\,,  \\ \la{F}
\j{U}\ARG &=& \sum_{n}\int_{0}^{\iy}\fr{dv}{\p}\,\mb{log$\,$det}\,C\ARG\,,\\
\la{U}
C^{ab}\ARG &=& \d^{ab}\,(v^{2} + n^{2}) + i\,t^{abc}\,(2ns^{c} + \n^{c})
+ t^{acd}t^{bed}\,s^{c}s^{e}\,. \la{C}
\ee{\en}
By means of a colour transformation one achieves the following parametrisation 
$$\V{s} = (0,\, 0,\, s)\,, \qquad \qquad \V{\n} = (\n\pp,\, 0,\,
\n\vv)\,.$$
The determinant is then equal to
\be{\eq} \la{Det}
\mb{det}\,C = X^{3} + 2s^{2}X^{2} + (s^{4} -\V{\n}^{2} - 4ns\,\n\vv)\,X
- \n\pp^{2}s^{2}
\ee{\eq}
with $X = v^{2} + n^{2}$. 

We denote by $\g_{a}\ (a = 1,2,3)$ the roots of the right--hand side of Ed. \rf{Det}
on $X$, which will be studied below. The formula 
\be{\eq} \la{Log}
\int_{0}^{\iy}\fr{dv}{\p}\,\log (1 + \fr{a}{v^{2}}) = a^{1/2}
\ee{\eq}
allows us to express the function \rf{U} as
\be{\en}
\mb{Re}\,\j{U}\ARg &=& \sum_{n}\Biggl.\Biggr.\bigl[ \sum_{a=1}^{3}(n^{2} -
\g_{a}[s,\V{\n},n])_{+}^{1/2} - 3\mo{n}\,\bigr] - \fr{1}{2}\,,  \la{ReU}\\
\mb{Im}\,\j{U}\ARg &=&
\sum_{n}\Biggl.\Biggr.\sum_{a=1}^{3}(\g_{a}[s,\V{\n},n] 
- n^{2})_{+}^{1/2}\,. \la{ImU}
\ee{\en}
Due to the periodicity and even parity  $\j{U}$ of $s$ one can consider it on the half--period
$0 \leq s \leq 1/2$. Right away we have to exclude the domain where there is a nonzero imaginary part. Obviously this is nonzero if,
$$\g_{a}[s,\vec{\n},0] > 0\,.$$
Thus, let us find when the equation 
\be{\eq} \la{x}
x^{2} + 2s^{2}\,x + s^{4} - \vec{\n}^{2} = \fr{\n\pp^{2}s^{2}}{x}
\ee{\eq}
possesses positive or complex solutions (see Fig.~5). 

It is easy to see that the lower branch of the hyperbolic curve
 $2''$ intersects the parabola 1 in two points,
giving rise to negative real roots 
$\g_{2}, \g_{3}$. It is important to emphasise the following circumstance. 
As soon as 
$\n\pp \not= 0$, the upper branch of the hyperbola 
 $2'$ intersects the parabola in a positive point 
 $\g_{1}$, leading to the instability. Due to this, the general case 
$\n\pp \not= 0$ is of no intrest.

Fig.~5 is fairly informative and even allows us to see the stable minimum 
$F\ARG$. This function is as smaller, as greater 
$\mo{\vec{\n}}$ is, whike $\mo{\g_{a}}$ --- is conversely greater. The negativity requirement 
imposes the additional restriction 
$\vec{\n}^{2} \leq s^{4}$,
which is saturated when $\mo{\vec{\n}} = \n\vv = s^{2} = \fr{1}{4}$.

\subsection*{Appendix C. $\quad$Some formulae with the special functions}

The Poisson summation 
\begin{displaymath}
\sum_{k = -\infty }^{\infty }f(2\pi k) 
= \frac{1}{2\pi }\sum_{k = -\infty }^{\infty }\int
_{-\infty }^{\infty }d\tau \,e^{-i\,k\tau }\,f(\tau )\,.
\end{displaymath}
Some integrals with the cylindrical functions \cite{Prud83}
\begin{displaymath}
\int   _{0}^{\infty }dy\,\frac{\cos(ky)}{(y^{2}+v^{2})^{\frac{1}{2}}}   =
\mbox{K}_{0}(kv)\,,\qquad
-\int _{v}^{\infty }dy\,\frac{\cos(ky)}{(y^{2}-v^{2})^{\frac{1}{2}}} =
\frac{\pi }{2}\mbox{Y}_{0}(kv)\,,
\end{displaymath}
\begin{displaymath}
\int _{o}^{u}v\,dv\,\mbox{K}_{0}(kv) 
= \frac{1}{k^{2}}(-ku\,\mbox{K}_{1}(ku)+1)\,,
\end{displaymath}
\begin{displaymath}
\frac{\pi }{2}\int _{0}^{u}v\,dv\,\mbox{Y}_{0}(kv) 
= \frac{1}{k^{2}}(\frac{\pi }{2}
ku\,\mbox{Y}_{1}(ku)+1)\,.
\end{displaymath}
The Hurwitz representation of the generalise  $\zeta$-function \cite{Bate53}
valid for \mbox{Re$\,s < 0$}
\begin{displaymath}
\zeta (s,v) = 2(2\pi )^{s-1}\,\Gamma (1-s)\sum_{k = 1}
^{\infty }n^{s-1}\,\sin(2\pi nv+\pi s/2)\,.
\end{displaymath}
We use the periodic continuation of $\zeta$-function by 
means of this formula and denote it as $\zeta _{H}$. 
For any integer $k$
\begin{displaymath}
\zeta _{H}(-\frac{1}{2},k) = -\frac{\zeta (3/2)}{4\pi }\,, 
\qquad  \quad \zeta _{H}(-\frac
{1}{2},k+\frac{1}{2}) = (1-1/ \sqrt{2})\frac{\zeta (3/2)}{4\pi }\,.
\end{displaymath}

}

\section*{}

\newpage
\section*{Figure Captions}

$\ $\\
\noindent
Fig. 1. Dependence of the real part of the effective potential on $u$ and $s$ at $a=0.01$. \\
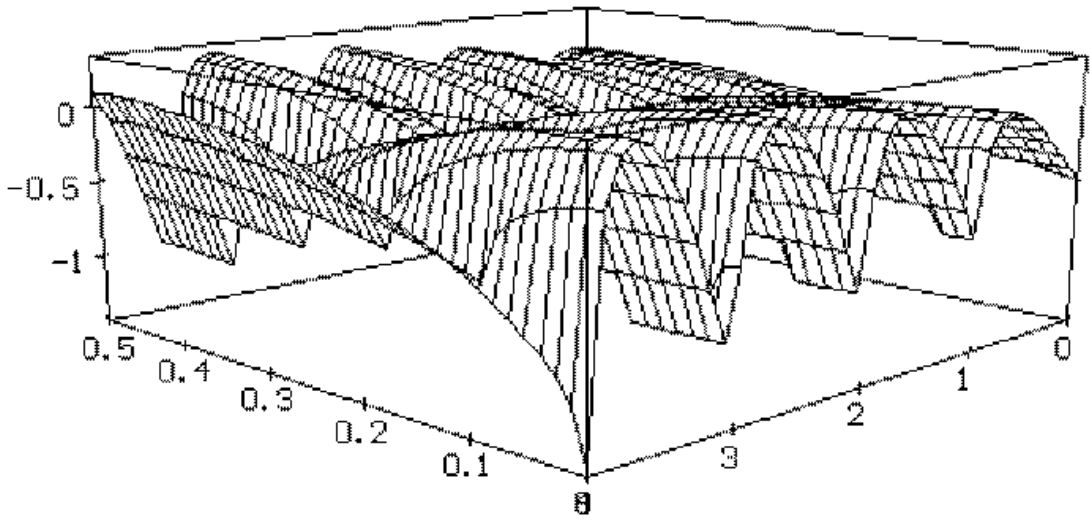
Fig. 2. Dependence of the imaginary part of the effective potential on $u$ and $s$. \\
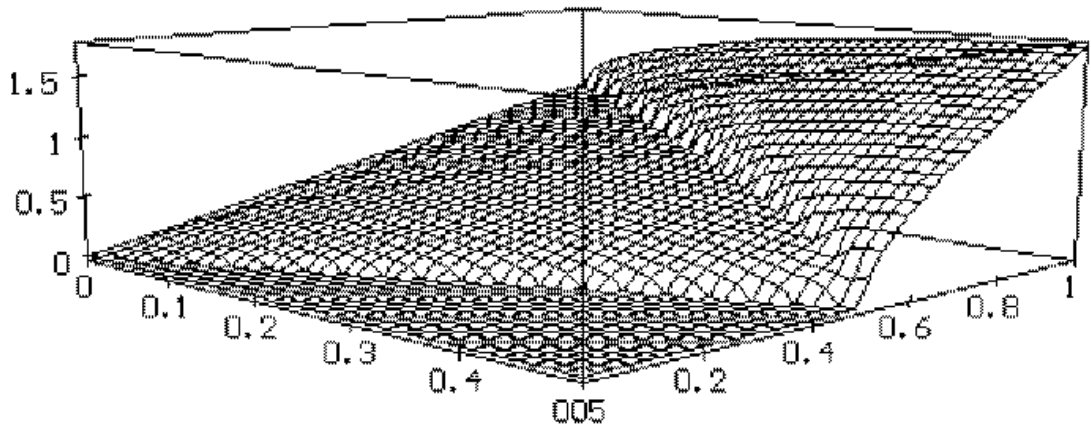
Fig. 3. Plot of the dependence of {\rm Re $F$ } on  
$u$ at fixed  $s$ at the phase transition ( $s = 0$, $a = 0.38$, $s = 0.5$, $a = 2$).\\
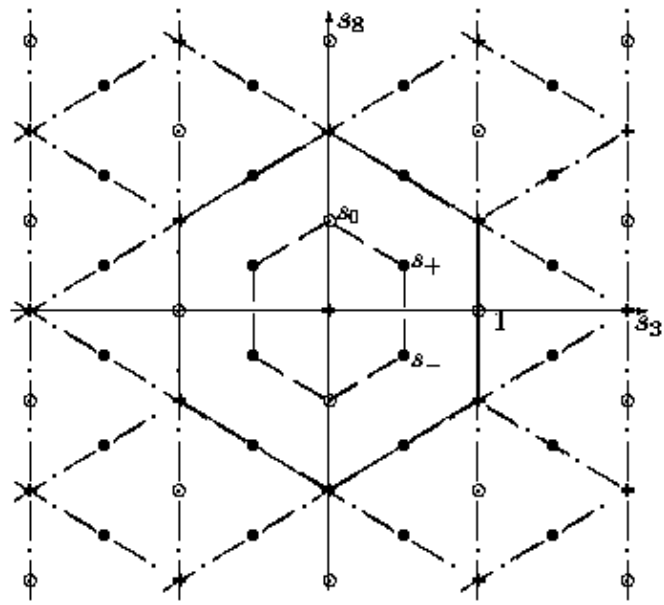
Fig. 4. Minima of the effective action in $SU(3)$ gluodynamics in terms of the variable $s=g\beta\sigma/(2\pi)$.\\
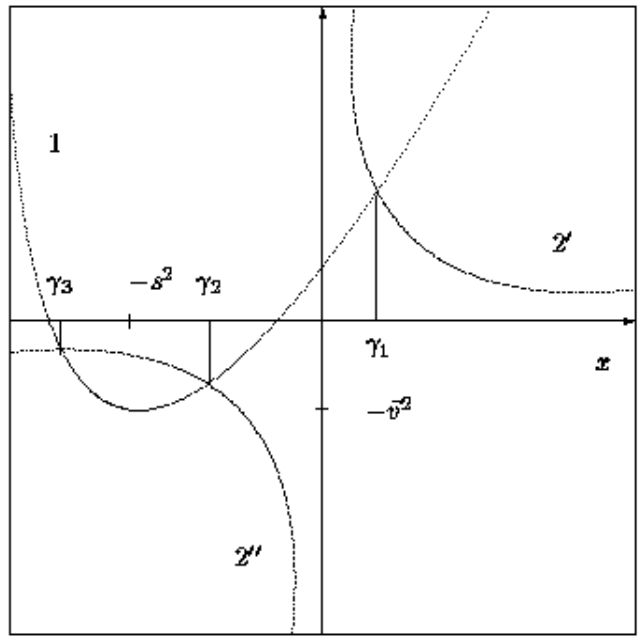
Fig. 5. Plots for determining the roots $\gamma_i$ of the 
equation $\mbox{det} C(n=0)=0$ for group $SU(2)$. 
Parabola 1 crosses $x$ axis in points $-s^2\pm \vert\vec{v}\vert$.


\newpage
\begin{thebibliography}{100}

\bibitem{Cutts78}
{D.~Cutts et al},  {\it Phys. Rev. Lett.}, vol.~41, p.~363, 1978.

\bibitem{Lyons80}
L.~Lyons,  preprint 38/80, {Oxford Univ. Nucl. Phys. Lab.}, 1980.

\bibitem{Grib92}
V.N.~Gribov, ``{\it Orsay lectures on confinement},'' preprint {92/60, 94/20},
  LPTHE, Orsay.

\bibitem{Yang54}
{C.N.~Yang, R.L.~Mills},  {\it Phys. Rev.}, vol.~69, p.~191, 1954.

\bibitem{Frit73}
{H.~Fritzsch, M.~Gell-Mann, H.~Leutwyler},  {\it Phys. Lett.}, vol.~B 74,
  p.~365, 1973.

\bibitem{Wein73}
S.~Weinberg,  {\it Phys. Rev. Lett.}, vol.~31, p.~494, 1973.

\bibitem{Gros73}
{D.J.~Gross, F.~Wilczek},  {\it Phys. Rev. Lett.}, vol.~30, p.~1343, 1973.

\bibitem{Polit73}
H.D.~Politzer,  {\it Phys. Rev. Lett.}, vol.~30, p.~1346, 1973.

\bibitem{Colem73}
{S.~Coleman, D.J.~Gross},  {\it Phys. Rev. Lett.}, vol.~31, p.~851, 1973.

\bibitem{Marc78}
{W.~Marciano, H.~Pagels},  {\it Phys. Rep.}, vol.~36, p.~137, 1978.

\bibitem{Bura80}
A.J.~Buras,  {\it Rev. Mod. Phys.}, vol.~52, p.~199, 1980.

\bibitem{Mand80}
S.~Mandelstam,  {\it Phys. Rep.}, vol.~67, p.~109, 1980.

\bibitem{Band81}
M.~Bander,  {\it Phys. Rep.}, vol.~75, no.~4, p.~205, 1981.

\bibitem{tHoo74}
G.'t~Hooft,  {\it Nucl. Phys.}, vol.~B 75, p.~461, 1974.

\bibitem{Poly75}
A.M.~Polyakov,  {\it Phys. Lett.}, vol.~B 59, p.~82, 1975.

\bibitem{Poly78}
A.M.~Polyakov,  {\it Phys. Lett.}, vol.~B 72, p.~477, 1978.

\bibitem{Poly87}
A.M.~Polyakov, {\it Gauge Fields and Strings}.
\newblock Chur: Harwood Acad. Publ., 1987.

\bibitem{Koga942}
{I.I.~Kogan, A.~Kovner}, ``{\it Compact $QED_{3}$ - a simple example of a
  variational calculation in a gauge theory.},'' preprint {UMN-TH-1317-94,
  TPI-MINN-94/37-T, hep-th@xxx/9410067}.

\bibitem{Amar90}
{M.G.~Amaral, M.E.~Pol},  {\it J. Phys.}, vol.~G 16, p.~1, 1990.

\bibitem{Itzy80}
{C.~Itzykson, J.-B.~Zuber}, {\it Quantum Field Theory}.
\newblock {McGraw--Hill, USA}, 1980.

\bibitem{Chen84}
{T.-P.~Cheng, L.-F.~Li}, {\it Gauge theories of elementary particles physics}.
\newblock {Clarendon Press, Oxford}, 1984.

\bibitem{Namb74}
Y.~Nambu,  {\it Phys. Rev.}, vol.~D 10, p.~4262, 1974.

\bibitem{Jacob74}
{M.~Jacob, ed.}, {\it Dual theory.}
\newblock {North--Holland, Amsterdam}, 1974.

\bibitem{Mande76}
S.~Mandelstam,  {\it Phys. Rep.}, vol.~C 23, p.~245, 1976.

\bibitem{Mande79}
S.~Mandelstam, in {\it {In: Proc.\ 1979 Int.\ Sym.\ on Lepton and Photon
  Interactions at High Enrgies, {\rm Fermilab, Batavia, Illinois, 1979}}},
  (H.~T.B.W.~Kirk, ed.).

\bibitem{THoo76}
G.'t~Hooft,  {\it Nucl. Phys.}, vol.~B 79, p.~276, 1976.

\bibitem{THoo79}
G.'t~Hooft,  {\it Nucl. Phys.}, vol.~B 153, p.~141, 1979.

\bibitem{Mand79}
S.~Mandelstam,  {\it Phys. Rev.}, vol.~D 19, p.~2391, 1979.

\bibitem{Maeda90}
{S.~Maedan, Y.~Matsubara, T.~Suzuki},  {\it Progr. Theor. Phys.}, vol.~84,
  no.~1, p.~130, 1990.

\bibitem{Monde92}
{H.~Monden et al},  {\it Phys. Lett.}, vol.~B 294, p.~100, 1992.

\bibitem{Suzu93}
T.~Suzuki,  {\it Phys. Rev. (Proc. Suppl.)}, vol.~B 30, p.~276, 1993.

\bibitem{Savv77}
G.~Savvidy,  {\it Phys. Lett.}, vol.~B 71, p.~133, 1977.

\bibitem{Matin78}
{S.G.~Matinyan, G.~Savvidy},  {\it Nucl. Phys.}, vol.~B 134, p.~539, 1978.

\bibitem{Niels78}
{N.K.~Neilson, P.~Olesen},  {\it Nucl. Phys.}, vol.~B 144, p.~376, 1978.

\bibitem{Niels79}
{H.B.~Neilson, M.~Ninomiya},  {\it Nucl. Phys.}, vol.~B 156, p.~1, 1979.

\bibitem{Ambj79}
{J.~Ambj\o rn, N.K.~Neilson, P.~Olesen},  {\it Nucl. Phys.}, vol.~B 152,
  p.~75, 1979.

\bibitem{Niels792}
{H.B.~Neilson, P.~Olesen},  {\it Nucl. Phys.}, vol.~B 160, p.~380, 1979.

\bibitem{Simon88}
Yu.A.~Simonov,  {\it Nucl. Phys.}, vol.~B 307, p.~512, 1988.

\bibitem{Calla78}
{G.G.~Callan, R.F.~Dashen, D.J.~Gross},  {\it Phys. Rev.}, vol.~D 17,
  p.~2717, 1978.

\bibitem{Calla791}
{G.G.~Callan, R.F.~Dashen, D.J.~Gross},  {\it Phys. Rev.}, vol.~D 19,
  p.~1826, 1979.

\bibitem{Calla792}
{G.G.~Callan, R.F.~Dashen, D.J.~Gross},  {\it Phys. Rev.}, vol.~D 20,
  p.~3279, 1979.

\bibitem{Witt79}
E.~Witten,  {\it Nucl. Phys.}, vol.~B 149, p.~285, 1979.

\bibitem{Mand79ds}
S.~Mandelstam,  {\it Phys. Rev.}, vol.~D 20, p.~3223, 1979.

\bibitem{Anish79}
{R.~Anishetty, M.~Baker, J.S.~Ball et al},  {\it Phys. Lett.}, vol.~B 86,
  p.~52, 1979.

\bibitem{Ball80}
{J.S.~Ball, F.~Zach},  {\it Phys. Lett.}, vol.~B 95, p.~273, 1980.

\bibitem{Oehme89}
R.~Oehme,  {\it Phys. Lett.}, vol.~B 232, p.~489, 1989.

\bibitem{Oehme90}
R.~Oehme,  {\it Phys. Rev.}, vol.~D 42, p.~4209, 1990.

\bibitem{Oehme93}
R.~Oehme,  {\it Mod. Phys. Lett.}, vol.~8, p.~1533, 1993.

\bibitem{Migd83}
A.~A.~Migdal,  {\it Phys. Rep.}, vol.~102, no.~4, p.~199, 1983.

\bibitem{Make79}
{Yu.~Makeenko, A.A.~Migdal},  {\it Phys. Lett.}, vol.~B 88, p.~135, 1979.

\bibitem{Gerv79}
{J.L.~Gervais, A.~Neveu},  {\it Phys. Lett.}, vol.~B 80, p.~255, 1979.

\bibitem{rusIros90}
G. S. Irosnikov, {\it JETP} (Journal of Experimental and Theoretical Physics, Russia), vol. 97, p. 424, 1990. 

\bibitem{rusIros91}
G. S. Irosnikov, {\it JETP} (Journal of Experimental and Theoretical Physics, Russia), vol. 100, p. 45, 1991.

\bibitem{tHoo74N}
G.'t~Hooft,  {\it Nucl. Phys.}, vol.~B 72, p.~461, 1974.

\bibitem{Witt79N}
E.~Witten,  {\it Nucl. Phys.}, vol.~B 160, p.~57, 1979.

\bibitem{Kugo793}
{T.~Kugo, I.~Ojima},  {\it Progr. Theor. Phys.}, vol.~66, p.~1, 1979.

\bibitem{Kugo78}
{T.~Kugo, I.~Ojima},  {\it Progr. Theor. Phys.}, vol.~60, no.~6, p.~1869,
  1978.

\bibitem{Kugo792}
{T.~Kugo, I.~Ojima},  {\it Progr. Theor. Phys.}, vol.~61, no.~2, p.~644,
  1979.

\bibitem{Hata93}
{H.~Hata, I.~Niigata},  {\it Nucl. Phys.}, vol.~B 389, p.~133, 1993.
\newblock hep-th/9405145.

\bibitem{Shinta84}
M.~Shintani,  {\it Phys. Lett.}, vol.~B 137, p.~220, 1984.

\bibitem{Kugo791}
{T.~Kugo, I.~Ojima},  {\it Progr. Theor. Phys.}, vol.~61, p.~294, 1979.

\bibitem{Wils74}
K.~Wilson,  {\it Phys. Rev.}, vol.~D 10, p.~2445, 1974.

\bibitem{Kogu75}
{J.~Kogut, L.~Susskind},  {\it Phys. Rev.}, vol.~D 11, p.~395, 1975.

\bibitem{Sussk79}
L.~Susskind,  {\it Phys. Rev.}, vol.~D 20, p.~2610, 1979.

\bibitem{Creu802}
M.~Creutz,  {\it Phys. Rev.}, vol.~D 21, no.~8, p.~2308, 1980.

\bibitem{Lerr81}
{L.~McLerran, B.~Svetitsky},  {\it Phys. Lett.}, vol.~B 98, p.~195, 1981.

\bibitem{Kuti81}
{K.~Kuti, J.~Polonyi, K.~Szlachanyi},  {\it Phys. Lett.}, vol.~B 98,
  p.~199, 1981.

\bibitem{Enge81}
{J.~Engels, F.~Karsch, H.~Satz et al},  {\it Phys. Lett.}, vol.~B 101,
  p.~89, 1981.

\bibitem{Enge82}
{J.~Engels, F.~Karsch, H.~Satz},  {\it Nucl. Phys.}, vol.~B 205, p.~545,
  1982.

\bibitem{Mari89pr}
E.~Marinari,  {\it Phys. Rep.}, vol.~184, no.~2-4, p.~131, 1989.

\bibitem{Svet86}
B.~Svetitsky,  {\it Phys. Rep}, vol.~132, no.~1, p.~1, 1986.

\bibitem{Svet82}
{B.~Svetitsky, L.~Yaffe},  {\it Nucl. Phys.}, vol.~B 210, no.~[FS6],
  p.~423, 1982.

\bibitem{Svet822}
{B.~Svetitsky, L.~Yaffe},  {\it Phys. Rev.}, vol.~D 26, p.~963, 1982.

\bibitem{Jaco86}
{S.~Jacobs, M.~G.~Olson, C.~Suchyta~III},  {\it Phys. Rev.}, vol.~D 33,
  p.~3338, 1986.

\bibitem{Mari89}
E.~Marinari,  {\it Nucl. Phys. (Proc. Suppl.)}, vol.~B, no.~9, p.~209,
  1989.

\bibitem{Kars92}
F.~Karsch, ``{\it Deconfinement and chiral symmetry restoration on the lattice},'' in
  {\sl {QCD 20 Years Later, {\rm 9-13 Jun 1992, Aachen}}}, ({P.M.~Zerwas,
  H.A.~Kastrup}, ed.), pp.~717--747, {World Scientific, Singapore}, 1993.

\bibitem{Koga941}
{I.I.~Kogan, A.~Kovner}, ``{\it A variational approach to the QCD wave functional:
  Dynamical mass generation and confinement.},'' preprints {LA-UR-94-2727,
  PUPT-1492, hep-th@xxx/9408081}.

\bibitem{rusSvesVar}
N. A. Sveshnikov, 1995 (unpublished).

\bibitem{Khve94}
{A.~Khvedelidze, V.~Pervushin}, ``{\it Zero mode of Gauss constraint in gaugeless
  reduction of Yang--Mills theory},'' preprint {JINR E2-94-332}.

\bibitem{Lunev92}
F.A.~Lunev,  {\it Phys. Lett.}, vol.~B 295, p.~99, 1992.

\bibitem{Lavel93}
{M.~Lavelle, D.~McMullan}, ``{\it On quark confinement},'' preprints {MZ-TH/93-03,
  DIAS-STP-93-04}.

\bibitem{Dahl85}
K.J.~Dahlem,  {\it Z. Phys.}, vol.~C 29, p.~553, 1985.

\bibitem{Mandul88}
J.E.~Mandula,  {\it Phys. Lett.}, vol.~B 210, p.~117, 1988.

\bibitem{Anish84}
R.~Anishetti,  {\it J. Phys.}, vol.~G 10, p.~423, 1984.

\bibitem{Belya90}
{V.M.~Belyaev, V.L.~Eletzky},  {\it Z. Phys.}, vol.~C 45, p.~355, 1990.

\bibitem{Enqv90}
{K.~Enqvist, K.~Kajantie},  {\it Z. Phys.}, vol.~C 47, p.~291, 1990.

\bibitem{Polon90}
{J.~Polonyi, S.~Vazques},  {\it Phys. Lett.}, vol.~B 240, p.~183, 1990.

\bibitem{Belya91}
V.M.~Belyaev,  {\it Phys. Lett.}, vol.~B 254, p.~153, 1991.

\bibitem{rusSkal92}
V. V. Skalozub, preprint IP-92-12E, Institute of Theoretical Physics, Kiev, 1992. 

\bibitem{Fock37}
V.~A.~Fock,  {\it Sow. Phys}, vol.~12, no.~HEFT 4, p.~404, 1937.

\bibitem{Schw51}
J.~Schwinger,  {\it Phys. Rev.}, vol.~82, p.~664, 1951.

\bibitem{Dura82}
{L.~Durand, E.~Mendel},  {\it Phys. Rev.}, vol.~D 26, p.~1368, 1982.

\bibitem{Shif80}
M.~Shifman,  {\it Nucl. Phys.}, vol.~B 173, p.~13, 1980.

\bibitem{Shut89}
D.~Sh\mbox{\"{u}}tte,  {\it Phys. Rev.}, vol.~D 40, p.~2090, 1989.

\bibitem{SveTim92PL}
{N.A.~Sveshnikov, E.G.~Timoshenko},  {\it Phys. Lett.}, vol.~B 289,
  p.~423, 1992.

\bibitem{SveTim91Pr}
{N.A.~Sveshnikov, E.G.~Timoshenko}, ``{\it Confinement phase transition mechanism
  of SU(2)-gluodynamics},'' preprint 91-140, IHEP, Serpukhov, 1991.

\bibitem{Tim94Pro}
E.G.~Timoshenko, ``{\it Dynamics of gauge fields in finite domain,}'' in {\sl {VIII
  Workshop on High Energy Physics and QFT, {\rm 17-23 Sep 1993, Zvenigorod}}},
  (B.B.~Levchenko, ed.), (Moscow), pp.~213--217, NPI MSU, MSU Publishing Co.,
  1994.

\bibitem{Morc85}
{G.~Morchio, F.~Strocchi},  {\it Commun. Math. Phys.}, vol.~99, p.~153,
  1985.

\bibitem{Morc87}
{G.~Morchio, F.~Strocchi},  {\it J. Math. Phys.}, vol.~28, no.~8, p.~1912,
  1987.

\bibitem{Morc87jmp}
{G.~Morchio, F.~Strocchi},  {\it J. Math. Phys}, vol.~28, no.~3, p.~622,
  1987.

\bibitem{rusZakh68}
V. E. Zakharov, {\it Zh. Prikl. Mekh. Tekhn. Fiz.} (Russia) vol. 2, p. 86, 1968. 

\bibitem{Busl86}
{V.S.~Buslaev, L.D.~Faddeev, L.A.~Takhtajan},  {\it Phys.}, vol.~D 18,
  p.~255, 1986.

\bibitem{Lewi86}
{D.~Lewis, J.~Marsden, R.~Montgomery and T.~Ratiu},  {\it Physica}, vol.~D
  18, p.~391, 1986.

\bibitem{Regg74}
{T.~Regge, C.~Teitelboim},  {\it {Ann. Phys. (N.Y.)}}, vol.~88, p.~286,
  1974.

\bibitem{Solov92}
V.O.~Soloviev,  {\it Phys. Lett.}, vol.~B 292, p.~30, 1992.

\bibitem{Bala92}
{A.P.~Balachandran, G.~Bimonte, K.S.~Gupta, A.~Stern},  {\it Int. J. Mod.
  Phys.}, vol.~A 7, p.~4655, 1992.

\bibitem{Solov93}
V.O.~Soloviev,  {\it J. Math. Phys.}, vol.~34, p.~5747, 1993.

\bibitem{Kiri76}
A.~A.~Kirillov, {\it Elements of the theory of representations}.
\newblock NY: Springer, 1976.

\bibitem{Emch72}
G.G.~Emch, {\it Algebraic methods in Statistical Mechanics and Quantum Field
  Theory}.
\newblock Wiley--Interscience, 1972.

\bibitem{Brat79}
{O.~Brately, D.W.~Robinson}, {\it Operator Algebras and Quantum Statistical
  Mechanics}.
\newblock Vol.~1,2, N. Y.: Springer Verlag, 1979, 1981.

\bibitem{Bona89}
P.~Bona,  {\it J. Math. Phys.}, vol.~30, no.~12, p.~2994, 1989.

\bibitem{Best90}
{P.~Besting, D.~Sch\mbox{\"{u}}tte},  {\it Phys. Rev.}, vol.~D 42, p.~594,
  1990.

\bibitem{SveTim95IJMP}
{N.A.~Sveshnikov, E.~G.~Timoshenko}, ``Boundary effects in the gauge field
  theories at finite temperature,'' 1995.
\newblock {\it Int. J. Mod. Phys.}, (to be published). \\
Consequently appeared in:  {\it Phys. Rev.} vol. D~58, No 8, pp. 085024 -085024-9 (1998).

\bibitem{Glim81}
{J.~Glimm, A.~Jaffe}, {\it Quantum Physics. A functional Integral Point of
  View}.
\newblock N. Y.: Springer--Verlag, 1981.

\bibitem{SveTim92Pr}
{N.A.~Sveshnikov, E.G.~Timoshenko}, ``{\it Confinement phase transition mechanism in
  gluodynamics,}'' preprint 92-31, IHEP, Serpukhov, 1992.

\bibitem{Meza87}
{M.~Mezard, G.~Parisi, M.A.~Virasoro}, {\it Spin Glass Theory and Beyond}.
\newblock Singapore: World Scientific, 1987.

\bibitem{Binn92}
{J.J.~Binney, N.J.~Dowrick, A.J.~Fisher, M.E.~Newman}, {\it The theory of
  critical phenomena}.
\newblock {Clarendon Press, Oxford}, 1992.

\bibitem{Itzy89}
{C.~Itzykson, J.-M.~Drouffe}, {\it Statistical Field Theory}.
\newblock {Cambridge Univ. Press}, 1989.

\bibitem{Amit78}
D.J.~Amit, {\it Field Theory, Renormalisation Group and Critical Phenomena}.
\newblock {World Scientific, Singapore}, 1978.

\bibitem{rusTim93}
E.G. Timoshenko. ``{\it Algebraic confinement and Wilson loop in SU(N) gluodynamics.}''{\sl Phys. Atom. Nucl.} [Journal of Nuclear Physics (Russia), pp. 277-284], vol. 56, No 11, pp. 1613-1616, 1993.

\bibitem{Tim92NPI}
E.G.~Timoshenko, ``{\it Wilson Loop for SU(N)-gluodynamics},'' preprint 92-17/266,
  NPI MSU, Moscow, 1992.

\bibitem{Bass84}
{A.~Basseto, I.~Lazzizera, R.~Soldati},  {\it Nucl. Phys.}, vol.~B 236,
  p.~319, 1984.

\bibitem{Prud83}
A.P. Prudnikov, Yu. A. Brychkov, O.I. Marichev, {\it Integrals and series.} Transl.  by N.M. Queen.  N. Y.: Gordon and Breach Science Publishers,1986.

\bibitem{Bate53}
{H.~Bateman, A.~Erdelyi}, {\it Higher Transcedental Functions}.
\newblock Vol.~1, MC Graw--Hill, 1953.

\bibitem{TimProtv1}
{N.A.~Sveshnikov, E.G.~Timoshenko}, {\it Problems on High Energy
Physics and Field Theory} -- Proceedings of the XV Workshop, IHEP,
Protvino, 1995, No.~162.

\bibitem{TimProtv2}
{E.G.~Timoshenko}, {\it Problems on High Energy
Physics and Field Theory} -- Proceedings of the XV Workshop, IHEP,
Protvino, 1995, No.~169.

\end{thebibliography}
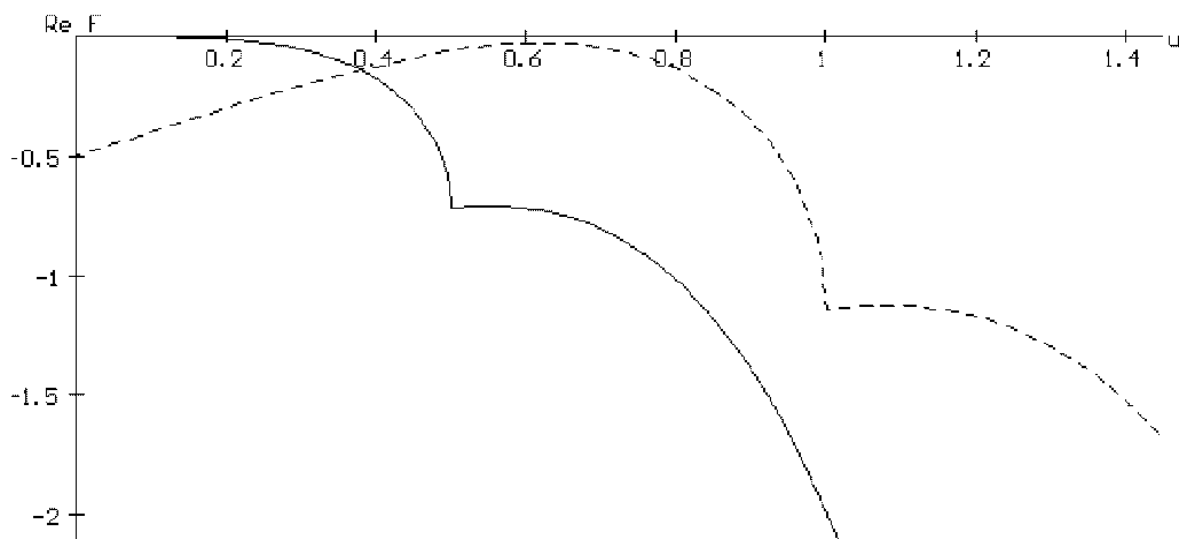
\end{document}